\documentclass[%
reprint,
superscriptaddress,
preprintnumbers,
amsmath,amssymb,
aps,
rmp,
]{revtex4-2}

\makeatletter
\let\NAT@sort\z@
\let\NAT@cmprs\z@
\makeatother

\usepackage[utf8]{inputenc}
\usepackage[english]{babel}
\usepackage{graphicx}
\usepackage{dcolumn}
\usepackage{bm}
\usepackage{multirow, array, tabularx, booktabs, makecell}
\usepackage{physics}
\usepackage{amsthm}
\usepackage[dvipsnames]{xcolor}
\usepackage[normalem]{ulem}
\usepackage{enumitem}
\usepackage{rotating}
\usepackage{textcomp}
\usepackage[colorlinks=true,linkcolor=blue,citecolor=blue,filecolor=blue,urlcolor=blue]{hyperref}
\usepackage{epigraph}
\usepackage{amsmath}
\usepackage{amssymb}
\usepackage{tcolorbox}
\usepackage{physics,dsfont}
\definecolor{darkviolet}{rgb}{0.58, 0.0, 0.83}
\definecolor{mygreen}{rgb}{0.0, 0.5, 0.0}
\usepackage{pifont}
\newcommand{\cmark}{\ding{52}}
\newcommand{\xmark}{\ding{56}}

\begin{document}
%
\preprint{\texttt{APCTP Pre2026 - 007, IFT-UAM/CSIC-26-117}}

%
\title{Quantum Chaos and Spread of States in Krylov Subspace: A Topical Review}

\author{Hyun-Sik Jeong}
\affiliation{\mbox{Asia Pacific Center for Theoretical Physics, Pohang 37673, Korea}}
\affiliation{\mbox{Department of Physics, Pohang University of Science and Technology, Pohang 37673, Korea}\vspace{0.03cm}}

\author{Juan F. Pedraza\vspace{0.3cm}}
\affiliation{\mbox{Instituto de F\'isica Te\'orica UAM/CSIC, Calle Nicol\'as Cabrera 13-15, 28049 Madrid, Spain}}

\begin{abstract}
Krylov state complexity, or spread complexity, has emerged as a sharp and versatile diagnostic of quantum chaos, information spreading, and many-body dynamics. Built from the Lanczos algorithm and grounded in the optimal-basis theorem, Krylov complexity thereby provides a robust spectroscopic window into quantum dynamics. A central theme is the characteristic overshoot observed in chaotic systems: a complexity peak in which chaotic evolution drives the state deeper into the Krylov chain than in integrable systems before relaxing to equilibrium. This behavior, tied to random-matrix universality classes of spectral statistics, is illustrated across a broad range of models, including quantum billiards, quantum spin chains, and variants of the SYK model. We also discuss proposed holographic descriptions of Krylov complexity in Einstein gravity, and conclude by outlining future directions and open problems, including time-dependent systems and quantum-field-theoretic formulations. A \texttt{Mathematica} notebook is provided for numerical exploration of Krylov complexity and spectral statistics across models.
\end{abstract}

\maketitle
\tableofcontents

%
\section{Introduction}\label{}
\begingroup
\begin{flushleft}
{\renewcommand{\arraystretch}{1.5} 
\begin{tabularx}{\columnwidth}{@{} p{0.29\columnwidth} X @{}}
\multirow{1}{*}{\includegraphics[width=\linewidth]{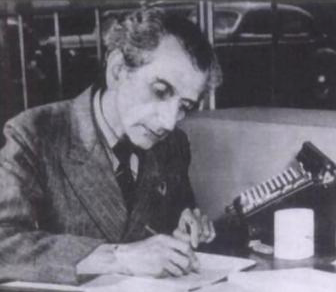}} & 
\small \textit{\underline{Cornelius Lanczos}}: To obtain a solution in very few steps means nearly always that one has found a way that does justice to the inner nature of the problem. \\
& \hfill \footnotesize -- March 9, 1947 \\ [1.8ex]
\multirow{1}{*}{\includegraphics[width=\linewidth,trim={0 1.6cm 0 1.5cm},clip]{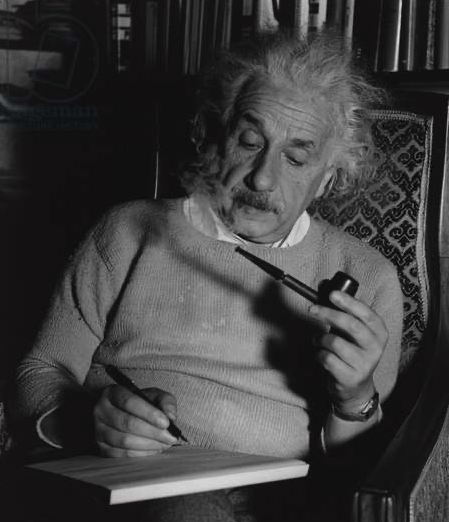}} & 
\small \textit{\underline{Albert Einstein}}: Your remark on the importance of adapted approximation methods makes very good sense to me, and I am convinced that this is a fruitful mathematical aspect, and not just a utilitarian one.\vspace{-2mm} \\
& \hfill \footnotesize -- March 18, 1947 \\
\end{tabularx}
\vspace{2mm}
\hrule
\vspace{1mm}
\hfill \footnotesize \textit{Correspondence between C. Lanczos and A. Einstein}}
\end{flushleft}
\endgroup
\vspace{-5mm}
%
\subsection{Many-body chaos in quantum systems}
Quantum chaos is a ubiquitous phenomenon with far-reaching implications, from the foundations of statistical mechanics to the cutting edge of quantum information theory~\cite{Gutzwiller_1990,GUHR1998189,2000DanielBraun,hilborn2000chaos,Haake_2018}. It provides a theoretical framework for thermalization in many-body systems and a powerful language for information scrambling, especially in high-energy physics and black hole dynamics. Yet, while classical chaos is sharply diagnosed by sensitive dependence on initial conditions, quantum many-body chaos remains harder to characterize. Its diagnosis requires complementary tools, combining spectral statistics with probes of dynamical growth.

The traditional characterization of quantum chaos is rooted in the Bohigas-Giannoni-Schmit (BGS) conjecture~\cite{Bohigas:1983er,Guhr:1997ve,Bohigas}, which posits a deep connection between classically chaotic systems and random matrix theory (RMT)~\cite{Meh2004}. According to this paradigm, the spectra of chaotic quantum systems exhibit universal fluctuations described by the Gaussian orthogonal, unitary, or symplectic ensembles, manifesting in characteristic level repulsion and spectral rigidity. In many-body systems, the typical level spacing is exponentially small in system size. For instance, in matrix-like systems with $N^2$ degrees of freedom, a thermal spectral window at inverse temperature $\beta$ has a typical spacing $\Delta E\sim \beta^{-1} e^{-N^2}$. Spectral probes of chaos therefore resolve correlations on very long Heisenberg time scales,
$\Delta t \sim t_H\sim \beta\, e^{N^2}$,
encoding the late-time structure of chaotic dynamics~\cite{Bohigas:1983er,Berry1985-mx,Muller:2004nb}.

By contrast, early-time quantum chaos is often diagnosed through the exponential growth of out-of-time-order correlators (OTOCs)~\cite{larkin1969quasiclassical,berman1978condition}, which occurs for time scales shorter than the scrambling time, $\Delta t\ll t_*\sim \beta\, \log N$. This growth is governed by a quantum Lyapunov exponent, bounded under broad assumptions by $\lambda_L \leq 2\pi k_B T/\hbar$, the Maldacena-Shenker-Stanford (MSS) bound~\cite{Maldacena_2016}. While spectral statistics and OTOCs serve as central diagnostics of quantum chaos, the precise connection between these early-time and late-time diagnostics remains a major open question, lying at the intersection of chaos, thermalization, and the fundamental limits of quantum dynamics.

Large-$N$ systems with a semiclassical description provide a particularly clear arena where progress has been made. In such systems, factorization and the large number of degrees of freedom generate a parametric hierarchy between time scales, allowing both diagnostics to be studied as distinct but related manifestations of chaos. Notably, some theories in this class saturate the MSS bound and are thus among the fastest scramblers of information in nature, a property famously realized by black holes~\cite{Shenker:2013pqa,Roberts:2014isa,deBoer:2017xdk}. Consequently, saturation of the bound has become a powerful diagnostic of holographic behavior~\cite{Perlmutter:2016pkf}, helping identify quantum systems that may admit a gravitational dual description within the Anti-de Sitter/Conformal Field Theory (AdS/CFT) correspondence~\cite{Maldacena:1997re,Witten:1998qj}. From the bulk perspective, the growth of OTOCs can be understood in terms of high-energy scattering of quanta near black hole horizons~\cite{Shenker:2014cwa}, which in turn imposes sharp constraints on the analytic structure of holographic correlators~\cite{Grozdanov:2017ajz,Blake:2018leo,Ahn:2025exp}.

Late-time probes reveal a complementary, spectral side of holographic chaos. In finite-volume holographic systems, the black hole microstate spectrum is expected to exhibit random-matrix correlations, diagnosed by the dip-ramp-plateau structure of the spectral form factor~\cite{Cotler:2016fpe}. The ramp encodes long-range spectral rigidity, while the plateau reflects the discreteness of the underlying spectrum. A semiclassical gravitational account of the ramp has been proposed in terms of double-cone saddles~\cite{Saad:2018bqo}. This perspective becomes especially sharp in two-dimensional Jackiw-Teitelboim (JT) gravity, where the gravitational path integral is related to a double-scaled random matrix ensemble~\cite{Saad:2019lba,Stanford:2019vob}. Together, these early- and late-time diagnostics position quantum chaos as a central organizing principle for understanding the microscopic dynamics of information scrambling and the emergence of spacetime geometry.

Despite these successes, standard diagnostics of quantum chaos remain incomplete. Spectral statistics are powerful but intrinsically tied to very long Heisenberg time scales, while OTOCs are most directly applicable in semiclassical large-$N$ regimes or in systems with a clean separation of scales. In generic finite-size quantum systems, and especially in systems without a classical limit, the relation between spectral universality, dynamical growth, and information spreading is much less transparent. This motivates the search for universal dynamical diagnostics with broader applicability.

This need becomes even sharper in open quantum systems, where coupling to an environment often leads to effective non-Hermitian dynamics~\cite{Ashida:2020dkc,Rivas_2012}. In such systems, the spectrum is generically complex, and the usual diagnostics based on real energy levels no longer apply~\cite{Grobe:1988zz,Grobe:1989aa}. The Grobe-Haake-Sommers (GHS) conjecture~\cite{Grobe:1988zz} provides a non-Hermitian analogue of the BGS framework, proposing that chaotic open systems exhibit Ginibre-type spectral statistics~\cite{Ginibre:1965zz} rather than standard Wigner-Dyson behavior~\cite{BerryTabor,Bohigas:1983er}. In this broader classification, Hermitian symmetry classes are mapped to non-Hermitian counterparts, including the AI$^\dagger$ and AII$^\dagger$ classes, while class A is described by the standard complex Ginibre ensemble~\cite{Hamazaki:2020kbp}. These ideas have proven useful in a variety of non-Hermitian many-body systems~\cite{Hamazaki:2018aa,Akemann:2019ab,Sa:2020ab,Li:2021aa,Garcia-Garcia:2021rle,Shivam:2023aa}.

Nevertheless, the GHS conjecture is not universally applicable, particularly when the relation between classical and quantum chaos is subtle~\cite{PhysRevLett.133.240404}. Recent studies indicate that Ginibre statistics alone may fail to capture the onset of chaos in certain dissipative systems with well-defined classical limits, suggesting that the classical-quantum correspondence requires a more refined treatment in the presence of non-Hermitian dynamics~\cite{Villasenor:2025eaz,Mondal:2025qdd,Rufo:2025mop,PhysRevResearch.7.013276}. This underscores the need for refined probes, such as the complex spacing ratio (CSR)~\cite{Sa:2020ab}, which extends the standard spacing-ratio statistic~\cite{Atas:2013aa} to complex spectra by probing radial level repulsion and angular correlations.

Taken together, the above developments point toward a central goal in modern quantum chaos: the identification of a universal, or at least broadly applicable, dynamical diagnostic of information spreading. Such a diagnostic should apply to both few-body and many-body systems, work with or without a classical limit, remain meaningful in Hermitian and non-Hermitian settings, align with established spectral signatures of chaos, and admit a natural interpretation in holographic systems. It should also go beyond purely spectral information by directly characterizing how quantum states or operators explore Hilbert space under time evolution.

Krylov state complexity, also known as spread complexity~\cite{Balasubramanian:2022tpr}, has recently emerged as a particularly promising candidate for this role. Built from the Lanczos algorithm and grounded in the optimal-basis theorem, it quantifies the spread of a quantum state along a dynamically generated Krylov chain. In doing so, it provides a bridge between spectral data and real-time quantum evolution, while remaining applicable to finite-dimensional systems where traditional large-$N$ probes may be less effective. Related notions, such as Krylov operator complexity~\cite{Parker:2018yvk,Caputa:2024vrn}, further connect this framework to operator growth, scrambling, and thermalization. These features make Krylov state complexity the central focus of this pedagogical review, in which we use elementary quantum-mechanical models of chaos as a starting point before turning to many-body systems, non-Hermitian dynamics, and holographic descriptions.

%
\subsection{Quantum complexity in Krylov subspace}

The notion of ``complexity'' in quantum physics is intuitive but not unique. Historically, it has been associated with the difficulty of preparing a state or implementing a unitary transformation, but different physical questions lead to different definitions; see Refs.~\cite{Baiguera:2025dkc,Chapman:2021jbh} for comprehensive reviews. In many-body physics and quantum information, complexity often aims to quantify how far a state or operator moves away from its initial configuration under time evolution. In this sense, it provides a dynamical measure of how an initial state $\ket{\psi_0}=\ket{\psi(t=0)}$ explores the available Hilbert space as information spreads across an increasing number of degrees of freedom.

To make this intuition precise, one must choose a way of organizing the Hilbert space into configurations of increasing complexity. Given an ordered orthonormal basis $\{\ket{\mathcal{B}_n}\}$, the index $n$ can be interpreted as a complexity label, and the complexity of a time-evolved state $\ket{\psi(t)}$ can be defined as the expectation value of this label. Equivalently, one can measure the average position, or spread, of the wave function in the chosen basis. The usefulness of this construction, however, depends crucially on the basis itself. This basis dependence is familiar from other notions of quantum complexity, including circuit and Nielsen complexity~\cite{Nielsen_2012}, where the result depends on various physical choices such as gate sets, cost functions, and tolerance thresholds. A more intrinsic diagnostic should instead be derived directly from the dynamics of the system.

The Krylov subspace provides precisely such a construction~\cite{Parker:2018yvk,Balasubramanian:2022tpr}. Originally introduced by Aleksey Krylov in the context of large-scale linear systems and eigenvalue problems~\cite{krylov1931}, the Krylov subspace associated with a Hamiltonian $H$ and an initial state $\ket{\psi_0}$ is generated by repeated action of $H$, namely by the sequence $\{\ket{\psi_0},H\ket{\psi_0},H^2\ket{\psi_0},\ldots,H^{n-1}\ket{\psi_0}\}$. Physically, this is the dynamically adapted subspace for Schrödinger evolution: it is the smallest subspace that captures the trajectory generated by the Hamiltonian from the chosen initial state. Because it is built from $H$ and $\ket{\psi_0}$ themselves, it provides an intrinsic coordinate system for tracking how the state spreads under time evolution.

In practice, the Krylov sequence must be converted into an orthonormal basis. A direct Gram-Schmidt procedure~\cite{Gram1883,Schmidt1908} is possible in principle, but it is computationally expensive in many-body systems, both because the Hilbert space is large and because each new vector must be orthogonalized against all previous ones. In contrast, for Hermitian Hamiltonian evolution, the Lanczos algorithm~\cite{Lanczos:1950zz} provides a far more efficient construction. It recasts the orthogonalization procedure as a three-term recurrence relation, recursively generating the Krylov basis $\{\ket{K_n}\}$.

This Lanczos recursion has a simple physical interpretation: it effectively maps the original many-body dynamics onto a one-dimensional semi-infinite tight-binding chain~\cite{Mattis_1981}, commonly called the Krylov chain. In this representation, the time-evolved state behaves as a wave packet hopping along the chain. The Krylov index $n$ labels the distance from the initial state and therefore serves as a natural proxy for complexity. Krylov state complexity, also known as spread complexity, is then defined as the average position of the wave packet along this chain, thereby quantifying how far the initial state has spread in this dynamically generated basis.

The Krylov basis is not merely convenient; it is optimal in a precise sense. The optimal-basis theorem~\cite{Balasubramanian:2022tpr} states that, among ordered orthonormal bases, the Krylov basis minimizes the spread of the time-evolved state, making Krylov complexity the minimal complexity within this class of basis-dependent spread measures. Thus, the resulting complexity does not reflect an arbitrary coordinate choice, but rather the intrinsic structure of Hamiltonian evolution.

The same Krylov machinery appears in two closely related settings. In the Schrödinger picture, repeated action of the Hamiltonian generates a Krylov basis for state evolution, leading to Krylov state complexity~\cite{Balasubramanian:2022tpr}. In the Heisenberg picture, repeated action of the Liouvillian generates a Krylov basis for operator evolution, leading to Krylov operator complexity~\cite{Parker:2018yvk}. Both constructions rely on the same mathematical framework of Krylov subspace and the Lanczos algorithm~\cite{krylov1931,Lanczos:1950zz,Viswanath_1994}, with connections between the two explored in~\cite{Caputa:2024vrn}. In this review, we focus primarily on Krylov state complexity, hereafter referred to simply as Krylov complexity. For comprehensive treatments of Krylov operator complexity, see Refs.~\cite{Nandy:2024evd,Rabinovici:2025otw}.

Krylov subspace methods provide a powerful language for modern many-body dynamics~\cite{Viswanath_1994,liesen2012krylov}. For quantum chaos, the central theme of this review, Krylov complexity has emerged as a dynamical diagnostic that bridges spectral statistics and OTOCs by tracking how states spread in real time. In chaotic systems, this spreading typically displays distinctive features, such as rapid growth, an overshoot or peak, and eventual saturation, while integrable systems often exhibit more constrained dynamics~\cite{Balasubramanian:2022tpr,Erdmenger:2023wjg,Camargo:2024deu,Baggioli:2024wbz,Huh:2024ytz}. The same framework is also useful computationally, where Krylov subspace methods underlie efficient numerical and quantum algorithms~\cite{Bharti:2021aa,Cortes:2022aa,Kirby2023exactefficient}.

This connection between real-time spreading and spectral diagnostics has also made Krylov complexity increasingly relevant in holography. On the one hand, it offers a dynamical probe of many-body chaos that can be compared directly with spectral statistics, OTOCs, and other measures of information spreading. On the other hand, several proposals relate Krylov complexity to gravitational observables, including proper momentum, wormhole growth, and effective radial dynamics in black hole geometries~\cite{Rabinovici:2023yex,Lin:2022rbf,Heller:2024ldz,Ambrosini:2024sre,Jian:2020qpp,Caputa:2021sib,Fu:2025kkh,Jeong:2026iac}. Lower-dimensional settings relating Jackiw-Teitelboim (JT) gravity~\cite{Jackiw:1984je,Teitelboim:1983ux} and its deformations to variants of the Sachdev-Ye-Kitaev (SYK) model~\cite{Sachdev:1992fk,Kitaev2015Talk}, provide particularly concrete examples in which information-theoretic growth can be compared with geometric notions of black-hole interior growth~\cite{Stanford:2014jda,Belin:2021bga,Myers:2024vve,Caceres:2025myu}. Krylov complexity therefore fits naturally into the broader holographic program relating quantum information, computation, and chaos with notions of spacetime emergence~\cite{Czech:2017ryf,Pedraza:2021mkh,Pedraza:2021fgp,Pedraza:2022dqi,Carrasco:2023fcj}. See Refs.~\cite{Chapman:2021jbh,Chen:2021lnq,Altland:2026tog} for comprehensive reviews on related subjects.

The remainder of this manuscript develops Krylov state complexity as a diagnostic of quantum chaos across a variety of models and holographic settings. Section~\ref{SECII} introduces the formal Krylov framework, including the Lanczos algorithm, the recursive construction of the Krylov basis, wave-function dynamics in Krylov space, survival amplitudes, and the relation to spectral statistics. Section~\ref{SECIII} presents the elementary models used throughout the review to benchmark chaos, ranging from random matrix ensembles and quantum billiards to spin chains and variants of the Sachdev-Ye-Kitaev model. Section~\ref{SECIV} then studies Krylov complexity in standard Hermitian quantum systems in detail, emphasizing characteristic chaos indicators such as the complexity peak, the Ehrenfest time, and the transition from integrability to chaos, while also developing connections between Lanczos coefficients and spectral probes. Section~\ref{SECV} extends the formalism to non-Hermitian quantum systems, introducing singular-value methods and the bi-Lanczos algorithm to address complex spectral statistics and test the robustness of Krylov complexity in such settings. Section~\ref{SECVI} summarizes concrete holographic applications of Krylov complexity, including its relation to wormhole growth, proper momentum in local quenches, and modern applications of the brick-wall framework in AdS/CFT. Section~\ref{SECVII} concludes with a summary of key lessons, open questions, and future directions.

We also provide a supplementary \texttt{Mathematica} notebook as an ancillary file to facilitate numerical exploration of the Lanczos algorithm, Krylov complexity, and spectral statistics across models.

%
\section{Krylov complexity: formalism}\label{SECII}

%
\subsection{Krylov basis and Lanczos algorithm}
Let us begin by reviewing the Krylov subspace method for approximating dynamics of quantum state at given Hamiltonian. Consider an initial state $\ket{\psi_0}:=\ket{\psi(0)}$ in a $D_{\mathcal H}$-dimensional Hilbert space $\mathcal{H}=\mathbb{C}^{D_{\mathcal H}}$, evolving under a time-independent Hamiltonian $H$:
\begin{align}\label{}
\begin{split}
\ket{\psi(t)} = e^{-i H t} \ket{\psi_0} \,,
\end{split}
\end{align}
which can be expressed as a linear combination of $\{\ket{\psi_0}, H \ket{\psi_0}, H^2 \ket{\psi_0}, \cdots\}$. In such a $D_{\mathcal H}$-dimensional Hilbert space, the direct exponentiation of the Hamiltonian becomes computationally prohibitive as the system size grows such as the many-body systems, i.e., ``curse of dimensionality". Nevertheless, the Krylov subspace method~\cite{krylov1931} provides a powerful workaround by approximating the dynamics within a smaller $D_{\rm K}$-dimensional subspace ($D_{\rm K} \leq D_{\mathcal H}$) that captures the essential dynamics of the initial state. 

The $D_{\rm K}$-dimensional Krylov subspace, $\mathcal{K}_{D_{\rm K}}(H, \ket{\psi_0})$, is defined as the span of the sequence of vectors generated by repeated applications of the Hamiltonian on the initial state such that
\begin{align}\label{}
\begin{split}
\mathcal{K}_{D_{\rm K}} = \text{span} \{ \ket{\psi_0} , H \ket{\psi_0} , \cdots, H^{D_{\rm K}-1} \ket{\psi_0} \} \,.
\end{split}
\end{align}
Physically, $\mathcal{K}_{D_{\rm K}}$ represents the effective region of the Hilbert space accessible to the state through the sequential action of the Hamiltonian. As time progresses, the probability amplitude ``spreads" from the initial state into higher-order Krylov vectors, effectively probing the structure of the Hilbert space connected to $\ket{\psi_0}$ via $H$.

To perform explicit calculations further, one requires an orthonormal basis $\mathcal{B}_{D_{\rm K}}= \{ \ket{K_0} , \ket{K_1} , \cdots, \ket{K_{D_{\rm K}-1}} \}$ for $\mathcal{K}_{D_{\rm K}}$. While a standard Gram-Schmidt process~\cite{Gram1883,Schmidt1908} could be used, the Lanczos algorithm~\cite{Lanczos:1950zz} exploits the hermiticity of $H$ to achieve this via a three-term recurrence relation: 
\begin{align}\label{RRC}
\begin{split}
b_{n+1} \ket{K_{n+1}} = \left(H-a_{n}\right)\ket{K_{n}} - b_{n}\ket{K_{n-1}} \,,
\end{split}
\end{align}
where $a_n= \bra{K_n}H\ket{K_n}$ and $b_n$ is the normalization constant: these are called Lanczos coefficients $\{a_n, b_n\}$. We discuss how the above three-term recurrence relation should be extended for the non-Hermitian systems in Section \ref{SECV}, i.e., the bi-Lanczos algorithm.
\begin{tcolorbox}[colback=gray!10, colframe=gray!50, title=\textbf{Lanczos algorithm}, coltitle=white]
Given an initial state (after the normalization), $\ket{\psi_0}$, evolving unitarily under a Hermitian Hamiltonian, $H$, the orthonormal Krylov basis $\{\ket{K_0} , \ket{K_1} , \cdots, \ket{K_{D_{\rm K}-1}}\}$ can be constructed via the Lanczos algorithm:
\begin{enumerate}
\item[\underline{\small{Step 1}}.\,] $b_0 := 0\,, \quad |K_{-1}\rangle := 0$ \,,
\item[\underline{\small{Step 2}}.\,] $|K_0\rangle := |\psi_0\rangle\,,\quad a_0=\langle K_0|H|K_0\rangle$,
\item[\underline{\small{Step 3}}.\,] For $n\geq1$:\\
$|\mathcal{A}_n\rangle=(H-a_{n-1})|K_{n-1}\rangle-b_{n-1}|K_{n-2}\rangle$ \,,
\item[\underline{\small{Step 4}}.\,] Set $b_n=\sqrt{\langle \mathcal{A}_n|\mathcal{A}_n\rangle}$ \,,
\item[\underline{\small{Step 5}}.\,] If $b_n=0$ stop; \\
otherwise set the orthonormal basis as $|K_n\rangle=b_n^{-1}|\mathcal{A}_n\rangle\,,~a_n=\langle K_n|H|K_n\rangle$, and go to {\small{Step 3}}.
\end{enumerate}
\end{tcolorbox}

Nevertheless, this recurrence procedure \eqref{RRC} ensures that $\ket{K_n}$ is automatically orthonormal to all previous vectors $\ket{K_0}, \cdots, \ket{K_{n-1}}$, drastically reducing the computational and memory cost. Thus, the Lanczos algorithm constructs the orthonormal basis, \textit{Krylov basis}, $\mathcal{B}_{D_{\rm K}}= \{ \ket{K_0} , \ket{K_1} , \cdots, \ket{K_{D_{\rm K}-1}} \}$, which spans the Krylov subspace $\mathcal{K}_{D_{\rm K}}$. A key computational advantage of this approach is its reliance on a three-term recurrence relation: unlike the standard Gram-Schmidt process, which requires orthogonalization against the entire preceding basis, each new vector $\ket{K_n}$ in the Lanczos sequence only needs to be orthogonalized with respect to its two immediate predecessors, $\ket{K_{n-1}}$ and $\ket{K_{n-2}}$.

One can further construct the reduction operator $\mathbb{K}_{D_{\rm K}}$ and $\mathbb{K}_{D_{\rm K}}^{\dagger}$ with the generated Krylov bases:
\begin{align}
\begin{split}
\mathbb{K}_{D_{\rm K}} &=
\begin{bmatrix}
\bra{K_0} \\
\bra{K_1} \\
\vdots \\
\bra{K_{D_{\rm K}-1}} 
\end{bmatrix} \,,\\
\mathbb{K}_{D_{\rm K}}^{\dagger} &= 
\begin{bmatrix} 
\ket{K_0}, & \ket{K_1},&  \cdots&, \ket{K_{D_{\rm K}-1}}
\end{bmatrix}  \,,
\end{split}
\end{align}
where $\ket{K_{n}} = (\cdots)^{T}$. The operator $\mathbb{K}_{D_{\rm K}}$ acts as a reduction operator that projects the full $D_{\mathcal H}$-dimensional Hilbert space dynamics onto the $D_{\rm K}$-dimensional Krylov subspace. Conversely, $\mathbb{K}_{D_{\rm K}}^{\dagger}$ serves as an embedding (or reconstruction) operator, mapping the effective states in the Krylov chain back to the original Hilbert space.

In this basis, the Hamiltonian is represented by a tridiagonal matrix $T = \mathbb{K}_{D_{\rm K}} \, H \, \mathbb{K}_{D_{\rm K}}^{\dagger}$:
\begin{align}\label{}
T=
\begin{pmatrix}
a_0 & b_1 & 0   & \cdots & 0 \\
b_1 & a_1 & b_2 & \ddots & \vdots \\
0   & b_2 & a_2 & \ddots & 0 \\
\vdots & \ddots & \ddots & \ddots & b_{D_{\rm K}-1} \\
0 & \cdots & 0 & b_{D_{\rm K}-1} & a_{D_{\rm K}-1}
\end{pmatrix}\,.
\end{align}
This structure reveals a profound physical mapping: the dynamics of a complex quantum system are approximated by an excitation propagating along a 1D tight-binding chain, called \textit{Krylov chain}, which can also be seen from \eqref{RRC}. By definition, any initial state is mapped to the first site of the chain, $\mathbb{K}_{D_{\rm K}}\ket{\psi_0}=(1,0,\cdots,0)^T$, and the evolution corresponds to the particle hopping to subsequent sites.

We make a few technical remarks on the Lanczos algorithm. The original Lanczos algorithm is easily susceptible to significant numerical instability~\cite{Parlett1979TheLA,Simon1984TheLA,Parlett_1998}. Because each new Krylov element is constructed from only the two preceding ones, finite-precision errors propagate and accumulate across iterations, leading to a rapid loss of orthogonality. While high-precision arithmetic can mitigate this~\cite{Viswanath_1994}, it is often prohibitively expensive in terms of memory and computational efficiency.

To address this, one can employ the Full Orthogonalization algorithm~\cite{Golub1972TheLA,Parlett_1998,Rabinovici:2020ryf,Hashimoto:2023swv}. This method ensures orthonormality up to machine precision by explicitly reorthogonalizing each newly constructed Krylov element against all previously generated elements at every iteration. Several algorithmic refinements have been developed to address these limitations, most notably Selective Orthogonalization~\cite{Parlett1979TheLA}, Partial Reorthogonalization~\cite{Simon1984TheLA}, and the Implicitly Restarted Lanczos Algorithm~\cite{BENNER199775}.
\begin{tcolorbox}[colback=gray!10, colframe=gray!50, title=\textbf{Full Orthogonalization algorithm}, coltitle=white]
Define the diagonal matrix $\mathcal{D}$ with the eigenvalues of the Hamiltonian $H$. Given an initial state $\ket{\psi_0}$ evolving under $H$, the orthonormal Krylov basis $\{\ket{K_0} , \ket{K_1} , \cdots, \ket{K_{D_{\rm K}-1}}\}$ is constructed as follows:
\begin{enumerate}
\item[\underline{\small{Step 1}}.\,] $b_0 := 0\,, \quad |K_{-1}\rangle := 0$ \,,
\item[\underline{\small{Step 2}}.\,] $|K_0\rangle := |\psi_0\rangle\,,\quad a_0=\langle K_0|\mathcal{D}|K_0\rangle$ \,,
\item[\underline{\small{Step 3}}.\,] For $n\geq1$:\\
$|\mathcal{A}_n\rangle=(\mathcal{D}-a_{n-1})|K_{n-1}\rangle-b_{n-1}|K_{n-2}\rangle$ \,,
\item[\underline{\small{Step 4}}.\,] Reassign the element to remove accumulated errors:\\ $\ket{\mathcal{A}_n} \rightarrow \ket{\mathcal{A}_n} - \sum_{m=0}^{n-1} \, \braket{K_m}{\mathcal{A}_n} \, \ket{K_m}$ \,,
\item[\underline{\small{Step 5}}.\,] Set $b_n=\sqrt{\langle \mathcal{A}_n|\mathcal{A}_n\rangle}$ \,,
\item[\underline{\small{Step 6}}.\,] If $b_n=0$ stop; \\
otherwise set the orthonormal basis as $|K_n\rangle=b_n^{-1}|\mathcal{A}_n\rangle\,,~a_n=\langle K_n|\mathcal{D}|K_n\rangle$, and go to {\small{Step 3}}.
\end{enumerate}
This algorithm preserves machine-precision orthonormality by reorthogonalizing each new iterate against all preceding basis vectors.
\end{tcolorbox}

Last but not least, it is instructive to note that matrix tridiagonalization $T$ (i.e., the Lanczos coefficients) can be efficiently achieved, using a method proposed by Householder~\cite{Householder1958UnitaryTO}, which employs a sequence of similarity transformations known as \textit{Householder reflections} ~\cite{Dumitriu:2002ntg}. This approach is capable of reducing any Hermitian matrix to a tridiagonal form. As demonstrated by~\cite{Sau_1979}, this procedure is equivalent to the Lanczos algorithm when the latter is initialized with the canonical starting vector $(1,0,\cdots,0)^T$. 

Consequently, to utilize Householder reflections for computing Lanczos coefficients from an arbitrary starting vector, an initial unitary rotation must be performed to transform the basis such that the desired vector takes this canonical form. This method is widely known as the \textit{Hessenberg decomposition}~\cite{Golub1979AHM}, which casts a general matrix into upper-Hessenberg form; for Hermitian matrices, this result simplifies to a complete tridiagonal structure. The procedure is typically more efficient than the Lanczos algorithm and is implemented in \texttt{Mathematica} via the built-in command, \texttt{HessenbergDecomposition}.

%
\subsection{Wave function dynamics and Krylov complexity}
Within the $D_{\rm K}$-dimensional Krylov subspace the time-evolved state $\ket{\psi(t)}$ is represented as a linear combination of the basis vectors with a tridiagonal matrix $T = \mathbb{K}_{D_{\rm K}} \, H \, \mathbb{K}_{D_{\rm K}}^{\dagger}$:
\begin{align}\label{EQK1}
\begin{split}
\ket{\psi(t)} = \mathbb{K}_{D_{\rm K}}^{\dagger} \,e^{-i T t}\, \mathbb{K}_{D_{\rm K}} \ket{\psi_0} = \sum_{n=0}^{D_{\rm K}-1} \psi_n(t) \ket{K_n} \,,
\end{split}
\end{align}
where the coordinate vector $\psi_n(t)$ is defined as $e^{-i T t}\, \mathbb{K}_{D_{\rm K}} \ket{\psi_0} := (\psi_{0}(t), \psi_{1}(t) ,\cdots,\psi_{D_{\rm K}-1}(t))^T$. The summation form arises naturally from the definition of matrix-vector multiplication. When the expansion operator $\mathbb{K}_{D_{\rm K}}^{\dagger}$ (whose columns are the basis vector $\ket{K_n}$) acts on the coordinate vector, it performs a linear combination of the Krylov basis vectors weighted by their respective amplitudes.
\begin{figure}[]
\centering
     \includegraphics[width=8.8cm]{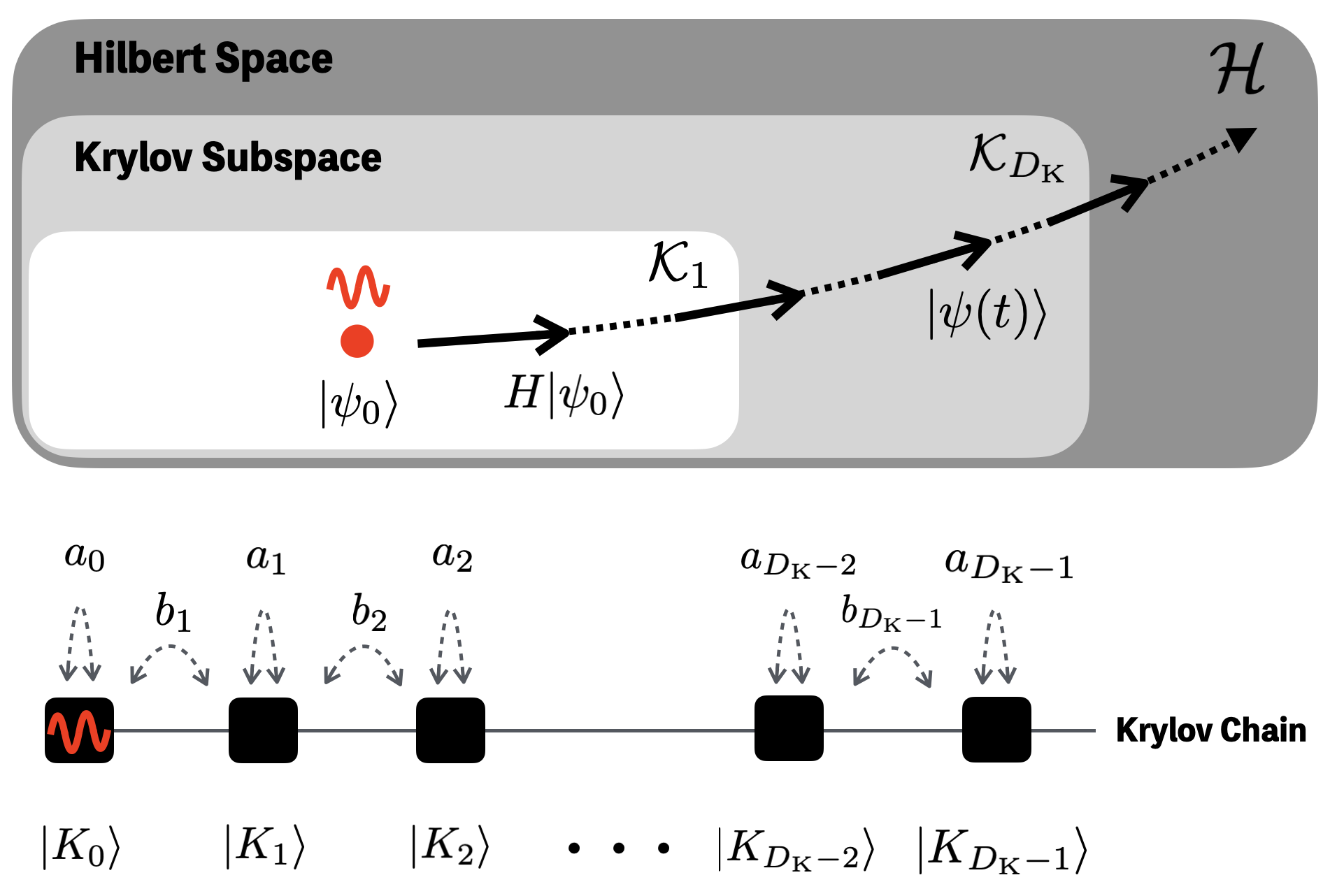}
 \caption{Schematic dynamics of the initial state $\ket{\psi_0}$ within $D_{\rm K}$-dimensional Krylov subspace $\mathcal{K}_{D_{\rm K}}$ under the given Hamiltonian $H$. Any initial state $\ket{\psi_0}$ is mapped or projected to the first site ($n=0$) of the Krylov chain, $\mathbb{K}_{D_{\rm K}}\ket{\psi_0}=(1,0,\cdots,0)^T$, and the evolution corresponds to the particle hopping to subsequent sites.}\label{KryFigure}
\end{figure}

In essence, Eq. \eqref{EQK1} provides a clear physical interpretation by mapping the complex many-body dynamics onto a simple one-dimensional tight-binding chain. In this picture, the initial state $\psi_0$ is represented as a single excitation localized at the first site of the Krylov lattice, $(1,0,\cdots,0)^T$, which then spreads through the chain as time evolves. The tridiagonal matrix $T$ governs this motion, where the diagonal elements $a_n$ function as onsite potentials and the off-diagonal elements $b_n$ act as hopping amplitudes that drive the state to populate higher-order Krylov basis vectors. Once the evolution is computed within this reduced $D_{\rm K}$-dimensional space, the operator $\mathbb{K}_{D_{\rm K}}^{\dagger}$ acts as a bridge, mapping the result back to the full $D_{\mathcal H}$-dimensional Hilbert space. We show the schematic picture of these in Fig. \ref{KryFigure}.
This Krylov subspace method’s remarkable efficiency stems from the fact that it replaces the often-impossible task of exponentiating a massive Hamiltonian $H$ with the far more economical exponentiation of the small tridiagonal matrix $T$, allowing us to capture the essential physics with a fraction of the computational cost.

The evolution of an initial state within the Krylov basis \eqref{EQK1} is characterized with by defining $\ket{K_0}=\ket{\psi(0)}$, we establish the initial condition $\psi_{n}(t=0) = \delta_{n0}$. Also, the principle of unitarity ensures the normalization of the state at all
times $\sum_{n=0}^{D_{\rm K}-1} \,|\psi_n(t)|^2 = 1$.

Specifically, the dynamics are governed by the Schrödinger equation, $i \, \partial_t \ket{\psi(t)} = H \ket{\psi(t)}$ (with $\hbar=1$). When combined with the three-term recurrence relation \eqref{RRC}, the amplitudes $\psi_n(t)$ follow a discrete evolution equation resembling a one-dimensional tight-binding model: 
\begin{align}\label{SCHRO}
\begin{split}
i \, \partial_t \psi_n(t)  = a_n \psi_n(t) + b_{n+1} \psi_{n+1}(t) + b_{n} \psi_{n-1}(t)  \,. 
\end{split}
\end{align}
In this framework, Krylov complexity (or spread complexity)~\cite{Balasubramanian:2022tpr} is defined as
\begin{align}\label{SCDEF}
\begin{split}
C(t) := \sum_{n=0}^{D_{\rm K}-1} \, n \abs{\psi_n(t)}^2  = \sum_{n=0}^{D_{\rm K}-1} \, n \abs{\braket{K_n}{\psi(t)}}^2  \,,
\end{split}
\end{align}
where it quantifies the average position of a time-evolving quantum state as it spreads over the Krylov chain, where $n$ serves as the discrete position index. Thus, by mapping the state's trajectory within the Hilbert space into Krylov subspace, the Krylov basis captures the efficient dynamics of quantum state.

Krylov complexity has emerged as a powerful diagnostic tool across various domains of physics. It has provided significant insights into, for instance, quantum chaos~\cite{Rabinovici:2022beu,Erdmenger:2023wjg,Balasubramanian:2022tpr,Balasubramanian:2023kwd,Gautam:2023bcm,Camargo:2024deu,Bhattacharjee:2024yxj,Baggioli:2024wbz,Huh:2024ytz,Aguilar-Gutierrez:2026jjv}, quantum scars~\cite{Bhattacharjee:2022qjw,Nandy:2023brt}, topological phase transitions \cite{Caputa:2022eye,Caputa:2022yju}, random matrix theories \cite{Bhattacharyya:2023grv,Balasubramanian:2022dnj,Balasubramanian:2023kwd}, saddle-dominated scrambling~\cite{Afrasiar:2022efk,Huh:2023jxt, Zhou:2024rtg,Aguilar-Gutierrez:2025hbf}, localization and thermalization~\cite{Alaoui:2023cdo,Alishahiha:2024rwm,Menzler:2024atb,Cohen:2024ngg}, quantum measurements~\cite{Gill:2023umm, Bhattacharya:2023yec}, open quantum systems \cite{Bhattacharya:2023zqt,Carolan:2024wov,Baggioli:2025knt}, quantum walks \cite{Jeevanesan:2023ogo}, quantum gravity and the AdS/CFT correspondence \cite{Lin:2022rbf,Rabinovici:2023yex,Heller:2024ldz,Fu:2025kkh,Caputa:2024sux}. See also \cite{Nandy:2024evd,Rabinovici:2025otw} for more references therein.

While quantum complexity could technically be defined in any basis $\mathcal{B} := \{\ket{B_n}\}$, the Krylov basis is mathematically appealing supported by the optimal basis theorem~\cite{Balasubramanian:2022tpr}. Consider a general cost functional:
\begin{align}
\begin{split}
C_\mathcal{B}(t) = \sum_n \, c_n \,\abs{\braket{B_n}{\psi(t)}}^2 \,,
\end{split}
\end{align}
where the weights $c_n$ are positive and monotonically increasing. The optimal basis theorem states that for $c_n=n$, the Krylov basis is the one that minimizes
this cost functional:
\begin{align}
\begin{split}
    C(t) = \min_\mathcal{B} C_\mathcal{B}(t) \,,
\end{split}
\end{align}
where a comprehensive proof is established in \cite{Balasubramanian:2022tpr}. Essentially, this functional minimization identifies the Krylov basis as the most efficient representation. In discrete-time scenarios this optimality holds at all times~\cite{Nahum:2016muy,Chan:2017kzq,Skinner:2018tjl}, making the Krylov basis a natural choice for quantifying the dynamical growth of state complexity.

%
\subsection{TFD, survival amplitude, and spectral form factor}
Most notably in the context of Krylov complexity~\cite{Balasubramanian:2022tpr}, the thermofield double (TFD) states~\cite{Maldacena:2001kr} have emerged as a useful initial state for investigating random matrix theory (RMT) behavior within state-dependent probes of quantum chaos. Because RMT signatures are fundamentally statements about the full energy spectrum (specifically level-spacing statistics), the TFD state can be advantageous. Unlike single-eigenstate probes, the TFD state incorporates information from the entire spectrum, making it one of the most natural setups for bridging dynamical properties of chaos with traditional time-independent spectral statistics.

The TFD state is defined in a doubled Hilbert space consisting of two identical copies of a system, denoted as left (L) and right (R). It is expressed as:
\begin{align}\label{TFD}
\begin{split}
\ket{{\text{TFD}}} := \frac{1}{\sqrt{Z(\beta)}}\sum_{n} e^{- \beta E_n/2} \ket{n_\text{L}} \otimes\ket{n_\text{R}} \,,
\end{split}
\end{align}
where $\beta$ is the inverse temperature, $E_n$ are the eigenvalues of the Hamiltonian $H$, and $Z(\beta) = \sum_n e^{-\beta E_n}$ is the partition function. Here, $H_{\text{L}}$ and $H_{\text{R}}$ act independently on their respective Hilbert spaces such that $H_{\text{L}, \text{R}} \ket{n_{\text{L}, \text{R}}} = E_n \ket{n_{\text{L}, \text{R}}}$

From a quantum information perspective, the TFD state is the purification of the thermal Gibbs state. Tracing out the degrees of freedom of one boundary (e.g., system R) yields a reduced density matrix that describes a thermal ensemble at temperature $T=1/\beta$. In the context of holography, the TFD state is dual to a two-sided eternal black hole in AdS space, where the two boundaries of the asymptotic spacetime correspond to the left and right CFTs \cite{Maldacena:2001kr}. This duality has motivated extensive research into its entanglement structure \cite{Chapman:2018hou}, wormhole geometries \cite{Maldacena:2017axo,Gao:2016bin}, and applications in quantum teleportation and simulation \cite{Wu:2018nrn,Zhu:2019bri,Brown:2019hmk,Nezami:2021yaq,Schuster:2021uvg}.

While the TFD state is invariant under the combined evolution $H_{\text{L}}-H_{\text{R}}$, it evolves non-trivially under the action of a single Hamiltonian $H$ (or effectively $(H_{\text{L}}+H_{\text{R}})/2$). The time-evolved TFD state is then given by $\ket{\psi(t)} = e^{-i H t} \ket{{\text{TFD}}}$.

A key related quantity of interest is the \textit{survival amplitude}, $S(t)$, which measures the overlap between the initial state and its evolution:
\begin{align}\label{}
\begin{split}
S(t) := \braket{\psi(t)}{\psi_0} = \psi_0(t) \,,
\end{split}
\end{align}
where the moments, $\mu_n := \lim_{t\rightarrow0} d^n S(t)/dt^n$, are directly related to the Lanczos coefficients $\{a_n, b_n\}$ used to compute Krylov complexity, for instance $\mu_1 = i a_0$, $\mu_2 = - a_0^2 - b_1^2,\, \cdots$~\cite{Balasubramanian:2022tpr}. 

For an initial TFD state, the survival amplitude can be further expressed in terms of the partition function with a complex argument~\cite{Balasubramanian:2022tpr,Caputa:2024vrn}:
\begin{align}\label{SFFSre}
\begin{split}
S(t) = \frac{Z_{\beta-i t}}{Z_\beta} = \sqrt{\text{SFF}(t)}\,.
\end{split}
\end{align}
This identification demonstrates that the spectral form factor (SFF)~\cite{Guhr:1997ve,Cotler:2016fpe}, a standard dynamical diagnostic of spectral statistics with RMT, is equivalent to the survival probability of the TFD state. This link is pivotal for studying quantum speed limits~\cite{delCampo:2017bzr}.

Utilizing the TFD state as the reference initial state can also establish a direct correspondence between the SFF and Krylov complexity. The qualitative evolution of the SFF in chaotic systems follows a distinct slope-dip-ramp-plateau structure. Krylov complexity is conjectured to complement this behavior through a quadratic growth-linear growth-peak-plateau sequence. Notably, the late-time saturation of both quantities is mathematically linked. At infinite temperature $\beta=0$, the long-time average of the SFF relates to the system dimension and the saturation value of Krylov complexity $C(t=\infty)$ as follows:
\begin{align}\label{SFFCRE}
\lim_{T\rightarrow \infty } \frac{1}{T}\int_{0}^{T}\text{SFF}(t) = \frac{1}{D_{\mathcal H}} =\frac{1}{1+2C(t=\infty)} \,,
\end{align}
where the saturation value becomes as $C(t=\infty) \approx {(D_{\rm K} - 1)}/{2}$~\cite{Cotler:2016fpe,Erdmenger:2023wjg,Rabinovici:2020ryf,Rabinovici:2022beu}, providing a direct bridge between spectral statistics and the spread of Krylov complexity. For the maximally entangled TFD state considered here, the Krylov space spans the full Hilbert space, $D_{\rm K}=D_{\mathcal H}$.

%
\section{Elementary models in quantum mechanics for chaos}\label{SECIII}

%
\subsection{Random matrix theories: \textit{the} universal theories}
Random matrix theory (RMT) provides the foundational statistical framework for identifying universal features in quantum chaotic systems~\cite{Guhr:1997ve,MEHTA1960395,Akemann:2011csh,Dyson:1962oir}. It posits that the fine-grained energy fluctuations of Hamiltonian of quantum systems are not random in an uncorrelated sense, but rather follow the statistical laws of large random matrices~\cite{Meh2004,Dyson:1962es,Bohigas:1983er}. This correspondence is formalized by the Bohigas-Giannoni-Schmit (BGS) conjecture~\cite{Bohigas:1983er,Guhr:1997ve,Bohigas}, which bridges the gap between classical chaos and quantum spectral signatures.

The statistical behavior of a quantum system is determined primarily by its underlying symmetries. In RMT, these are categorized into three fundamental Gaussian ensembles, distinguished by the Dyson index $\beta$: the Gaussian Orthogonal Ensemble (GOE; $\beta=1$), Gaussian Unitary Ensemble (GUE; $\beta=2$), and Gaussian Symplectic Ensemble (GSE; $\beta=4$). This symmetry-based classification underscores the universality of RMT, making it an essential framework for capturing the statistical behavior of quantum chaotic systems.

To isolate universal chaotic fluctuations from system-specific features (such as the average density of states), the energy eigenvalues must undergo a process called unfolding~\cite{Brody:1981cx}. This procedure involves a local rescaling of the spectrum so that the mean level spacing is normalized to unity: $\langle{s}\rangle=1$. Once unfolded, the spectrum is typically analyzed through two lenses: short-range correlations characterized by the nearest-neighbor spacing distribution $p(s)$, long-range correlations characterized by measures such as number variance or spectral rigidity.
\begin{figure*}[]
\centering
\begin{minipage}{0.23\textwidth}
\centering
\includegraphics[width=\textwidth]{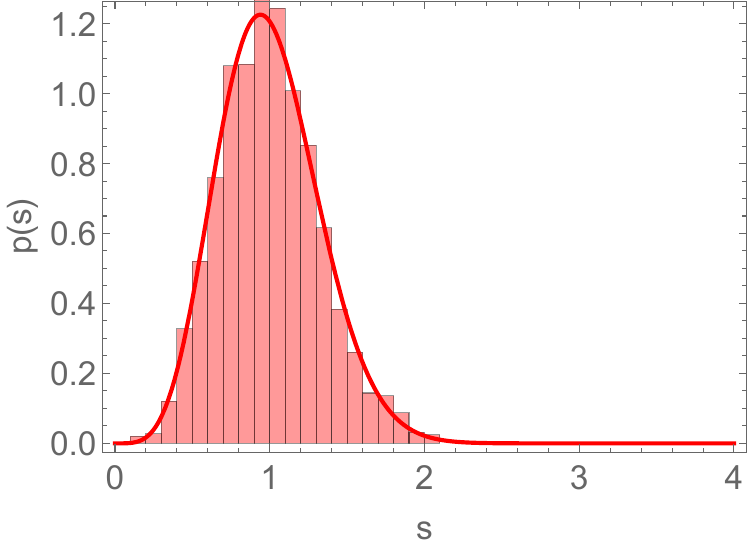}
\end{minipage}
\quad
\begin{minipage}{0.23\textwidth}
\centering
\includegraphics[width=\textwidth]{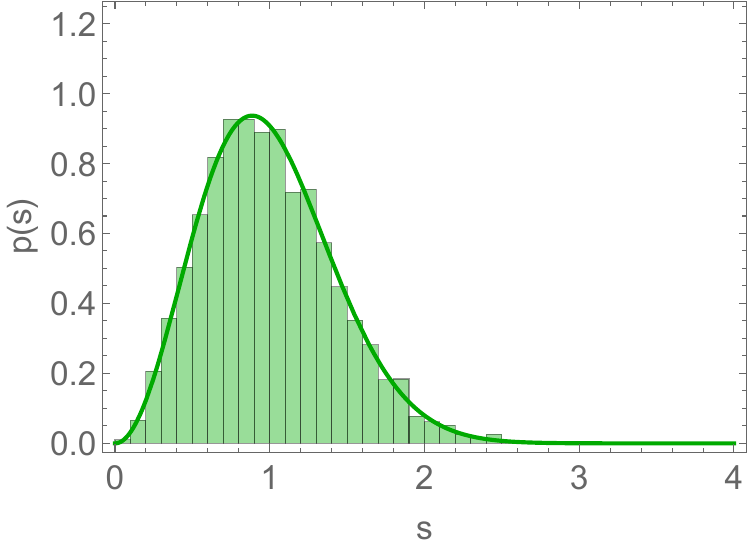}
\end{minipage}
\quad
\begin{minipage}{0.23\textwidth}
\centering
\includegraphics[width=\textwidth]{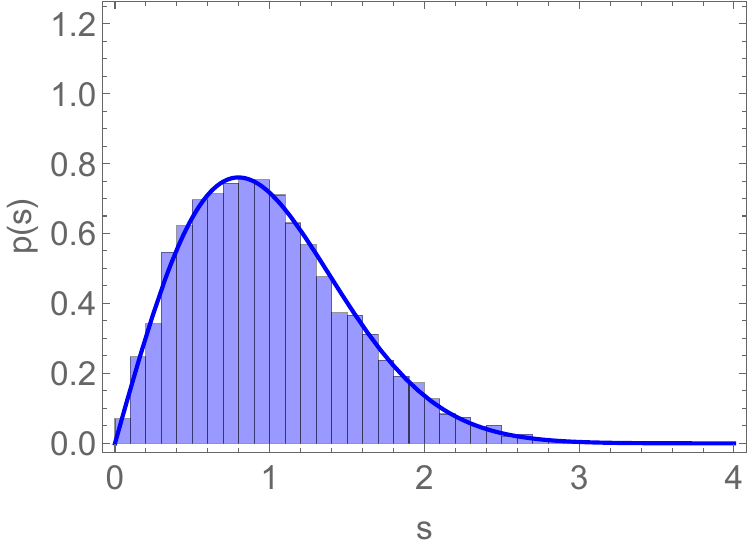}
\end{minipage}
\quad
\begin{minipage}{0.23\textwidth}
\centering
\includegraphics[width=\textwidth]{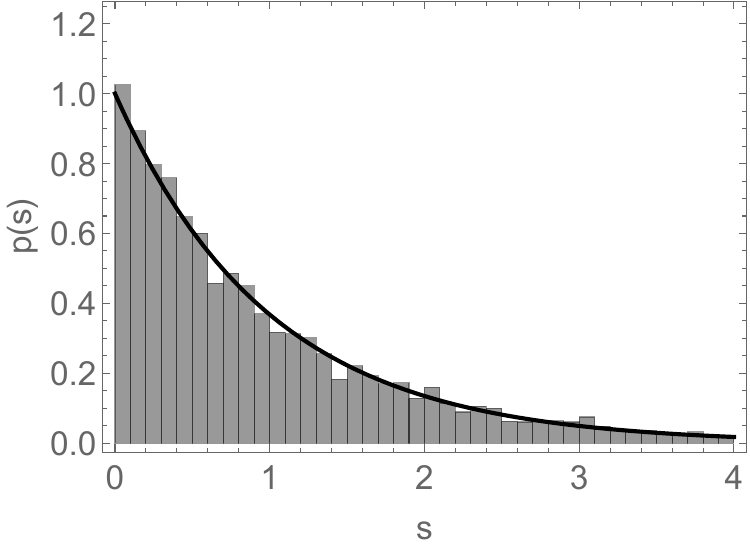}
\end{minipage}
\caption{Nearest-neighbor level spacing distributions. The probability density $p(s)$ is shown for the three standard random matrix ensembles (GSE, GUE, and GOE) (red, green, blue) and the integrable Poisson system (black). Histograms represent numerical data obtained by diagonalizing random matrices of dimension $200$, with statistics averaged over $50$ independent realizations. Solid colored lines denote the corresponding analytical predictions: the exact the Wigner surmise for the Gaussian ensembles \eqref{RMTLSD} and Poisson distribution \eqref{PS}.}\label{LSDSIM}
\end{figure*}
\begin{figure*}[]
\centering
\begin{minipage}{0.23\textwidth}
\centering
\includegraphics[width=\textwidth]{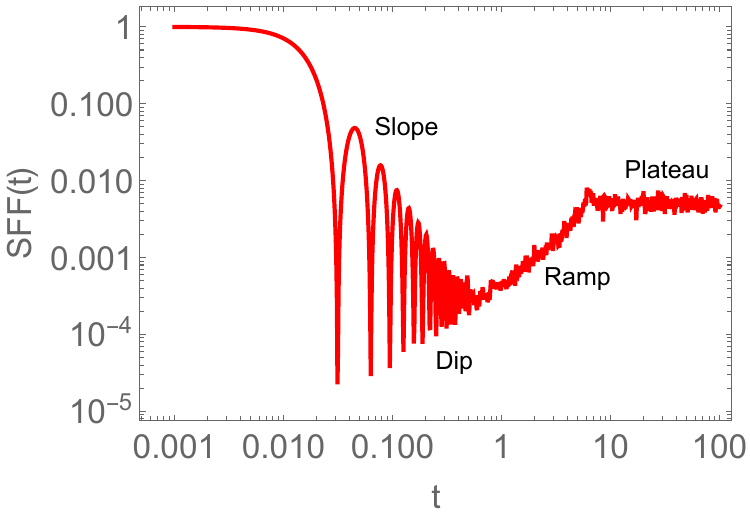}
\end{minipage}
\quad
\begin{minipage}{0.23\textwidth}
\centering
\includegraphics[width=\textwidth]{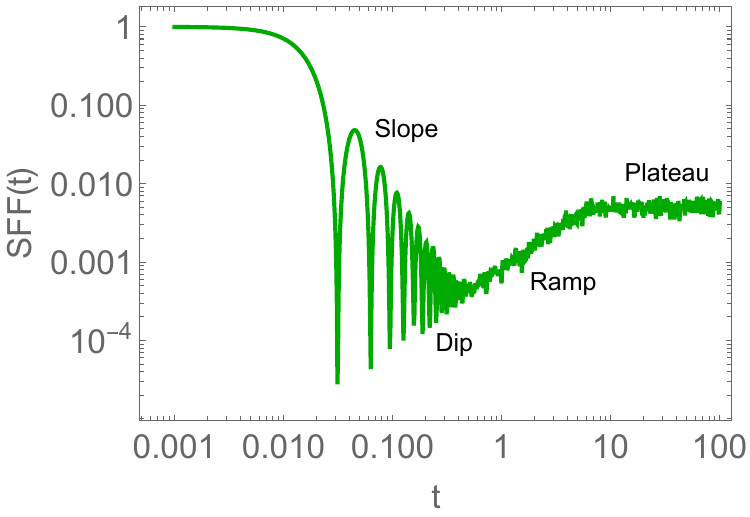}
\end{minipage}
\quad
\begin{minipage}{0.23\textwidth}
\centering
\includegraphics[width=\textwidth]{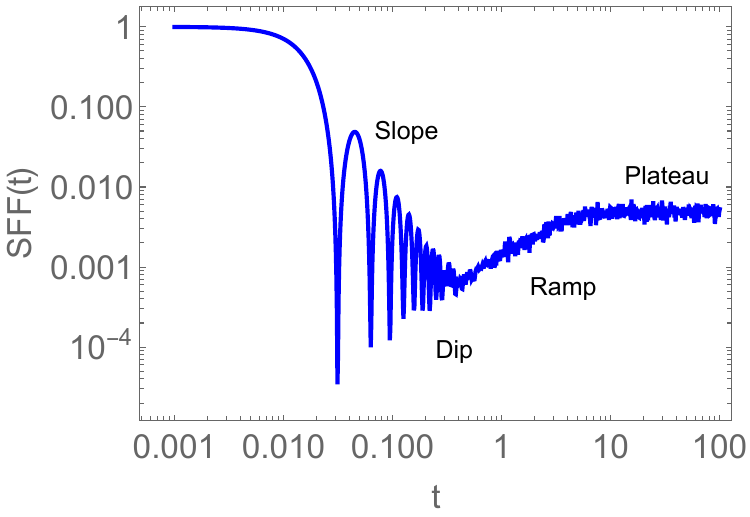}
\end{minipage}
\quad
\begin{minipage}{0.23\textwidth}
\centering
\includegraphics[width=\textwidth]{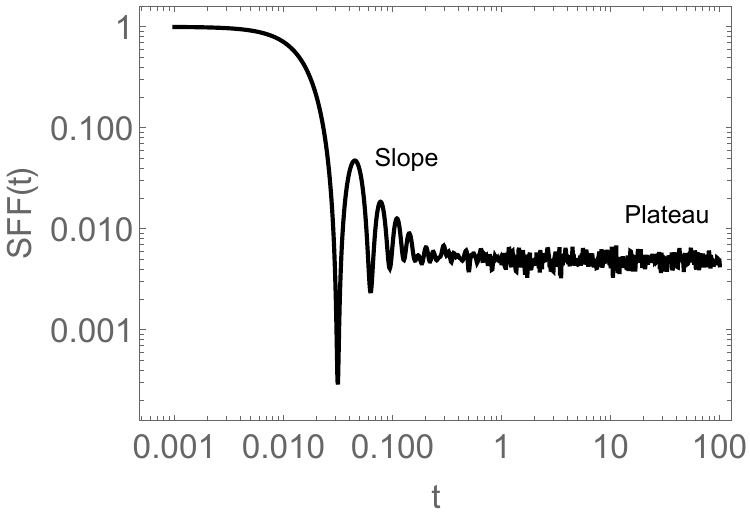}
\end{minipage}
\caption{Spectral form factor for the random matrix ensembles (GSE, GUE, and GOE) (red, green, blue) and the integrable Poisson system (black). The SFF is evaluated at infinite temperature ($\beta=0$) and plotted on a log-log scale. The numerical curve is generated from matrices of size $200$, averaged over $50$ realizations. The evolution displays the defining dynamical signatures of quantum chaos for the random matrix ensembles, i.e., the ramp structure: an initial early decay governed by the average density of states, an interruption at the dip time, a subsequent linear ramp reflecting two-level correlations, and a final saturation into a constant plateau.}\label{SFFSIM}
\end{figure*}

More specifically, a primary diagnostic of quantum chaos is level repulsion, the tendency for energy levels to avoid one another, i.e., the level spacing distribution. This is quantitatively described by the Wigner surmise~\cite{Wigner_1951,Dyson:1962oir,Dyson:1962es}:
\begin{align}\label{}
\begin{split}
p(s) = a(\beta) \, s^{\beta} \, e^{-b(\beta) s^2} \,.
\end{split}
\end{align}
The constants $a(\beta)$ and $b(\beta)$ are determined by normalization ($\int p(s) \,\dd s = 1$) and the unit mean spacing constraint ($\int  s \, p(s) \, \dd s = 1$). For the three universality classes, the specific distributions are: 
\begin{align}\label{RMTLSD}
\begin{split}
p_{\text{\tiny{GOE}}} &= \frac{\pi}{2} s \, e^{-\frac{\pi}{4}s^2} \,,\quad 
p_{\text{\tiny{GUE}}} = \frac{32}{\pi^2} s^2 \, e^{-\frac{4}{\pi}s^2} \,,\\
p_{\text{\tiny{GSE}}} &= \frac{2^{18}}{3^{6} \pi^3} s^4 \, e^{-\frac{64}{9\pi}s^2} \,.
\end{split}
\end{align}
In contrast, integrable systems lack these correlations. Their energy levels are statistically independent, following a Poisson distribution~\cite{BerryTabor}:
\begin{align}\label{PS}
\begin{split}
p_{\text{Poisson}}(s) = e^{-s} \,. 
\end{split}
\end{align}
Unlike the Wigner surmise, the Poisson distribution peaks at $s=0$, meaning levels can cluster arbitrarily close to one another. The transition from Poisson \eqref{PS} to Wigner-Dyson statistics \eqref{RMTLSD} is the standard hallmark of the onset of quantum chaos.

While the level spacing distribution $p(s)$ probes energy-domain correlations, the spectral form factor (SFF) acts as a time-dependent diagnostic of spectral statistics. It is defined via the analytic continuation of the partition function $Z_\beta$:
\begin{align}\label{}
\begin{split}
\text{SFF}(t) := \frac{\left|Z_{\beta-i t}\right|^2}{\left|Z_\beta\right|^2} = \frac{1}{{Z_\beta}^2}\sum_{n,m}e^{-i(E_n-E_m)t}e^{-\beta (E_n+E_m)} \,.
\end{split}
\end{align}
The SFF tracks how spectral correlations evolve over time. A signature of RMT behavior in chaotic systems is the emergence of a ``linear ramp" structure at late times~\cite{Brezin:1997aa,Cotler:2016fpe}. On a log-log plot, this manifests as a slope of approximately $1$. Any significant deviation from this linear growth suggests non-universal or non-chaotic dynamics, making the SFF a robust tool for verifying the presence of RMT correlations in complex quantum systems.

\texttt{Mathematica} serves as a robust platform for the numerical simulation and analysis of RMT ensembles: 
1) GOE: \texttt{Gaussian\allowbreak Orthogonal\allowbreak Matrix\allowbreak Distribution}, 
2) GUE: \texttt{Gaussian\allowbreak Unitary\allowbreak Matrix\allowbreak Distribution}, 
3) GSE: \texttt{Gaussian\allowbreak Symplectic\allowbreak Matrix\allowbreak Distribution} 
and 4) the Poisson system with \texttt{WignerSemicircleDistribution} or \texttt{NormalDistribution}. 
By utilizing built-in random matrix generators or custom routines, one can construct large-scale matrices that satisfy the specific symmetry constraints of the GOE (real symmetric), GUE (complex Hermitian), and GSE (quaternion Hermitian). Once the eigenvalues are extracted, an unfolding procedure is applied to normalize the mean level spacing to unity, allowing for a direct comparison between numerical data and theoretical predictions. We present statistical properties of energy spectra for different universality classes of RMT obtained via numerical simulations: level spacing distribution in Fig. \ref{LSDSIM} and the spectral form factor in Fig. \ref{SFFSIM}.

%
\subsection{Quantum billiards: classical-quantum correspondence}
Next, we provide a quick review of elementary quantum mechanical models for chaos in both one-body and many-body systems. Specifically, we examine quantum billiards as the prototypical one-body system, followed by an analysis of three many-body spin-chain models: the mixed-field Ising model, the Heisenberg XXZ model, and the Sachdev-Ye-Kitaev (SYK) model.

The subsequent section examines the chaotic signatures of these models. Building on the previous definitions, we will present their spectral statistics alongside an analysis of their Krylov complexity to provide a comprehensive view of their dynamical properties.

We begin by reviewing low-dimensional quantum mechanical models that exhibit classical chaos. A prototypical example in one-body systems is the billiard systems, which serves as a paradigm for studying the intersection of classical and quantum chaos. In particular, we consider the stadium billiard \cite{Sinai_1970,Bunimovich_1975,Bunimovich_1979,Bunimovich_1991,Benettin:1978aa}. The Hamiltonian for this system is defined as:
\begin{align}\label{BILLMO}
\begin{split}
  H &= p_x^2 + p_y^2 + V_{\text{stad}}(x,y) \,,
\end{split}
\end{align}
where the potential $V_{\text{stad}}(x,y)$ vanishes within the domain $\Omega$ and is infinite elsewhere:
\begin{align}\label{}
\begin{split}
V_{\text{stad}}(x,y) =  
   \begin{cases}
       0 & (x,y)\in \Omega \\
       \infty & \textrm{else}
   \end{cases}
\,.
\end{split}
\end{align}
\begin{figure}[]
\centering
     \includegraphics[width=8.0cm]{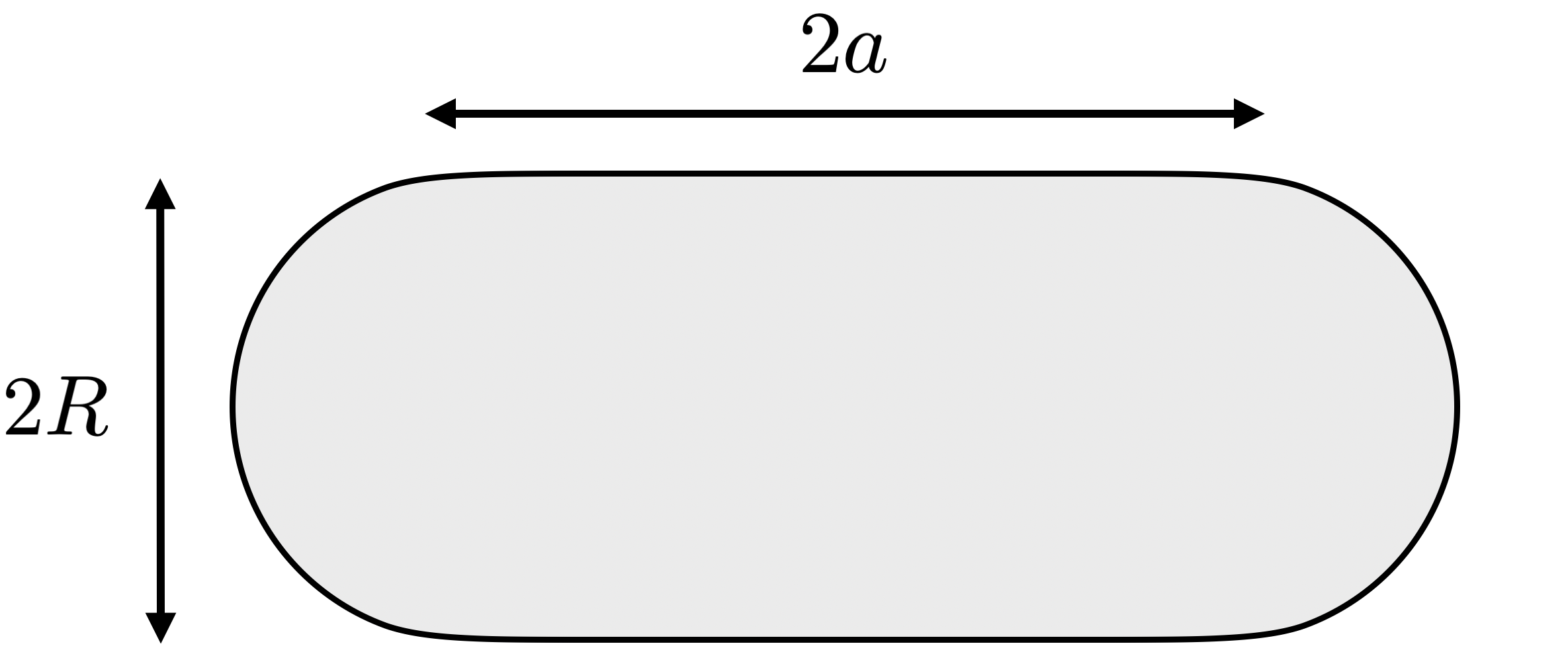}
 \caption{The geometry of the stadium billiard with the area $A = \pi R^2 + 4 a R$.}\label{skaa}
\end{figure}
The geometry of the stadium, illustrated in Fig. \ref{skaa}, consists of two semicircles of radius $R$ joined by two parallel straight lines of length $2a$. Consequently, the total area $A$ of the domain is given by
\begin{align}\label{}
\begin{split}
 A = \pi R^2 + 4 a R \,.
\end{split}
\end{align}
A critical feature of this model is the deformation parameter $a/R$. When $a/R=0$, the system reduces to a circular billiard, which is strictly integrable and characterized by a vanishing Lyapunov exponent. However, for any $a/R\neq0$, the system becomes non-integrable (chaotic), exhibiting a finite Lyapunov exponent in its classical trajectories \cite{BenettinStochastic,McDonaldSpectrum,CasatiSpectra}.

\begin{figure}[t!]
 \centering
{\includegraphics[width=7.1cm]{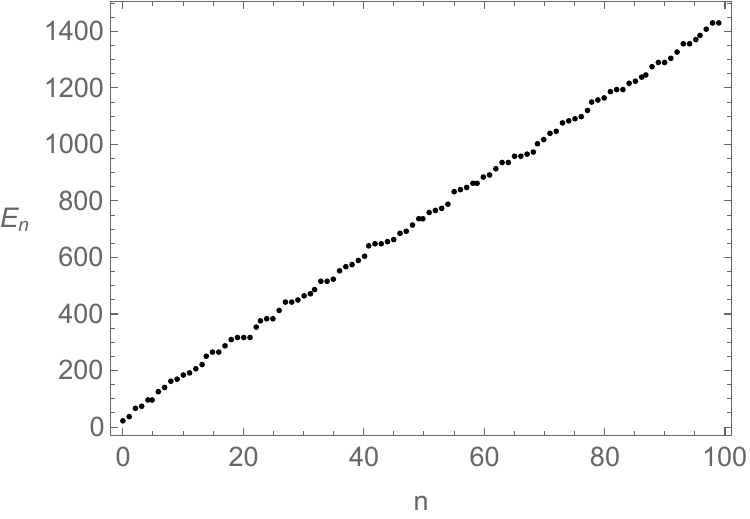}}
{\includegraphics[width=7.1cm]{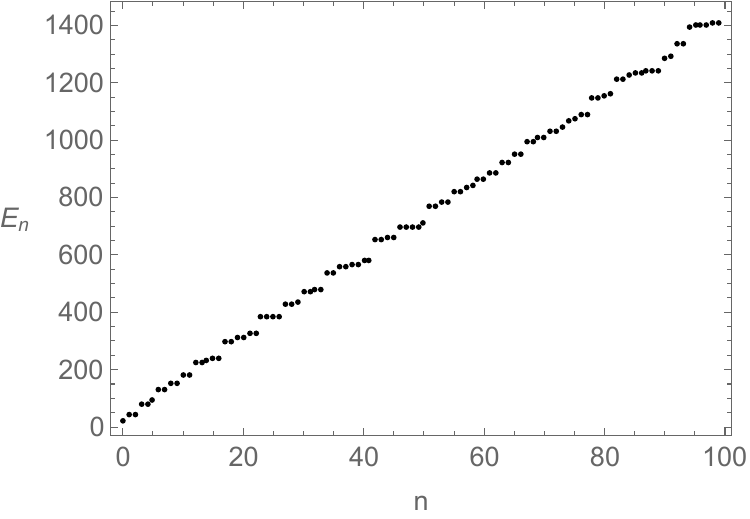}}
\caption{Eigenvalues of the quantum billiard systems with $a/R=1$ (upper panel) and $a/R=0$ (lower panel).}\label{EIGENSTA}
\end{figure}

To investigate the quantum dynamics of the billiard system, we solve the time-independent Schrödinger equation:
\begin{align}\label{SCHREQ}
\begin{split}
  - \nabla^2 \psi_n (x, y)  + V_{\text{stad}}(x,y)  \psi_n (x, y) = E_n \psi_n (x, y) \,,
\end{split}
\end{align}
where the potential is defined as in \eqref{BILLMO}. By numerically solving this eigenvalue problem, we obtain the energy spectra $E_n$. Figure \ref{EIGENSTA} illustrates the resulting energy eigenvalues for both the chaotic stadium billiard ($a/R=1$) and the integrable circular billiard ($a/R=0$).
The chaotic nature of this system has been further characterized using quantum Lyapunov exponents derived from OTOCs \cite{Hashimoto:2017oit}. Additionally, the energy level statistics align with the GOE for stadium billiards, while circle billiards follow a Poisson ensemble \cite{Sinai_1970,Bunimovich_1975,Bunimovich_1979,Bunimovich_1991,Benettin:1978aa,Backer_2007}. The obtained energy levels will be used to compute the Krylov complexity of the TFD state.

%
\subsection{Quantum spin chains: local interactions}
We now turn to many-body systems exhibiting quantum chaos. Specifically, we examine two spin-chain models known for their well-studied transitions between integrability and chaos: the mixed-field Ising model and the next-to-nearest-neighbor, i.e., local interactions, deformation of the Heisenberg XXZ model.

The mixed-field Ising model describes a one-dimensional spin chain subject to both longitudinal $h_z$ and transverse $h_x$ magnetic fields. Its Hamiltonian is defined as
\begin{align}\label{Ising}
H = -\sum^{N-1}_{i=1} S^{z}_{i}S^{z}_{i+1} - \sum^N_{i=1}\left( h_x S^{x}_{i} + h_z S^{z}_{i} \right) \,,
\end{align}
where $S^{x}_i$ and $S^{z}_i$ are spin-$1/2$ operators acting on site $i$, given by
\begin{align}\label{spinops}
S^{k}_{i}=\left(\mathds{1}_{2}\right)^{\otimes(i-1)}\otimes \sigma_{k}\otimes\left(\mathds{1}_{2}\right)^{\otimes(N-i)}~.
\end{align}
Here, $k\in\lbrace x,y,z\rbrace$ denotes the standard Pauli matrices. Under open boundary conditions for a chain of $N$ sites, the Hilbert space dimension is $D_{\mathcal H}=2^N$.
\begin{figure}[t!]
 \centering
{\includegraphics[width=7.5cm]{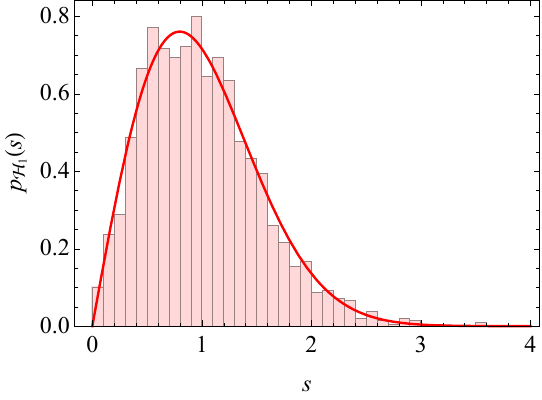}}
{\includegraphics[width=7.5cm]{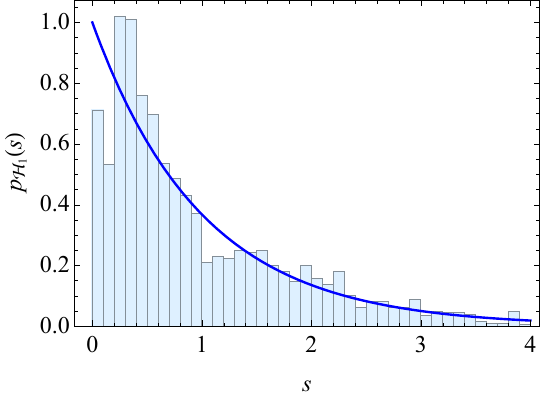}}
\caption{Nearest-neighbor level spacing distribution in the mixed-field Ising model \eqref{Ising} for the even symmetry sector $H_1$ with $N=12$ sites when $(h_x,h_z) = (-1.05,0.5)$ (upper panel) and $h_x,h_z) = (-1,0.001)$ (lower panel).}\label{NNSD}
\end{figure}

The spectral statistics of this model are shown to exhibit both chaotic and integrable behavior~\cite{Bauls_2011, Craps:2019rbj}, depending on the choice of longitudinal and transverse fields. Following the literature, we consider two representative cases
\begin{align}
(h_x,\,h_z) = 
\begin{cases}
\,\,  (-1.05,\,0.5) \,, \quad\, (\text{Chaotic case})  \\
\,\, (-1,\,0) \,. \qquad\quad\,\, (\text{Integrable case}) 
\end{cases}
\end{align}
Regardless of the field strengths $(h_x,h_z)$, the Hamiltonian possesses a global parity symmetry $[\hat{H}, \hat{\Pi}]=0$. The parity operator $\hat{\Pi}$ is defined via the product of permutation operators $\hat{P}_{i,j}$
\begin{equation} \label{eq:parity}
    \hat{\Pi}=
    \begin{cases}
       \hat{P}_{1,N} \, \hat{P}_{2,N-1} \, \cdots \, \hat{P}_{\frac{N}{2},\frac{N+2}{2}} \,,\,\,\quad \text{for}\,\, N \,\, \text{even}\\
       \hat{P}_{1,N} \, \hat{P}_{2,N-1} \, \cdots \, \hat{P}_{\frac{N-1}{2},\frac{N+3}{2}} \,,\,\,\, \text{for}\,\, N \,\, \text{odd}
    \end{cases}
\end{equation}
and 
\begin{equation}
    \hat{P}_{i,j}=\frac{1}{2}\left(\mathds{1}_{d}+S_i^xS_j^x+S_i^yS_j^y+S_i^zS_j^z \right) \,,
\end{equation}
where it swaps the spin configurations at sites $i$ and $j$. Physically, $\hat{\Pi}$ acts as a spatial reflection about the center of the chain, mapping a state to its mirror image (e.g., $\hat{\Pi} \,\,\vert\!\!\uparrow \uparrow \uparrow \downarrow \rangle =\vert\!\!\downarrow \uparrow \uparrow \uparrow  \rangle$). From such configurations, one can construct eigenstates of the parity operator, $\vert \psi_{\pm}\rangle = \frac{1}{\sqrt{2}}\left( \vert \!\uparrow \uparrow \uparrow \downarrow \rangle \pm \vert\! \downarrow \uparrow \uparrow \uparrow  \rangle \right)$, with eigenvalues $\pm 1$.

The symmetry properties of the Hamiltonian allow for its decomposition into two distinct parity sectors: the even parity ($H_1$) and odd parity ($H_2$) blocks. Under these conditions, the Hamiltonian is successfully block-diagonalized. Nevertheless, at the integrable limit where $h_z=0$, the system develops an additional symmetry. Specifically, the energy spectrum becomes invariant under a global spin-flip operation across the chain. This occurs because the spin-flip operator, $\Pi_n S_x^{(n)}$, commutes with the Hamiltonian when $h_z=0$~\cite{Craps:2019rbj}. To maintain a consistent number of symmetry sectors and avoid the numerical complexities of the emergent symmetry, we perform calculations at a point close to the exact integrable limit: $(h_x,h_z)=(-1,0.001)$.

Correct spectral analysis necessitates block-diagonalizing the Hamiltonian according to its conserved charges. This step is critical because spectral correlations, such as level repulsion, are only physically meaningful when evaluated within a single symmetry sector~\cite{DAlessio:2016aa,Craps:2019rbj}.

As illustrated in Fig. \ref{NNSD}, we present the level spacing distributions for the $H_1$ symmetry block (the $H_2$ block shows the same level spacing distribution). The chaotic case follows the Wigner-Dyson distribution, while the integrable case displays a Poisson distribution, characteristic of uncorrelated energy levels.
\begin{figure*}[]
\centering
\begin{minipage}{0.32\textwidth}
        \centering
        \includegraphics[width=\textwidth]{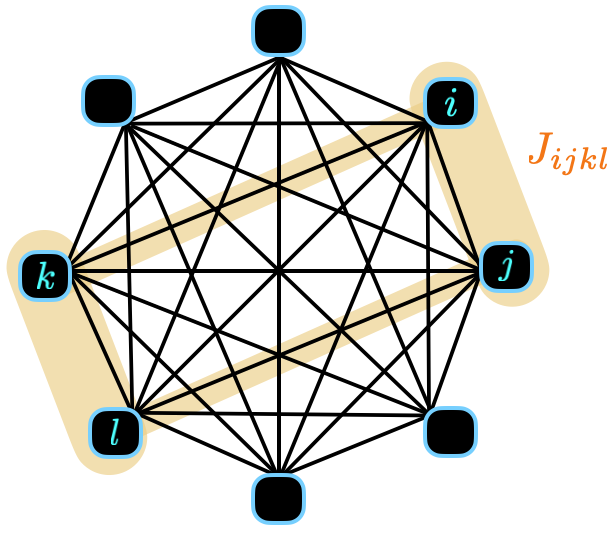}
\end{minipage}
\qquad\qquad
\begin{minipage}{0.29\textwidth}
        \centering
        \includegraphics[width=\textwidth]{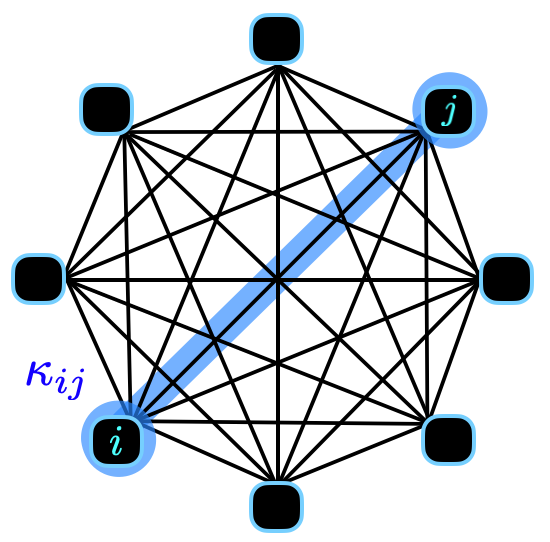}
\end{minipage}
\vspace{-0.3cm}
\caption{Schematic representation of the $q=4$ (left) and $q=2$ (right) SYK models. The mass-deformed SYK model Eq. \eqref{GSYK} is constructed from the combination of these two interaction scales.}\label{SYKSCHE}
\end{figure*}

Another paradigmatic class of spin chains characterized by local interactions is the Heisenberg XXZ model and its non-integrable deformations~\cite{Samaj:2013yva}. In integrable systems like the standard XXZ chain, the energy eigenvalues are uncorrelated, resulting in a Poisson distribution for the level spacing statistics. Integrability in such systems can be systematically broken by introducing next-to-nearest-neighbor couplings~\cite{Gubin:2012pxo,Rabinovici:2022beu}. The Hamiltonian for this deformed system is defined as: $H = H_{XXZ} + H_{NNN}$. The individual components are given by:
\begin{align} \label{eq-hamiltonianXXZ}
    H_{XXZ} &= \sum^{N-1}_{i=1}\,J\, (S^{x}_{i}\,S^{x}_{i+1} + S^{y}_{i}\,S^{y}_{i+1}) + J_{zz}\, S^{z}_i \,S^{z}_{i+1}\,,\cr
    H_{NNN} &=  \sum^{N-2}_{i=1} \, J_c\, S^{z}_{i}\,S^{z}_{i+2}\,,
\end{align}
where $S^{k}_{i}$ represents the spin-$1/2$ operators at site $i$, and $J_c$ denotes the next-to-nearest-neighbor coupling strength. The Hilbert space dimension remains $2^N$.

For our numerical analysis, we fix $(J, \, J_{zz})=(1, \, 0.5)$ and distinguish between the two regimes via the coupling $J_c$:
\begin{align}
(J, \, J_{zz}, \, J_c) = 
\begin{cases}
\,\,  (1, \, 0.5, \, 1) \,, \qquad (\text{Chaotic case})  \\
\,\,  (1, \, 0.5, \, 0) \,. \qquad (\text{Integrable case}) 
\end{cases}
\end{align}
This model preserves parity symmetry $[H,\hat{\Pi}]=0$ for all values of $J$, $J_{zz}$ and $J_c$. Furthermore, it conserves the total magnetization in the $z$-direction:$M_z =\sum_{i=1}^N S_i^z$. For a review of all the symmetries, see ~\cite{Joel:2013tpz}.

By fixing the chain length to $N=15$ and restricting our analysis to $M_z=-5/2$ sector, one can check the spectral properties of the system. Within both parity sectors, one finds the similar level spacing distributions as in Fig. \ref{NNSD} aligning with Wigner-Dyson distribution in the chaotic case, and the Poisson in the integrable case.

%
\subsection{Sachdev-Ye-Kitaev models: all-to-all interactions}
The Sachdev-Ye-Kitaev (SYK) model~\cite{Sachdev:1992fk,Kitaev2015Talk} and its variants represent a paradigmatic class of many-body systems with all-to-all interactions for investigating quantum chaos and the transition to integrability. These models serve as essential toy models in high-energy and condensed matter physics, particularly for their connections to holography and black hole physics~\cite{Rosenhaus:2018dtp, Trunin:2020vwy, Chowdhury:2021qpy} in the context of AdS/CFT correspondence.

In this section, we focus on two representative deformations: the mass-deformed SYK model~\cite{Song_2017,Eberlein_2017,Garcia-Garcia:2017bkg} and the sparse SYK model~\cite{Xu:2020shn,Garc_a_Garc_a_2021}. Both have been extensively characterized using level statistics, OTOCs, and more recently, Krylov complexity, for instance see~\cite{Song_2017,Eberlein_2017,Garcia-Garcia:2017bkg,Xu:2020shn,Garc_a_Garc_a_2021,Nosaka2018,Kim:2020mho,Garcia-Garcia:2020dzm,Lunkin:2020tbq,Nandy:2022hcm,Menzler:2024atb,C_ceres_2022,Orman:2024mpw,Caceres:2023yoj,Garcia-Garcia:2023jlu,Chen:2024imd}.

The mass-deformed SYK model consists of $N$ Majorana fermions in $0+1$ dimensions. It extends the standard SYK$_4$ Hamiltonian by adding a random quadratic (mass) term. Its Hamiltonian is given
\begin{align}\label{GSYK}
H = \frac{1}{4!}\sum^N_{i,j,k,l=1}\, J_{ijkl}\, \chi_i\, \chi_j\, \chi_k\, \chi_l\, + \frac{i}{2!}\, \sum^N_{i,j=1}\,\kappa_{ij}\, \chi_i\,\chi_j \,,
\end{align}
where Majorana fermions $\chi_i$ satisfy the algebra $\{\chi_i,\, \chi_j\} = \delta_{ij}$, acting on a Hilbert space of dimension $2^{{N}/{2}}$. Coupling constants $J_{ijkl}$ and $\kappa_{ij}$ are Gaussian random variables with zero mean, whose standard deviations are $\sigma_J=\sqrt{6}J/N^{3/2}$ and $\sigma_\kappa=\kappa/\sqrt{N}$, respectively. See Fig. \ref{SYKSCHE} for the schematic representations of SYK models.
 
The dynamics of the system are governed by the competition between the quartic and quadratic terms. The undeformed SYK model ($\kappa\rightarrow0$) is maximally chaotic, saturating the MSS bound~\cite{Maldacena_2016}. Its spectral statistics follow the Altland-Zirnbauer symmetry classification~\cite{You:2016ldz}, where the specific RMT ensemble is strictly determined by $N$ (mod $8$), such as GOE for $N=0$ (mod $8$), GUE for $N=2, 6$ (mod $8$), GSE for $N=4$ (mod $8$). As the mass deformation $\kappa$ increases ($\kappa>\kappa_c$), the system transitions toward integrability.

Evidence for this transition is further substantiated by level statistics and OTOCs, specifically regarding the behavior of the Lyapunov exponent \cite{Garcia-Garcia:2017bkg, Nosaka2018, Lunkin:2020tbq, Nandy:2022hcm, Garc_a_Garc_a_2021}, though the interpretation remains a subject of ongoing debate \cite{Kim:2020mho, Garcia-Garcia:2020dzm}. Furthermore, analytical results \cite{Monteiro:2020buv} demonstrate that for $\kappa>\kappa_c$ , the system enters a many-body localized phase where all states exhibit Poisson spectral correlations.
\begin{figure}[t!]
 \centering
{\includegraphics[width=7.5cm]{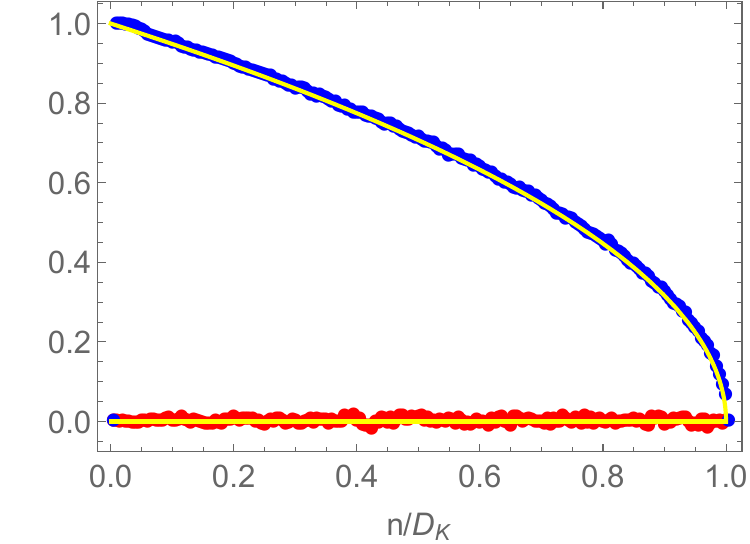}}
\caption{Mean Lanczos coefficients for the GUE with $N=D_{\rm K}=200$, averaged over $50$ realizations. Numerical results for $a_n$ (red dots) and $b_n$ (blue dots) are compared against the analytic predictions $a_n=0$ and $b_n=\sqrt{1-n}$ (yellow lines) as derived in \cite{Balasubramanian:2022dnj}.}\label{LANCRMT}
\end{figure}

Our second example focuses on the sparse SYK model \cite{Xu:2020shn, Garc_a_Garc_a_2021}, defined by the Hamiltonian:
\begin{align}\label{SSYK}
H = \frac{1}{4!}\sum^N_{i,j,k,l=1}\, x_{ijkl}\, J_{ijkl}\, \chi_i\, \chi_j\, \chi_k\, \chi_l \,,
\end{align}
with Majorana fermions $\chi_i$ and the Gaussian coupling $J_{ijkl}$ with zero mean and standard deviation
\begin{align}
    \sigma=\sqrt{\frac{6 J^2}{p N^{3}}} \,.
\end{align}
The sparsity parameter $x_{ijkl}$ is a random variable that equals $1$ with probability $p$, and $0$ with probability $1-p$: thus, $p$ determines the number of non-zero terms in the Hamiltonian. The specific relationship between the probability $p$ and the connectivity parameter $k$ is expressed as 
\begin{align}\label{k_order}
    kN=p\,\binom{N}{4} \,.
\end{align}

As $k$ decreases from the standard SYK limit ($p=1$), the model transitions from chaos to integrability. Numerical evidence from level statistics \cite{C_ceres_2022, Orman:2024mpw, Garc_a_Garc_a_2021} and the behavior of the Lyapunov exponent via OTOCs \cite{Caceres:2023yoj} suggests a critical threshold of $k_c\approx1$. Notably, Schwarzian dynamics and low-temperature gravitational features are maintained for $k$ values between $1/4$ and $4$ \cite{Xu:2020shn}, making the sparse SYK model a computationally efficient surrogate for exploring SYK-type physics~\cite{Xu:2020shn,Garc_a_Garc_a_2021,Orman:2024mpw} and traversable wormhole dynamics with quantum computings~\cite{Jafferis_2022,Caceres:2021nsa,Granet_2026}.

%
\section{Hermitian quantum system}\label{SECIV}

%
\subsection{Characteristic peak as an indicator for chaos}
Building upon the established framework of the Lanczos algorithm and the formalism of Krylov complexity, we now examine Krylov complexity as the probe of quantum chaos across several elementary quantum mechanical models. Our analysis commences with a random matrix theory, focusing on the Krylov complexity of a maximally entangled state, specifically, the TFD state at infinite temperature ($\beta=0$).

For the numerical simulations, we adopt the standard variances for the GOE, GUE, and GSE random matrices, set at $\sigma^2 = 2/N, 1/N$ and $1/(2N)$, respectively. To ensure statistical robustness, we utilize a Hilbert space dimension of $N=200$ averaged over $50$ independent realizations, selecting a non-degenerate block for the GSE case: the same setup for spectral statistics in Fig. \ref{LSDSIM}-\ref{SFFSIM}. We first characterize the behavior of the Lanczos coefficients; the mean values for the TFD state under a GUE Hamiltonian are illustrated in Figure \ref{LANCRMT}.
The Lanczos coefficient $a_n$ fluctuates around a mean value of zero, while $b_n$ vanishes as the ratio $n/D_{\rm K}$ approaches unity. This behavior signifies that the Krylov basis spans the entire Hilbert space ($D_{\mathcal H}=D_{\rm K}$), reflecting the capacity of unitary evolution to explore the full space: a characteristic feature of Lanczos coefficients across various ensembles. Notably, the analytical expressions for the mean coefficients ($a_n=0$ and $b_n=\sqrt{1-n}$) derived by \cite{Balasubramanian:2022dnj} via the density of states demonstrate excellent agreement with our numerical findings.

\begin{figure}[t!]
 \centering
{\includegraphics[width=7.5cm]{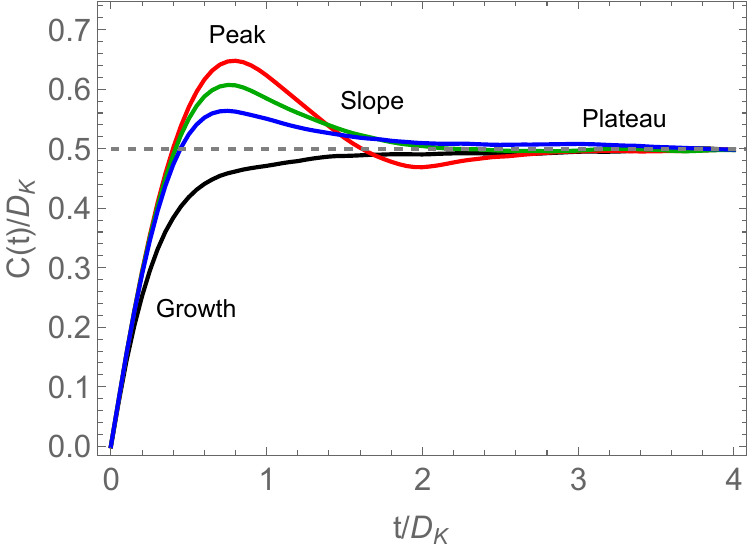}}
\caption{Krylov complexity for the GSE (red), GUE (green), and GOE (blue) ensembles with $N=200$ with $50$ realizations. The integrable Poisson system is also plotted (black). The peak magnitude is highest for the GSE, followed by the GUE and GOE cases, respectively.}\label{KrylovRMTFIG}
\end{figure}

The Krylov complexity, computed using the previously obtained Lanczos coefficients, is illustrated in Fig. \ref{KrylovRMTFIG}.
The complexity exhibits a characteristic four-stage dynamical evolution: an initial growth phase, followed by a slope, a peak, and a final plateau. This characteristic peak structure has been conjectured as a diagnostic signature of quantum chaos \cite{Balasubramanian:2022tpr}. This four-stage behavior mirrors the slope-dip-ramp-plateau structure commonly observed in the SFF, suggesting a profound correspondence between spectral statistics and the evolution of quantum state complexity.

Furthermore, the magnitude of the complexity peak serves as an indicator of the strength of eigenvalue correlations, with the GSE ensemble exhibiting the highest peak, followed by the GUE and GOE. Notably, the GSE case also displays a distinct suppression following the peak: a feature absent in the GOE and GUE results. This behavior likely parallels the sharp peak observed only in the SFF of GSE, unlike other ensembles~\cite{Guhr:1997ve,Cotler:2016fpe,Liu:2018hlr}.
\begin{figure*}[]
\centering
\begin{minipage}{0.18\textwidth}
\centering
\includegraphics[width=\textwidth]{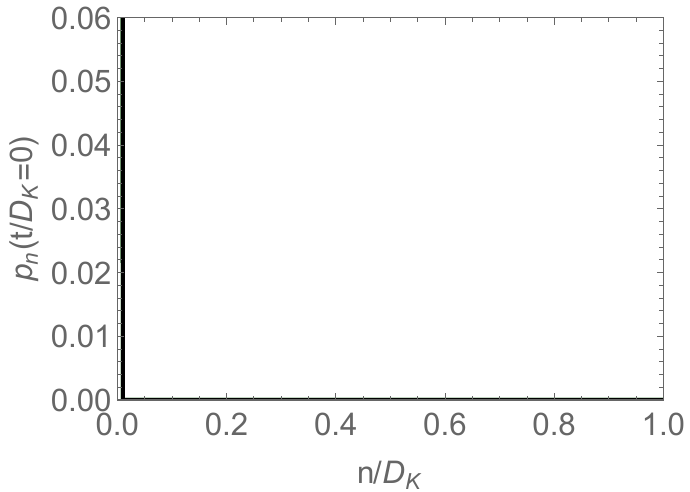}
\end{minipage}
\quad
\begin{minipage}{0.18\textwidth}
\centering
\includegraphics[width=\textwidth]{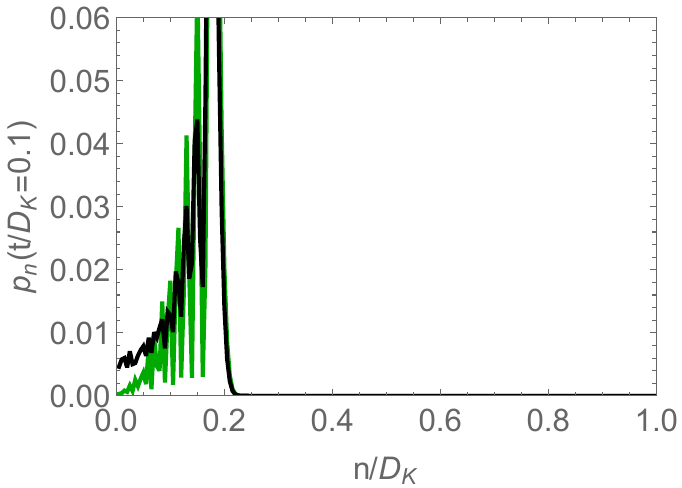}
\end{minipage}
\quad
\begin{minipage}{0.18\textwidth}
\centering
\includegraphics[width=\textwidth]{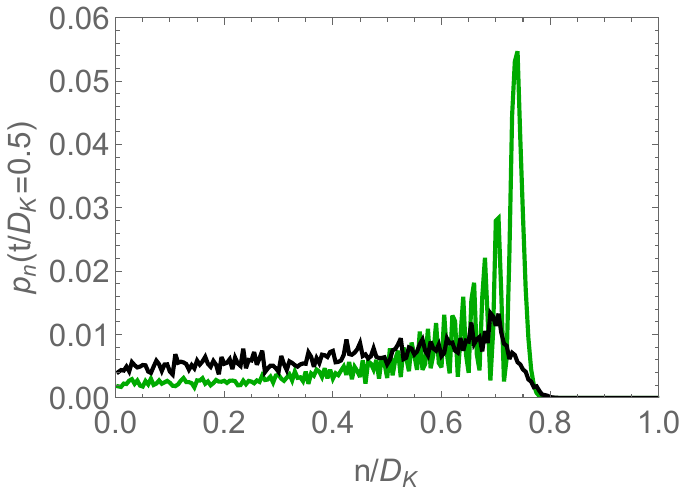}
\end{minipage}
\quad
\begin{minipage}{0.18\textwidth}
\centering
\includegraphics[width=\textwidth]{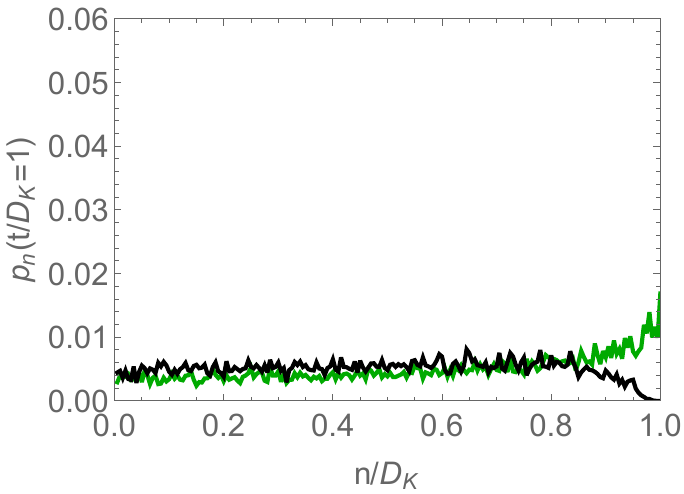}
\end{minipage}
\quad
\begin{minipage}{0.18\textwidth}
\centering
\includegraphics[width=\textwidth]{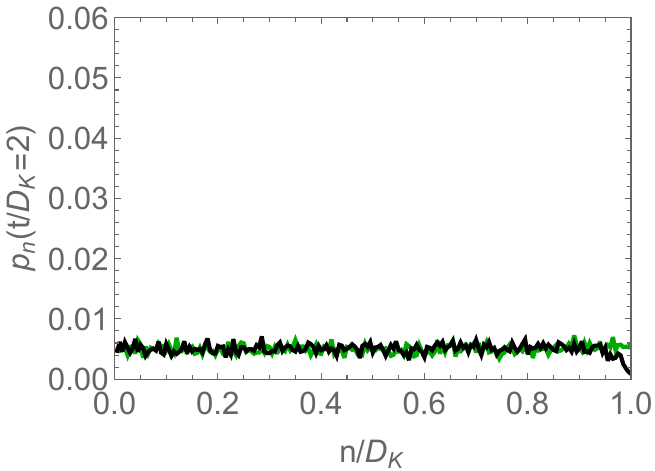}
\end{minipage}
\caption{Temporal snapshots of the probability distribution $p_n(t)$ within the Krylov basis for an evolving TFD state. Results are shown for a GUE (green) and Poissonian (black) Hamiltonian with $N=200$ at the time intervals indicated above each subpanel. The horizontal axis represents the normalized Krylov basis index, $n/D_{\rm K}\in[0,1]$, while the vertical axis denotes the probability of the state residing in a specific basis element.}\label{KrylovWaveFig}
\end{figure*}

It is worth noting that neither the initial growth nor the late-time saturation value of the Krylov complexity functions as a definitive indicator of chaos. Rather, it is the presence of the peak itself that is significant; for uncorrelated energy levels (i.e., the poisson theory), this peak vanishes \cite{Balasubramanian:2022tpr} (cf. the Poisson data in Fig. \ref{KrylovRMTFIG}). Finally, our numerical results are consistent with the analytical saturation value $C(t=\infty)=D_{\rm K}/2$ expected for a maximally entangled initial state \cite{Cotler:2016fpe,Erdmenger:2023wjg,Rabinovici:2020ryf,Rabinovici:2022beu}. In the following section, we provide an analytical treatment of this peak structure using $2\times2$ random matrix models.

It is also instructive to examine the time evolution of the wave function probability distribution within the Krylov basis, defined as $p_n(t) := \abs{\psi_n(t)}^2$. Figure \ref{KrylovWaveFig} illustrates the spread of this distribution from $t=0$ through the late-time saturation regime. Initially ($t=0$), the wave function is localized on the first Krylov basis element.

As time progresses, the dynamics manifest as a probability wavefront that propagates from the initial state toward higher-dimensional basis elements. Upon reaching the boundary of the Krylov space, it undergoes a reflection. This bounce off the Hilbert space boundary accounts for the characteristic downward slope observed in the Krylov complexity following its peak. Ultimately, the wave function becomes fully delocalized across the basis. In the late-time limit ($t\rightarrow\infty$), the probability distribution becomes uniform, $p_n(t\rightarrow\infty)=1/D_{\rm K}$, leading to an average position, and thus a saturation complexity of $C(t\rightarrow\infty)=D_{\rm K}/2$.

The existence of the peak of Krylov complexity signifies that the wavefront has reached the ``edge" of the Krylov space, the furthest available states from the initial state, before spectral correlations or level repulsion force it to delocalize back toward the middle of the chain (cfr. GUE vs. Poissonian at $p_n(t/D_{\rm K}=1)$ in Fig. \ref{KrylovWaveFig}).

After the peak, chaotic systems enter a period of decline known as the ``slope" before settling into a long-lived plateau. The slope is perhaps the most sensitive regime for distinguishing between different random matrix ensembles (GOE, GUE, GSE). It could be fundamentally driven by spectral rigidity: the tendency of energy levels to be evenly spaced rather than clustered. Spectral rigidity, a key feature of RMT, prevents the wave function from remaining at the furthest reach of the Krylov chain. Instead, the correlations between levels create a ``back-pressure" that drives the wave function toward a more uniform distribution.

To demonstrate the robustness of the peak structure as a universal signature of quantum chaos, we evaluate the Krylov complexity for quantum billiards and various spin-chain models.
\begin{figure}[t!]
 \centering
{\includegraphics[width=7.5cm]{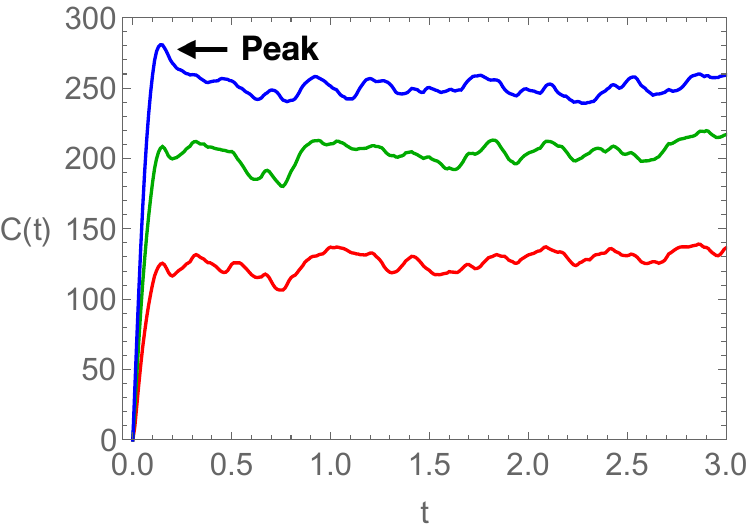}}
{\includegraphics[width=7.5cm]{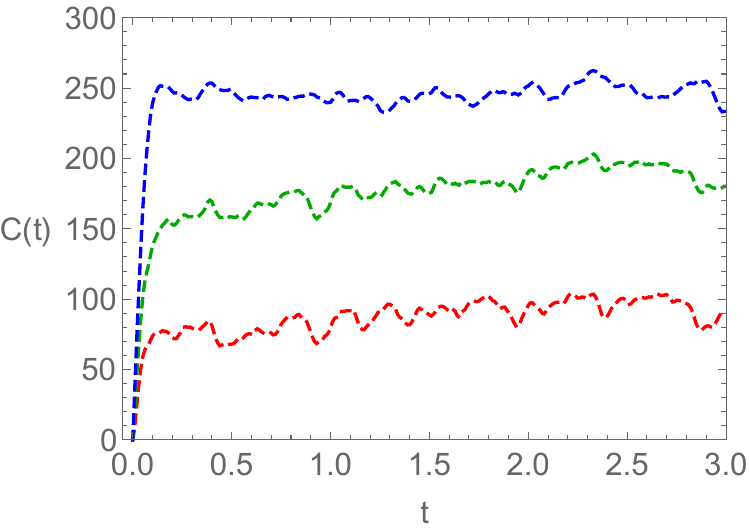}}
\caption{Krylov complexity for quantum billiard systems at inverse temperatures $\beta=0$ (blue), $0.001$ (green), and $0.003$ (red). The upper panel depicts the chaotic regime ($a/R=1$), while the lower panel illustrates the integrable case ($a/R=0$).}\label{SCQBS}
\end{figure}

The results for the billiard systems are presented in Fig. \ref{SCQBS}, showing the evolution across different temperature scales~\cite{Hashimoto:2023swv,Huh:2023jxt,Balasubramanian:2024ghv}. Specifically, the characteristic peak structure is suppressed as the temperature decreases (increasing $\beta$), a trend that is qualitatively consistent with the behaviors observed in the RMT ensembles~\cite{Balasubramanian:2022tpr}.

\begin{figure*}[]
\centering
\begin{minipage}{0.23\textwidth}
\centering
\includegraphics[width=\textwidth]{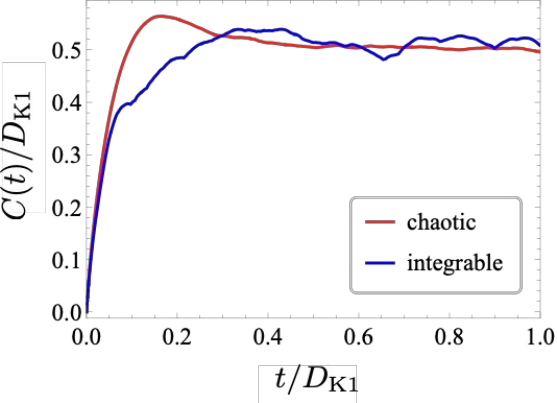}
\end{minipage}
\quad
\begin{minipage}{0.23\textwidth}
\centering
\includegraphics[width=\textwidth]{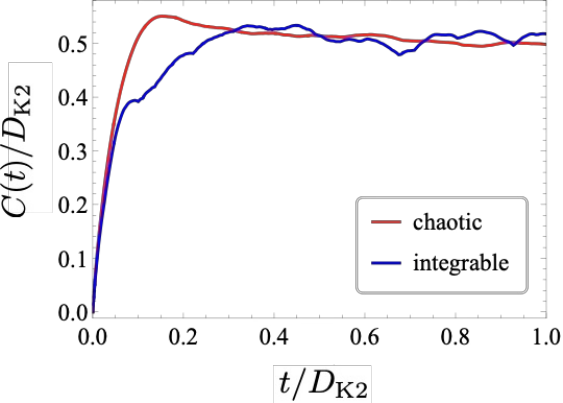}
\end{minipage}
\quad
\begin{minipage}{0.23\textwidth}
\centering
\includegraphics[width=\textwidth]{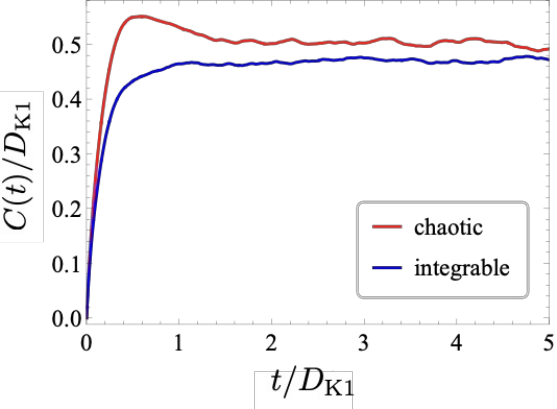}
\end{minipage}
\quad
\begin{minipage}{0.23\textwidth}
\centering
\includegraphics[width=\textwidth]{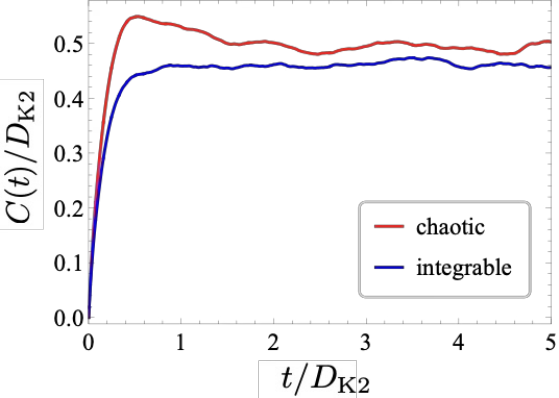}
\end{minipage}
\caption{The left two panels illustrate Krylov complexity for the mixed-field Ising model ($N=12$), while the right two panels display Krylov complexity for the XXZ spin chain with $N=15$ and $M_z=-5/2$. In all figures, $C/{D_{\rm K}}_{1, 2}$ denote the Krylov complexity calculated for distinct symmetry sectors (1 or 2).}\label{FIG_SPIN_CHAINS}
\end{figure*}

Furthermore, Fig.~\ref{FIG_SPIN_CHAINS} contrasts the complexity dynamics in many-body systems~\cite{Camargo:2024deu}. For the mixed-field Ising model ($N=12$), we observe a clear distinction between the chaotic and integrable phases within each symmetry sector. Similarly, for the XXZ spin chain with next-nearest-neighbor interactions ($N=15$, $M_z=-5/2$), the chaotic deformation exhibits the characteristic peak, which is absent in the integrable limit. More generally, resolving Krylov dynamics into fixed symmetry sectors leads to symmetry-resolved Krylov complexity~\cite{Caputa:2025mii,Kotta:2026utn,Das:2026gko}. Such a decomposition is also natural when comparing Krylov complexity with spectral diagnostics, which must be evaluated within fixed symmetry sectors. A more detailed analysis of the chaos-to-integrability transition, particularly in SYK models, is presented in subsequent sections.
%

%
\subsection{Spectral form factor and time-scale analysis}
There exists an intriguing observation regarding the shared dynamical features of the SFF and Krylov complexity. While the SFF is characterized by its well-known slope-dip-ramp-plateau structure, Krylov complexity of chaotic systems exhibits an initial growth, slope, a characteristic peak, and late-time saturation \cite{Balasubramanian:2022tpr, Erdmenger:2023wjg}. This correspondence suggests a deep link between spectral statistics and the growth of state complexity, a finding we further investigate in our analysis of various spin-chain models~\cite{Camargo:2024deu}.
\begin{figure}[t!]
 \centering
{\includegraphics[width=7.5cm]{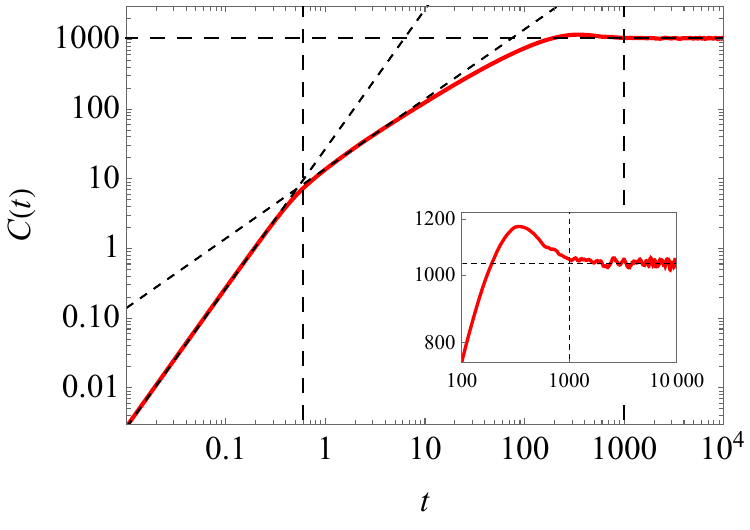}}
{\includegraphics[width=7.5cm]{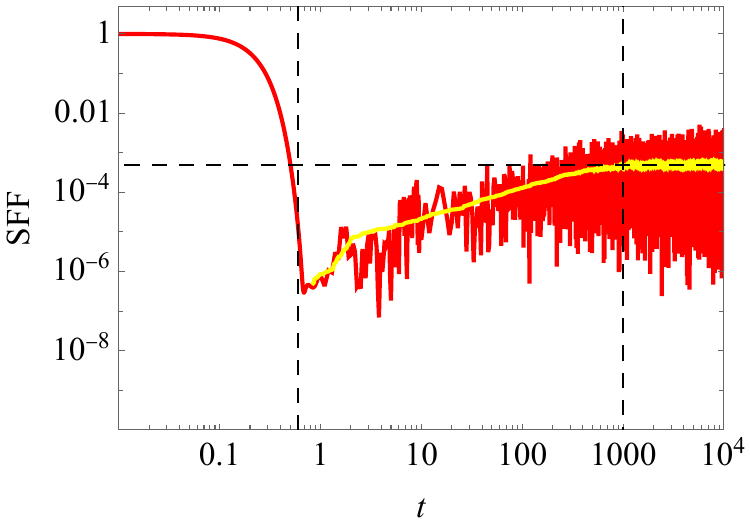}}
\caption{Log-log evolution of Krylov complexity and the SFF for the mixed-field Ising model in the chaotic regime ($\beta=0, N=12$). The data corresponds to the positive parity sector $H_1$ with ${D_{\rm K}}_1=2080$. The yellow curve represents a moving average employed to suppress numerical fluctuations and highlight the underlying dynamical trends.}\label{MFI_spread_Log}
\end{figure}
\begin{figure}[t!]
 \centering
{\includegraphics[width=7.5cm]{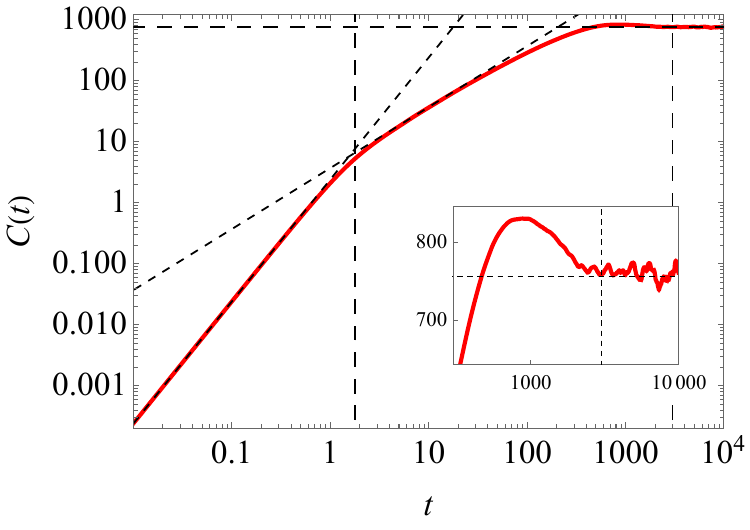}}
{\includegraphics[width=7.5cm]{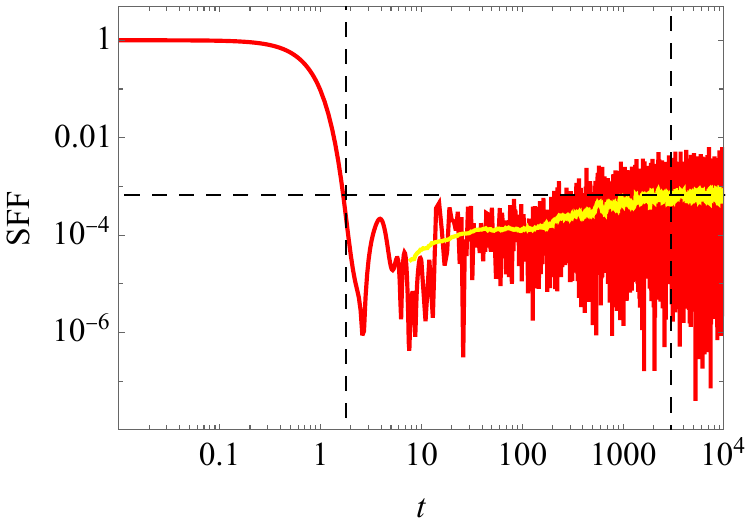}}
\caption{Log-log evolution of Krylov complexity and the SFF for the XXZ spin chain model in the chaotic regime  ($\beta=0, N=15$) of the sector $H_1$ with ${D_{\rm K}}_1=1512$ and $M_z=-5/2$.}\label{XXZ_Log_spread}
\end{figure}

Figures \ref{MFI_spread_Log}-\ref{XXZ_Log_spread} illustrate the Krylov complexity and SFF for the mixed-field Ising and Heisenberg XXZ models. Those results align with the propositions in \cite{Erdmenger:2023wjg}, suggesting a deep correspondence between the transition timescales of these two observables. For instance, one can find that both the timescale of the Krylov complexity saturation and the SFF plateau scale linearly with the Hilbert space dimension $D_{\mathcal H}$.

The emergence of a ``clear" ramp in the SFF, highlighted by the moving average (yellow line), appears to coincide with the onset of the Krylov complexity's peak structure following its linear growth phase. Notably, in the integrable limit, the SFF lacks a clear ramp and the Krylov complexity fails to exhibit a peak, reinforcing the link between these features and quantum chaos.
\begin{figure*}[]
\centering
\begin{minipage}{0.29\textwidth}
        \centering
        \includegraphics[width=\textwidth]{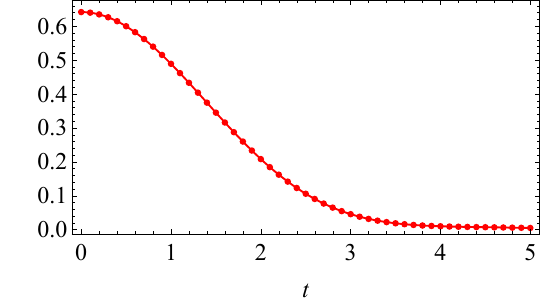}
\end{minipage}
\quad
\begin{minipage}{0.29\textwidth}
        \centering
        \includegraphics[width=\textwidth]{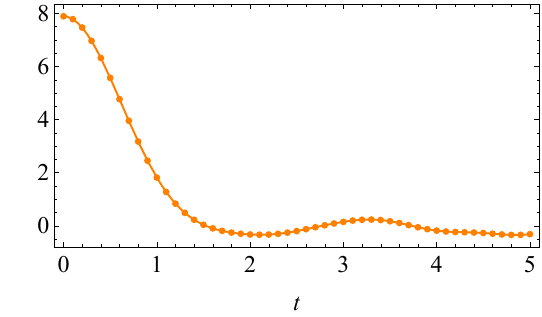}
\end{minipage}
\quad
\begin{minipage}{0.29\textwidth}
        \centering
        \includegraphics[width=\textwidth]{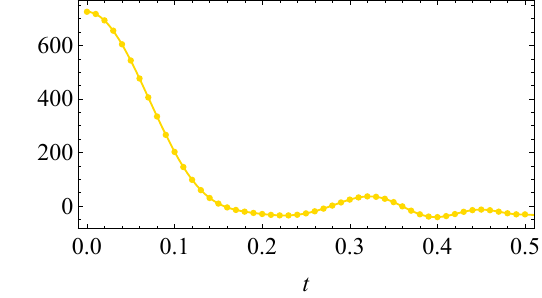}
\end{minipage}

\begin{minipage}{0.31\textwidth}
        \centering
        \includegraphics[width=\textwidth]{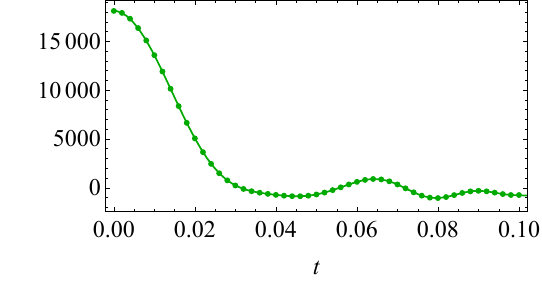}
\end{minipage}
\quad
\begin{minipage}{0.31\textwidth}
        \centering
        \includegraphics[width=\textwidth]{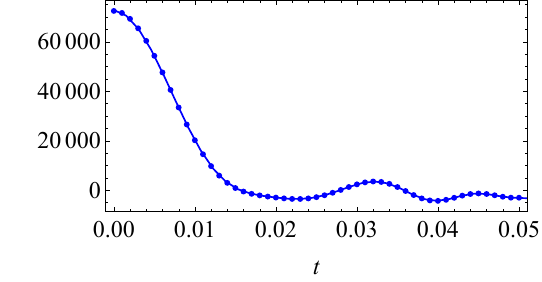}
\end{minipage}
\quad
\begin{minipage}{0.31\textwidth}
        \centering
        \includegraphics[width=\textwidth]{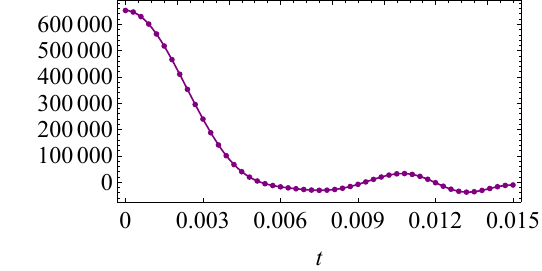}
\end{minipage}
\caption{Numerical validation of the Ehrenfest theorem \eqref{EHRTH2} in mass-deformed SYK models for various deformation parameters, interpolating SYK$_{q=4}$ and SYK$_{q=2}$ models: $\kappa=0$ (red), $1$ (orange), $10$ (yellow), $50$ (green), $100$ (blue), and $300$ (purple). Solid lines represent the numerical second time-derivative of the Krylov complexity (L.H.S.), while the dots denote the analytical expression derived from the Lanczos coefficients (R.H.S.).}\label{SMFIG5}
\end{figure*}

The SFF dynamics are characterized by two primary timescales, $t_\text{dip}$ and $t_\text{plateau}$ (indicated by vertical dashed lines). We distinguish three dynamical regimes:
(I) Initial Regime ($t \lesssim t_\text{dip}$): Krylov complexity undergoes quadratic growth, a behavior we shall analytically prove in the subsequent section, while the SFF exhibits its characteristic initial slope.
(II) Intermediate Regime ($t \gtrsim t_\text{dip}$): the complexity transitions to a linear growth phase, while the SFF enters the ramp phase, characterized by significant oscillations around a linear trend.
(III) Saturation Regime ($t \sim t_\text{plateau}$): Both observables reach their respective saturation values, with the complexity peak occurring shortly before the onset of the plateau.

Our analysis further revealed that $t_\text{plateau}$ remains comparable between the SFF and Krylov complexity even in integrable cases \cite{Camargo:2024deu}. This suggests that such a timescale is governed by the system's symmetry and dimensionality rather than the presence of chaos. Instead, the definitive signatures of chaos are the existence of the SFF ramp and the complexity peak. While saturation values can depend on the degeneracy in the spectra and the initial state, they are correlated for the maximally entangled state (e.g., infinite-temperature TFD state), as determined by Eq. \eqref{SFFCRE}.

Finally, we observe that $t_\text{dip} \sim \mathcal{O}(1)$ and $t_\text{plateau} \sim D_{\mathcal H}$ for these spin chain models, independent of system size. This stands in contrast to RMT predictions where $t_\text{dip} \sim \sqrt{D_{\mathcal H}}$  \cite{Cotler:2016fpe}. In RMT, the dip time is sensitive to the disconnected part of the SFF and the edge of the semicircular density of states (DOS). However, spin chains typically exhibit a Gaussian DOS, and this disparity likely accounts for the deviation in scaling behavior. The precise mechanism by which a Gaussian DOS leads to a size-independent dip time remains an open question for further investigation.

%
\subsection{Ehrenfest theorem and early-time growth}
A significant property of Krylov complexity is its adherence to the Ehrenfest theorem \cite{Erdmenger:2023wjg}:
\begin{align}\label{EHRTH}
\begin{split}
\partial^2_t C(t) = - \left[ \left[ C(t)\,, \mathcal{L} \right], \mathcal{L} \right] \,,
\end{split}
\end{align}
where $\mathcal{L} = H \otimes \mathbb{I}$ denotes the Liouvillian, constructed using the identity operator $\mathbb{I}$. By applying the evolution equation \eqref{SCHRO}, the Ehrenfest relation \eqref{EHRTH} can be reformulated in terms of the Lanczos coefficients and the state amplitudes:
\begin{align}\label{EHRTH2}
\begin{split}
\partial^2_t C(t) &= 2 \sum_n  \Big[ \left( b_{n+1}^2 - b_n^2 \right) \psi_n(t) \psi_n^{*}(t)  \\
 & \qquad + \left(a_{n+1} - a_{n}\right)b_{n+1}  \psi_{(n+1}(t) \psi_{n)}^{*}(t)  \Big]\,,
\end{split}
\end{align}
where we define the symmetric part as $\mathcal{T}_{(a} \tilde{\mathcal{T}}_{b)}:=\frac{1}{2}\left(\mathcal{T}_{a}\tilde{\mathcal{T}}_{b}+\mathcal{T}_{b}\tilde{\mathcal{T}}_{a}\right)$. Notably, \eqref{EHRTH2} is a universal relation valid for any system by construction.

In Fig.~\ref{SMFIG5}, we numerically verify the Ehrenfest theorem \eqref{EHRTH2} for the mass-deformed SYK model \eqref{GSYK}. The results demonstrate the precise correspondence between the second time derivative of the Krylov complexity and the specific combination of Lanczos coefficients and transition amplitudes dictated by the theorem.

Furthermore, one can derive the analytical behavior of the Krylov complexity in the early-time regime ($t\rightarrow0$). A related analysis regarding Krylov operator complexity can be found in \cite{Fan:2022xaa}. By imposing the boundary conditions at $t=0$ on Eq. \eqref{SCHRO}, we obtain
\begin{align}\label{ANST}
\begin{split}
   \psi_{n}(0) &= \delta_{n0} \,, \\
 \dot{\psi}_{n}(0) &= -i \left(b_1 \delta_{n1} + a_0 \delta_{n0}\right) \,, \\
 \ddot{\psi}_{n}(0) &= - \left(a_0^2 + b_1^2\right) \delta_{n0} - a_0 b_1 \delta_{n1} - b_1 b_2 \delta_{n2} \,,
\end{split}
\end{align}
which leads to the following initial conditions for the complexity and its derivatives
\begin{align}\label{ICSET}
\begin{split}
C(0) = 0 \,, \quad \dot{C}(0) = 0\,, \quad \ddot{C}(0) = 2 b_1^2 \,.
\end{split}
\end{align}
Combining these initial conditions with the definition of spread complexity \eqref{SCDEF}, we arrive at the early-time expansion
\begin{align}\label{ANAC}
\begin{split}
C(t) = b_1^2 t^2 + \mathcal{O}(t^3) \,.
\end{split}
\end{align}
This confirms that the Krylov complexity exhibits universal quadratic growth at early times. Interestingly, as also observed in the context of operator complexity \cite{Fan:2022xaa}, the first-order Lanczos coefficients $a_n$ do not contribute to this leading-order growth, rendering the initial dynamics dependent solely on $b_1$. The early time growth in figures \ref{MFI_spread_Log}-\ref{XXZ_Log_spread} are consistent with \eqref{ANAC}.

Notably, the quadratic early-time growth of Krylov complexity observed in Eq. \eqref{ANAC} is related to another quantum complexity: spectral complexity \cite{Iliesiu:2021ari, Erdmenger:2023wjg, Camargo:2024deu, Balasubramanian:2024ghv}. A detailed comparative analysis of Krylov and spectral complexity, particularly within the framework of spin-chain models, is provided in \cite{Camargo:2024deu}. Furthermore, while the Krylov complexity peak serves as a primary diagnostic for chaotic dynamics, the late-time saturation behavior of spectral complexity has also been suggested as a robust indicator of the chaos-to-integrability transition \cite{Camargo:2023eev, Balasubramanian:2024ghv}, associated with the level statistics.

%
\subsection{Lanczos coefficients as chaos probes}
\begin{figure}[h!]
\centering
{\includegraphics[width=7.5cm]{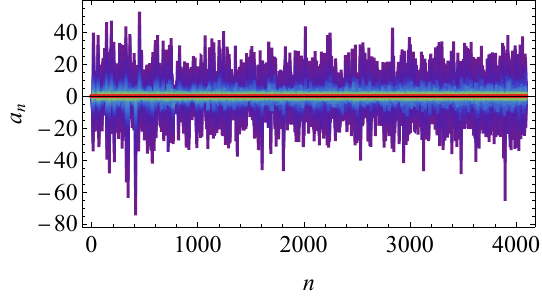}}
\hfill
{\includegraphics[width=7.5cm]{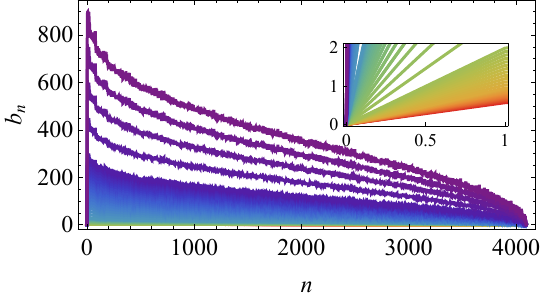}}
\caption{Lanczos coefficients $\{a_n,\,b_n\}$ for the mass-deformed SYK model across a range of deformation parameters. The colors indicate the strength of the deformation $\kappa = 0$ (red) to $\kappa =300$ (purple).} \label{SMFIG1}
\end{figure}

The Lanczos coefficients for the mass-deformed SYK model, computed via the Lanczos algorithm, are presented in Fig.~\ref{SMFIG1}. We observe that $a_n$ exhibits characteristic oscillations, while $b_n$ undergoes an initial growth (see inset), reaches a maximum, and subsequently decays as $n$ approaches the Hilbert space dimension. These behaviors are qualitatively consistent with results previously reported for the standard SYK model ($\kappa=0$) \cite{Balasubramanian:2022tpr} and various RMT ensembles \cite{Erdmenger:2023wjg}, where Lanczos coefficients can also be related to orthogonal-polynomial recursion coefficients~\cite{Qu:2025lgo,Pedraza:2026zji,Qu:2026dmv}.
\begin{figure}[]
\centering
\qquad {\includegraphics[width=7.0cm]{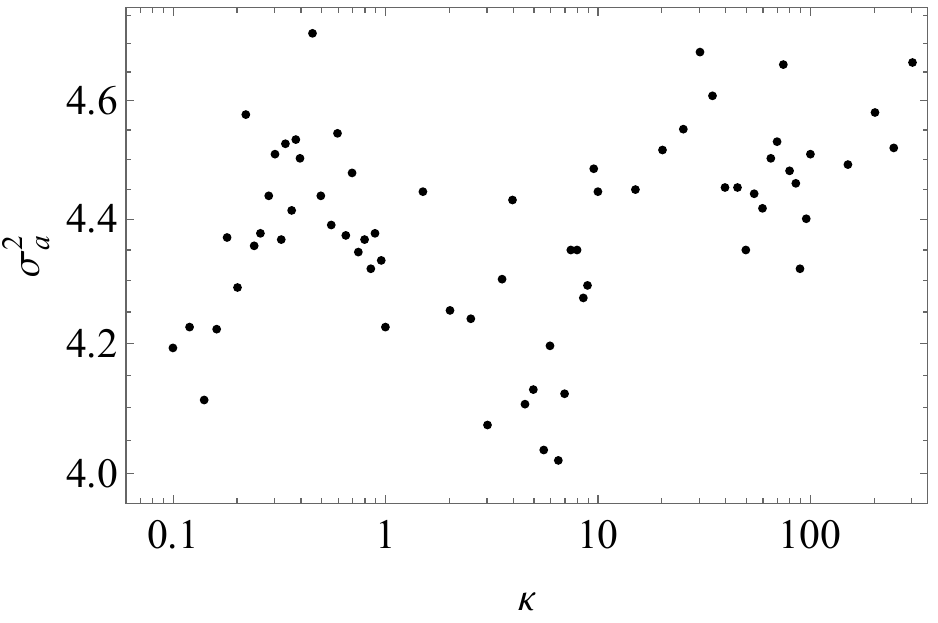}}
\qquad
{\includegraphics[width=7.6cm]{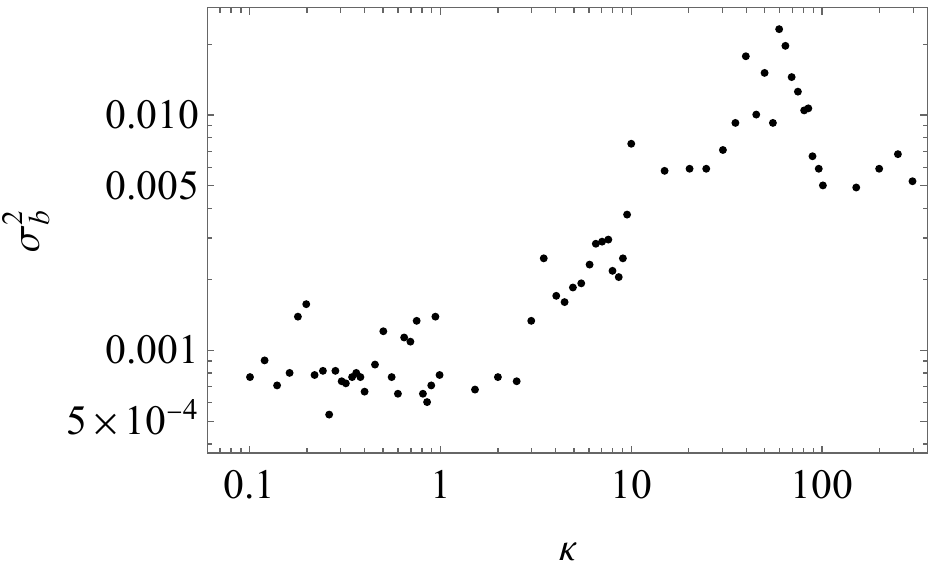}}
\caption{The variance of the Lanczos coefficients $\sigma_{a, b}^2$ vs. $\kappa$.} \label{SMFIG2}
\end{figure}
\begin{figure}[]
\centering
{\includegraphics[width=7.3cm]{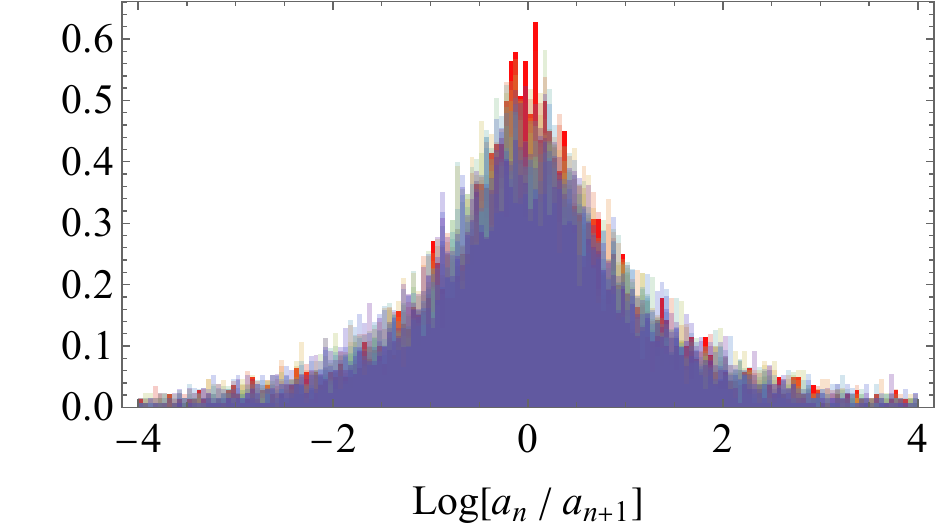}}
\hfill
{\includegraphics[width=7.5cm]{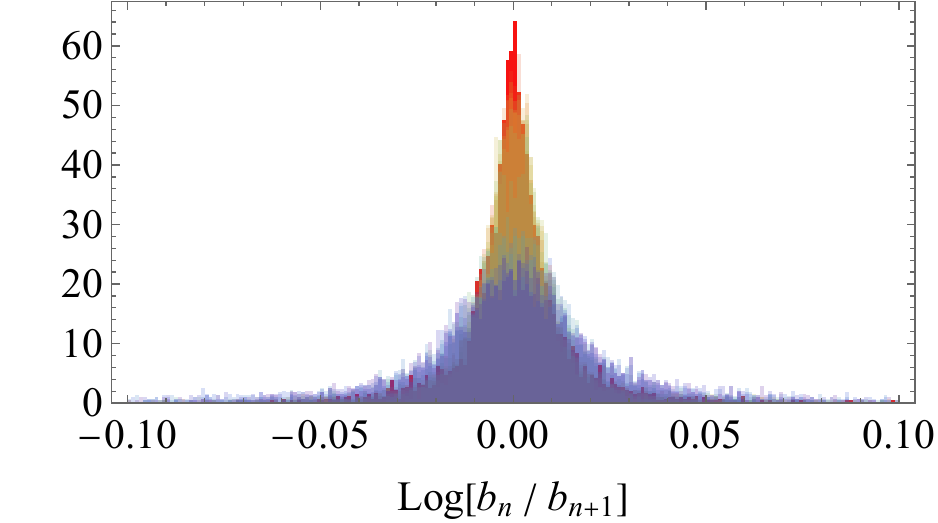}}
\caption{Histogram of the Lanczos coefficients $a_n$ (upper panel) and $b_n$ (lower panel) for values of $\kappa$ ranging from $\kappa = 0$ (red) to $\kappa =300$ (purple).} \label{SMFIG3}
\end{figure}

Furthermore, we evaluate the variance of the Lanczos coefficients, defined as \cite{Hashimoto:2023swv}:
\begin{align}\label{}
\begin{split}
    \sigma_{a}^2 &:= \text{Var} \left(x_i^{(a)}\right) \,, \quad x_i^{(a)} := \log \left( \frac{a_{2i-1}}{a_{2i}} \right) \,, \\
    \sigma_{b}^2 &:= \text{Var} \left(x_i^{(b)}\right) \,, \quad x_i^{(b)} := \log \left( \frac{b_{2i-1}}{b_{2i}} \right) \,.
\end{split}
\end{align}
For further context on these variances in the regime of operator complexity, see \cite{Rabinovici:2021qqt,Rabinovici:2022beu}. As illustrated in Fig.~\ref{SMFIG2}, the variance is significantly amplified in the integrable regime (large $\kappa$) compared to the chaotic regime (small $\kappa$): this feature is particularly evident in $\sigma_{b}^2$. These conclusions are further supported by the distribution histograms in Fig.~\ref{SMFIG3}, which visually contrast the coefficient distributions across different regimes.

%
\subsection{Chaos-integrability transition}
Following the framework established in \cite{Baggioli:2024wbz,Huh:2024ytz}, we further investigate the chaos-to-integrability transition in mass-deformed \eqref{GSYK} and sparse \eqref{SSYK} SYK models through the Krylov complexity. Our results demonstrate that the smooth crossover between chaotic and integrable regimes is consistently reflected in both traditional spectral diagnostics, such as the SFF and level spacing distributions, and the dynamical evolution of the characteristic Krylov complexity peak. This correspondence reinforces the utility of state complexity as a robust indicator of underlying Hilbert space dynamics.

\begin{figure}[]
\centering
{\includegraphics[width=7.5cm]{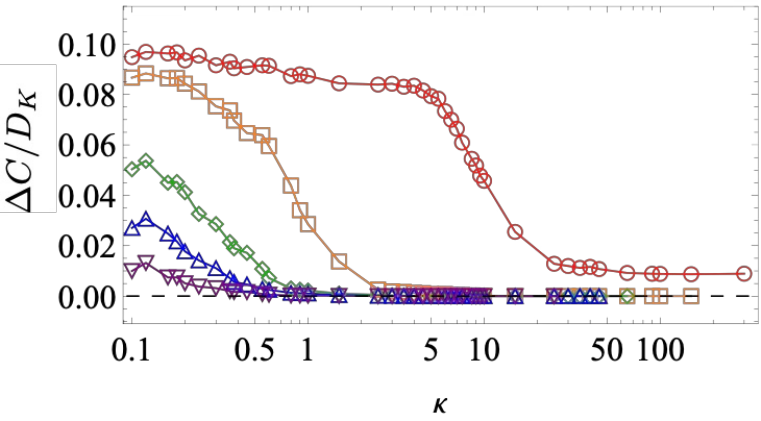}}
\hfill
{\includegraphics[width=7.5cm]{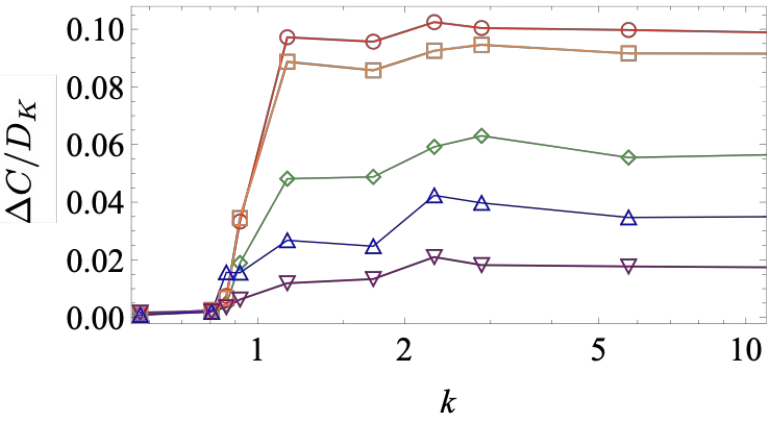}}
\caption{Size of peak value of Krylov complexity $\Delta C$, Eq. \eqref{eq:diff_peak}, for $\beta=0,\,1,\,3,\,5,\,10$ (red, orange, green, blue, purple): the upper panel for the mass-deformed SYK and lower panel for the sparse SYK model, respectively.} \label{KRYSYK1}
\end{figure}

To quantify the peak behavior as a diagnostic of quantum chaos, we define an order parameter, $\Delta C$, representing the difference between the peak value and the late-time plateau:
\begin{align}\label{eq:diff_peak}
\Delta C := C(t=t_{\text{peak}}) - C(t\rightarrow\infty) \,.
\end{align}
By construction, a non-vanishing $\Delta C$ characterizes the chaotic regime, whereas $\Delta C=0$ signifies the onset of integrability. See Fig.~\ref{KRYSYK1} for the results of SYK models.

For the mass-deformed SYK model at infinite temperature ($\beta=0$), $\Delta C$ exhibits a smooth transition from approximately $0.1$ at $\kappa=0$ to zero at large $\kappa$. The point at which $\Delta C$ vanishes aligns with the critical value $\kappa_c \approx 66$, consistent with previous findings from OTOC analysis and spectral statistics \cite{Garcia-Garcia:2017bkg}. Conversely, the sparse SYK model exhibits a transition to $\Delta C\approx 0$ at a critical sparsity of $k_c\approx1$, in agreement with $r$-parameter statistics reported in \cite{Garcia-Garcia:2020cdo}.

The behavior of the mass-deformed SYK model is highly sensitive to the inverse temperature $\beta$. As the system moves away from the infinite-temperature limit, the magnitude of $\Delta C$ in the chaotic phase decreases, indicating that the relative height of the peak is temperature-dependent. Furthermore, the point of continuous transition shifts toward lower values of $\kappa$, and the transition itself broadens from a sharp critical point into a wider crossover. This temperature-driven shift in the phase boundary is well-described by an empirical exponential function, $\kappa_c (\beta) \approx \kappa_c(0)\, e^{-\frac{2}{\pi} \beta}$. Physically, higher temperatures allow the system to explore a broader range of energy states, maintaining chaotic spectral features across a larger parameter space. This trend is consistent with alternative spectral probes~\cite{Garcia-Garcia:2017bkg} that indicate the integrable phase becomes more favorable as the temperature decreases.

In contrast to the mass-deformed case, the critical threshold for the sparse SYK model remains approximately constant at $k_c\approx1$ regardless of the value of $\beta$. This stability suggests that at $k\approx1$, the system reaches a maximum level of sparsity that eliminates level repulsion, rendering the Hamiltonian's lack of chaos insensitive to thermal probes. At sufficiently small $p$ (or $k<1$), a significant number of emergent discrete symmetries, including chiral symmetries, lead to exact spectral degeneracies~\cite{Garc_a_Garc_a_2021}. These degeneracies suppress the saturation value of the Krylov complexity in the TFD state~\cite{Erdmenger:2023wjg,Camargo:2024deu}, a behavior consistent with recent analyses using different initial states \cite{Jha:2024nbl}.

\begin{figure}[]
\centering
{\includegraphics[width=7.5cm]{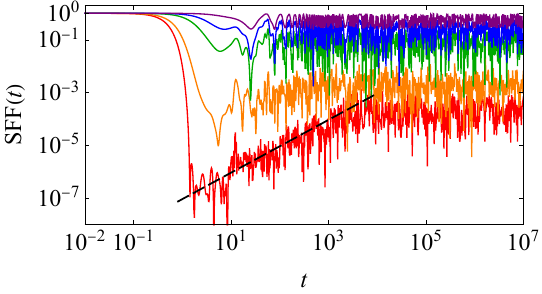}}
\hfill
{\includegraphics[width=7.5cm]{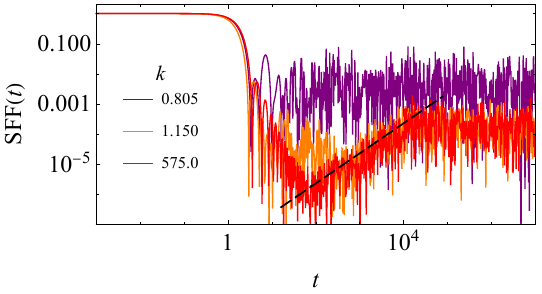}}
\caption{SFF in SYK models. The upper panel displays the mass-deformed SYK model with $\kappa=1$ at inverse temperatures $\beta=0,\,1,\,3,\,5,\,10$ (red, orange, green, blue, and purple). The lower panel illustrates the sparse SYK model at infinite temperature ($\beta=0$). In both cases, black dashed lines denote the linear ramp, signifying the growth of spectral correlations.} \label{KRYSYK2}
\end{figure}

Ultimately, $\Delta C$ proves to be a reliable order parameter for detecting chaotic-to-integrable transitions, showing a direct correspondence with the behavior of the SFF. As illustrated in Fig. \ref{KRYSYK2}, the linear ramp characteristic of the SFF vanishes in the regions where $\Delta C\rightarrow0$, consistent with Fig. \ref{KRYSYK1}. 

Complementing our analysis of the SFF, we now examine the chaos-to-integrability transition through level spacing statistics. In systems exhibiting mixed dynamics, the spectrum typically follows a variant of the Brody distribution \cite{Brody1973, Brody:1981aa}, which interpolates between the Poisson limit of integrable systems and the Wigner-Dyson statistics of RMT. A generalized form of this probability density function is given by \cite{Jafarizadeh:2012aa, Sabri:2012opl}, generalized Brody distribution:
\begin{equation}\label{GBD}
P(s) =  (b+1) \, c_b \, (\alpha \, s^b + \bar{\alpha}\, s^{b+1}) \, \text{exp}\left(-c_b \, s^{b+1} \right)  \,,
\end{equation}
where the parameters $\alpha$, $\bar{\alpha}$, and $c_b$ are defined as
\begin{align}\label{}
\begin{split}
\!\!\!\alpha = 1-\frac{
\left[\frac{\Gamma\left(\frac{b+2}{b+1}\right)}{c_b^\frac{1}{b+1}}\right]^2
-\frac{\Gamma\left(\frac{b+2}{b+1}\right)}{c_b^\frac{1}{b+1}}}{\left[\frac{\Gamma\left(\frac{b+2}{b+1}\right)}{c_b^\frac{1}{b+1}}\right]^2
-\frac{\Gamma\left(\frac{b+3}{b+1}\right)}{c_b^\frac{2}{b+1}}} \,, \,\,\,\,\,
\bar{\alpha} = \frac{
\frac{\Gamma\left(\frac{b+2}{b+1}\right)}{c_b^\frac{1}{b+1}}
-1}{\left[\frac{\Gamma\left(\frac{b+2}{b+1}\right)}{c_b^\frac{1}{b+1}}\right]^2
-\frac{\Gamma\left(\frac{b+3}{b+1}\right)}{c_b^\frac{2}{b+1}}} \,,
\end{split}
\end{align}
and 
\begin{equation}\label{GBDsub}
\!\!\!\!\!c_b = \left[\Gamma\left(\frac{b+2}{b+1}\right)\right]^{\,\epsilon (b+1)}\!\!\!\!\!\!\!\!\!, \quad 
\epsilon = 
\begin{cases}
 +1 \,\, \text{for Poisson $\leftrightarrow$ GOE,} \\
 -1 \,\, \text{for Poisson $\leftrightarrow$ GUE,}
\end{cases}
\end{equation}
The parameter $\epsilon$ determines the symmetry class: $\epsilon = +1$ yields the original Brody distribution~\cite{Brody1973,Brody:1981aa} interpolating between Poisson ($b=0$) and GOE ($b=1$), while $\epsilon = -1$ covers the transition from Poisson to GUE. The Brody parameter $b$ serves as a measure of the ``amount" of chaos.

To build intuition for how Krylov complexity characterizes mixed phases, we first consider a $2 \times 2$ matrix model \cite{Nieminen_2017}. For the GOE-Poisson transition, the Hamiltonian is:
\begin{align}\label{ACT1}
    H = 
\begin{pmatrix}
|\mathcal{A}|^d & \frac{1}{2}|\mathcal{B}|^d \\
\frac{1}{2}|\mathcal{B}|^d & 0 
\end{pmatrix} \,,
\end{align}
where $\mathcal{A}$ and $\mathcal{B}$ are Gaussian random variables with variance $\sigma$. The level spacing $s = E_{+} - E_{-}$, where $E_{\pm} = \frac{1}{2} \left[ |\mathcal{A}|^d \pm \sqrt{|\mathcal{A}|^{2d} + |\mathcal{B}|^{2d}} \right]$, follows the probability density for the spacing, $P(s)$:
\begin{align}\label{ee1}
\begin{split}
P &= \frac{2}{\pi \sigma^2} \int_{0}^{\infty} \dd \mathcal{A}\,\dd \mathcal{B}\,\, \text{exp}\left(-\frac{\mathcal{A}^2+\mathcal{B}^2}{2\sigma^2}\right)    \,\\
&= \frac{2}{\pi d^2 \sigma^2} \int_{0}^{\infty} \dd s \int_{0}^{\frac{\pi}{2}} \dd \phi \,\, s^{\frac{2}{d}-1} \cos^{\frac{1}{d}-1}\phi \, \sin^{\frac{1}{d}-1} \phi  \\ 
&\quad \times \text{exp}\left( - \frac{s^{\frac{2}{d}} \left(\cos^{\frac{2}{d}} \phi + \sin^{\frac{2}{d}} \phi \right)}{2\sigma^2} \right) \\
& \equiv \int_{0}^{\infty} \dd s \, P(s)\,,
\end{split}
\end{align}
that closely maps to the Brody distribution with $b = (2-d)/d$. Here, $d=1$ recovers the GOE limit ($b=1$), while $d=2$ corresponds to the Poisson limit ($b=0$).
See \cite{Huh:2024ytz} for a similar analysis of Poisson-GUE transition.
\begin{figure}[]
\centering
{\includegraphics[width=7.5cm]{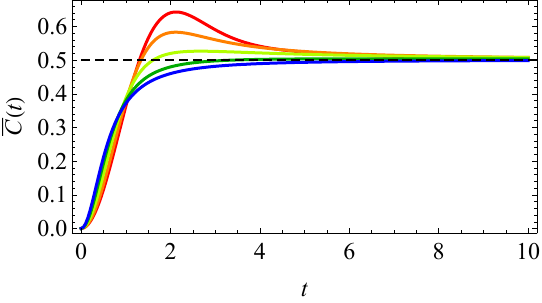}}
\hfill
{\includegraphics[width=7.5cm]{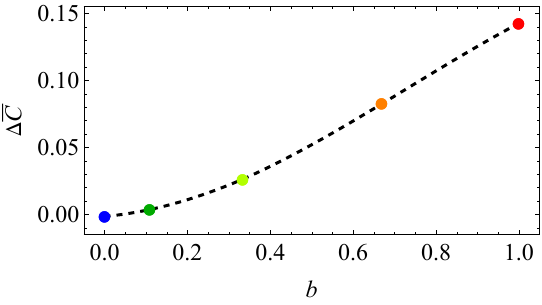}}
\caption{Upper panel: Averaged Krylov complexity for a matrix model interpolating between GOE and Poisson statistics \eqref{ACT1}, with $\sigma=1$ and $d=1,\, 1.2,\, 1.5,\, 1.8,\, 2$ (red,\, orange,\, yellow,\, green,\, blue). The dashed line represents the late-time plateau. Lower panel:  $\Delta \bar{C}$ defined in \eqref{eq:diff_peak} vs. Brody parameter $b$. The colored dots correspond to the same data shown in the upper panel.} \label{BRO1}
\end{figure}

For a $2\times2$ system in the TFD state, the Krylov complexity is analytically given by $C_{2\times2} =  \sin^2\left(\frac{s}{2} t \right)$ \cite{Caputa:2024vrn}. The ensemble-averaged complexity is then computed as
\begin{align}\label{ee3}
\bar{C} \equiv  \int_{0}^{\infty} C_{2\times2} \, P(s) \dd s \,.
\end{align}
Numerical results for the GOE-Poisson ($d\in[1,2]$) are shown in Fig. \ref{BRO1}. The complexity peak reaches its maximum in the chaotic RMT limits ($d=1$) and systematically diminishes as the system approaches the Poisson limit. These results highlight Krylov complexity as a probe of chaos-integrability transition consistent with the Brody distribution analysis.

\begin{figure}[]
\centering
{\includegraphics[width=7.5cm]{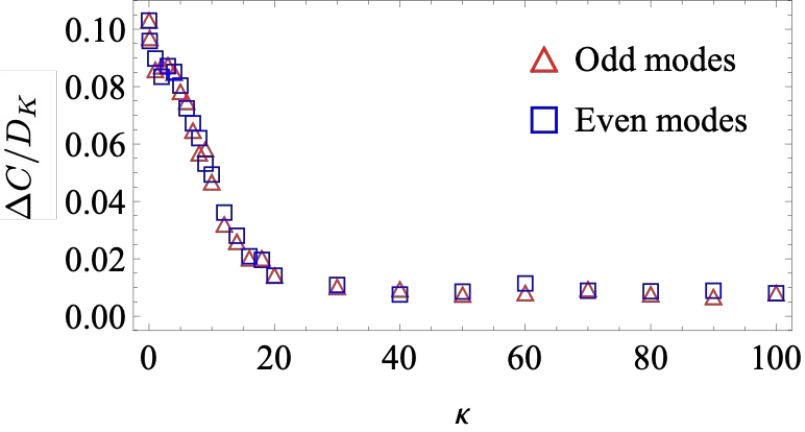}}
\hfill
{\includegraphics[width=7.5cm]{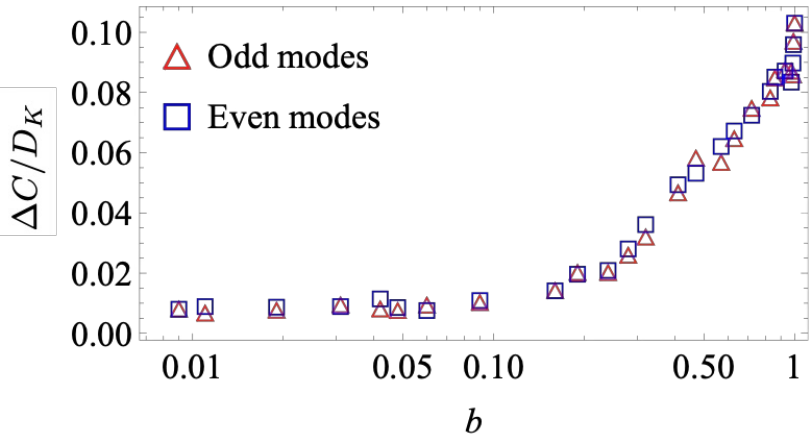}}
\caption{Upper panel: $\Delta C$ vs. mass parameter $\kappa$ in the deformed SYK model. This has the same information of the $\beta=0$ case in the upper panel of Fig. \ref{KRYSYK1}. Lower panel: $\Delta C$ vs. Brody parameter $b$. The smooth monotonic dependence between the two quantities demonstrate the reliability of the KCP ($\Delta C$) to characterize mixed phase systems.} \label{MDSYKPHASEFIG}
\end{figure}

Extending this analysis to the many-body regime, we examine the correlation between the Krylov complexity's peak and the Brody parameter in the mass-deformed SYK model. As shown in Fig. \ref{MDSYKPHASEFIG}, the KCP, defined as $\Delta C$ in \eqref{eq:diff_peak}, decreases monotonically as the mass deformation $\kappa$ increases.

Importantly, our results reveal a clear monotonic relationship between the KCP and the Brody parameter $b$. This alignment with traditional level statistics confirms that the KCP is a robust and reliable indicator of the chaos-to-integrability transition, effectively capturing the crossover from chaotic to integrable dynamics in both toy models and complex many-body systems. For a more exhaustive treatment of this transition via the Brody distribution, see the Supplemental Material of \cite{Huh:2024ytz}.
%

%
\subsection{Saddle-dominated scrambling}
While the OTOC is a well-established probe of quantum chaos, a positive Lyapunov exponent may be a necessary but not sufficient condition for quantum chaos. This is because unstable points or ``saddles" in classical phase space can also generate exponential divergence, a phenomenon known as saddle-dominated scrambling~\cite{Pilatowsky-Cameo:2019qxt,Xu:2019lhc,Rozenbaum:2019nwn,Hashimoto:2017oit,Hashimoto:2020xfr,Bhattacharyya:2020art}. Distinguishing this local instability from the global ergodicity characteristic of true quantum chaos is a critical challenge for dynamical probes.

It has recently been observed that Krylov complexity (both operator and state) also exhibits its conjectured signatures of chaos, exponential growth and the ``peak" structure, in systems characterized by saddle-dominated scrambling \cite{Bhattacharjee:2022vlt, Baek:2022pkt, Huh:2023jxt}. These results suggest that while Krylov complexity is a powerful diagnostic tool, it may not provide a sufficient condition for chaos without additional physical constraints.

To illustrate this, we review Krylov state complexity of the Lipkin-Meshkov-Glick (LMG) model \cite{Lipkin:1964yk, GLICK1965211}, an integrable system known to exhibit saddle-dominated scrambling. (For a related analysis, one may also consider the inverted harmonic oscillator~\cite{Huh:2023jxt} as another elementary model of saddle-dominated scrambling). The LMG Hamiltonian is expressed as:
\begin{align}\label{LMGH2}
\begin{split}
    H = \hat{x} + 2 \hat{z}^2 \,,
\end{split}
\end{align}
where $\{\hat{x}\,, \hat{y}\,, \hat{z}\} := \{\hat{S}_x/S\,, \hat{S}_y/S\,, \hat{S}_z/S\}$ are rescaled SU(2) spin operators with spin $S$. The commutation relations are governed by an effective Planck constant $\hbar_{\text{eff}} = 1/S$: $ [\hat{x}\,,\hat{y}] = i \hbar_{\text{eff}} \hat{z}$. In the large $S$ limit, $\hbar_{\text{eff}}\rightarrow0$, recovering the classical dynamics of the system \cite{Cotler:2017myn, Yin:2020oze}.

\begin{figure}[]
\centering
{\includegraphics[width=7.5cm]{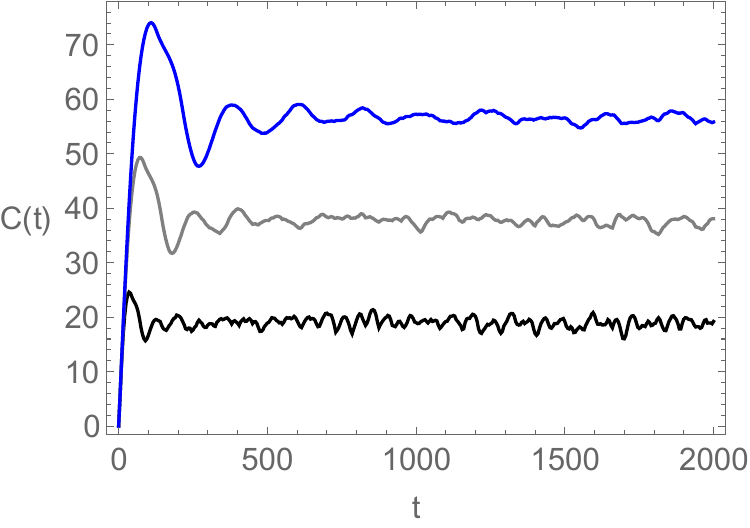}}
\hfill
{\includegraphics[width=7.5cm]{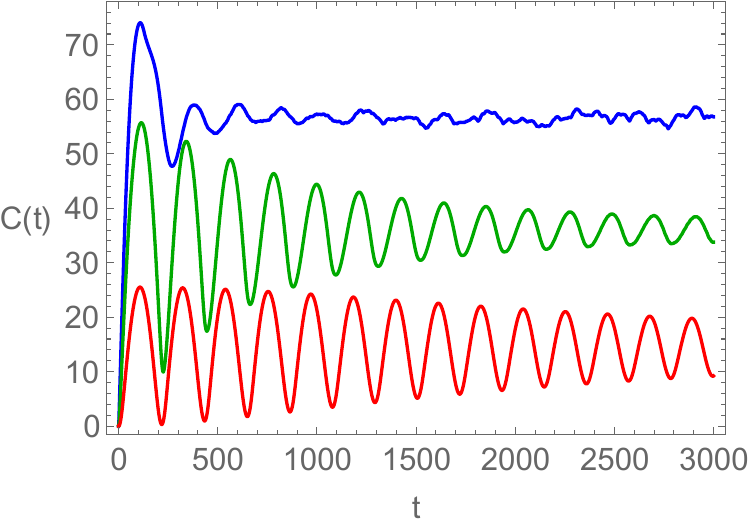}}
\caption{Upper panel: Krylov complexity at $\beta=0$ with $S = {25,50,75}$ (black, gray, blue). Lower panel:  Temperature dependence of Krylov complexity for $S=75$ at $\beta = {0\,,3\,,10}$ (blue, green, red).} \label{LMGfinitebetaFig}
\end{figure}

As shown in the upper panel of Fig. \ref{LMGfinitebetaFig}, the Krylov
complexity of the LMG model manifests dynamical features typically
associated with chaos: an initial ramp followed by a distinct
peak-and-slope structure and late-time saturation \cite{Balasubramanian:2022tpr}. This shows that conventional state
complexity, much like OTOCs and operator complexity
\cite{Xu:2019lhc, Bhattacharjee:2022vlt}, can exhibit apparent false
positives in systems with saddle-dominated scrambling.

A resolution of this issue was recently proposed through
\emph{unfolded Krylov complexity}, in which the spectrum is unfolded
before constructing the associated Krylov dynamics~\cite{Erdmenger:2026iga}.
This procedure removes the spurious peak associated with
saddle-dominated dynamics while retaining the universal behavior
characteristic of chaotic spectra, thereby restoring the discriminatory
power of Krylov complexity. An alternative proposal was put forward in
Ref. \cite{Camargo:2026szl} with logarithmic Krylov complexity in this context.

Finally, we consider the impact of finite temperature ($\beta$) on the complexity of the LMG model. Consistent with our observations in chaotic models such as the SYK model and quantum billiards, increasing $\beta$ significantly suppresses the Krylov complexity and ultimately eradicates the characteristic peak structure (Fig. \ref{LMGfinitebetaFig}, lower panel). This reinforces the notion that the peak as a signature of scrambling is most prominent in the high-temperature limit.
%

%
\subsection{States and density matrix operators}
A general quantum state is represented by a density matrix, modeling a classical statistical mixture of pure states. For pure states, there exists a well-defined one-to-one correspondence between kets in the Hilbert space and density matrix operators $\rho$ in the operator Hilbert space: $\rho = \ket{\psi} \bra{\psi}$.

In the Schrödinger picture, the evolution of a pure state can be formulated through either the Hamiltonian or Liouvillian formalism. In the former, time evolution is governed by the Schrödinger equation, while in the latter, it is generated by the Liouville-von Neumann equation: $\rho(t) = e^{-it\mathcal{L}} \rho(0)$, where $\mathcal{L}=[H, \cdot]$ is the Liouvillian superoperator.
Notably, both equations describe dynamical evolution within the Schrödinger picture of the underlying pure state $\rho(t) = \ket{\psi(t)} \bra{\psi(t)}$.

Using the TFD state as a maximally entangled pure state (i.e., $\ket{\psi(0)}$ is the TFD state), one can investigate the Krylov complexity for both the density operator, $C_K(t)$, and the state itself, $C_S(t)$ \cite{Caputa:2024vrn}. We highlight several key distinctions between these measures below.

An early-time analysis of the Lanczos coefficients reveals a specific scaling relationship: $b_{\text{operator}}^2 \,=\, 2 \, b_{\text{state}}^2$. This implies that in the early-time regime ($t\ll1$), the Krylov complexity associated with the density matrix evolves twice as fast as the state complexity: $C_K \,=\, 2 C_S$.

Furthermore, numerical simulations across various Hilbert space dimensions ($D_{\mathcal H}$) show that for small  sizes of random matrix models, $C_K$ and $C_S$ exhibit qualitatively similar features, including the characteristic peak structure in both complexities. However, in the large-$D_{\mathcal H}$ limit, the peak in $C_K$ becomes suppressed. Instead, in the late-time saturation regime for a finite $D_{\mathcal H}$-dimensional Hilbert space, the complexities satisfy the relation $C_K = D_{\mathcal H} \, C_S$.
 
Last but not least, the moment-generating function $G(t)$, which is central to determining the Lanczos coefficients, is defined differently for each complexity measure in terms of the survival (or return) amplitude $S(t)$ given in \eqref{SFFSre}: For operator complexity it has $G(t) = |S(t)|^2$ and for $G(t) =  S(t)$ for state complexity. These definitions reflect how the spectral properties of the Hamiltonian via $S(t)$ are mapped onto the respective Krylov chains, depending on the evolution of density matrix or state.

For a more comprehensive treatment of Krylov complexity for density matrices, including formulations involving modular Hamiltonians, where the mixed-state density matrix is expressed as the exponential of a modular Hamiltonian, the reader is referred to \cite{Caputa:2023vyr, Caputa:2024vrn}.

%
\section{Non-Hermitian quantum system}\label{SECV}
We have focused on closed quantum systems governed by Hermitian Hamiltonians, where energy eigenvalue is a real-valued observable and the time-evolution state is unitary. However, real-world quantum systems interact with their environment through dissipation, decay, and measurement. Such open systems can be often described by non-Hermitian Hamiltonians ($H\neq H^{\dagger}$)~\cite{Rivas_2012,Fazio:2024itp}. In this non-Hermitian setting, eigenvalues become complex, and the resulting dynamics behave fundamentally differently than their closed-system counterparts~\cite{Ashida:2020dkc,Rivas_2012,Bender_2007}. In this section, we explore how the framework of Krylov complexity of state extends to this non-Hermitian systems and show that it can remain a robust diagnostic of quantum chaos even in the non-Hermitian setting.

Initial progress in the study of Krylov complexity for open systems was established in~\cite{Bhattacharya:2023zqt}, which introduced a bi-Lanczos formalism for evaluating the Krylov complexity of operators. This operator-based framework is conceptually distinct from the state-based version; whereas state Krylov complexity is determined by evolution under a non-Hermitian Hamiltonian, operator Krylov complexity is governed by a Lindbladian superoperator derived from an underlying Hermitian Hamiltonian. Further investigations into operator/state Krylov complexity within the context of open quantum systems are provided, for instance, in~\cite{Liu:2022god,Bhattacharya:2022gbz,NSSrivatsa:2023pby,Bhattacharjee:2022lzy,Bhattacharyya:2023grv,Bhattacharjee:2023uwx,Zhou:2025ozx}.

%
\subsection{Non-Hermitian model and complex statistics}
The characterization of quantum chaos in closed systems relies fundamentally on spectral analysis. In non-Hermitian systems, the emergence of complex eigenvalues renders conventional Hermitian diagnostics inapplicable. To address this, several extensions of random matrix theory (RMT)~\cite{Meh2004} have been proposed to incorporate complex spectra~\cite{Grobe:1988zz,Hamazaki:2020kbp,Sa:2020fpf}, establishing a framework for identifying chaos and integrability in non-Hermitian quantum dynamics~\cite{Hamazaki:2018aa,Akemann:2019ab,Sa:2020ab,Li:2021aa,Garcia-Garcia:2021rle,Shivam:2023aa}.

In Hermitian systems, the transition from integrability to chaos is marked by a shift from Poisson to Wigner-Dyson statistics. In the non-Hermitian setting, this classification is extended through the Grobe-Haake-Sommers (GHS) conjecture~\cite{Grobe:1988zz}, which serves as the non-Hermitian counterpart to the Berry-Tabor and BGS conjectures~\cite{BerryTabor,Bohigas:1983er}.

For integrable non-Hermitian systems, eigenvalues typically follow a two-dimensional Poisson distribution. The nearest-neighbor spacing distribution is given by:
\begin{align}\label{2DPoidis}
   p(s) = \frac{\pi}{2} \, s \, e^{-\frac{\pi}{4}s^2}\,.
\end{align}
where $s_i = \min_{j \neq i} |E_i - E_j|$ represents the nearest-neighbor level spacing, measured as the standard Euclidean distance between eigenvalues $E_i$ and $E_j$ in the complex-energy plane~\cite{Hamazaki:2020kbp, Garcia-Garcia:2021rle}. Here, the linear repulsion $p(s)\approx s$ as $s\rightarrow0$ does not imply correlations but simply reflects the two-dimensional nature of the eigenvalue distribution in the complex plane.

In contrast, chaotic non-Hermitian quantum systems are described by Ginibre statistics~\cite{Ginibre:1965zz,Hamazaki:2020kbp}. For the Ginibre Unitary Ensemble (GinUE, or Class A), the spacing distribution is
\begin{align}\label{GinUEdis}
   p(s) = \left( \prod_{k=1}^{\infty} \frac{\Gamma\left(1+k,s^2\right)}{k!}  \right) \sum_{j=1}^{\infty} \frac{2 s^{2j+1} e^{-s^2}}{\Gamma\left(1+j,s^2\right)} \,,
\end{align}
where $\Gamma\left(1+k,s^2\right)$ is the incomplete gamma function. Chaotic systems exhibit a universal cubic level repulsion, $p(s) \approx s^3$, as $s \to 0$~\cite{Grobe:1988zz,Grobe:1989aa,Akemann:2019aa}. While Dyson’s Hermitian classification identifies three symmetry classes (A, AI, AII) with distinct statistics (GUE, GOE, GSE), the non-Hermitian Ginibre ensembles were originally thought to share a single universal spacing distribution~\cite{Ginibre:1965zz}. However, recent developments have identified distinct behaviors in classes A, AI$^{\dagger}$, and AII$^{\dagger}$ by considering alternative symmetries involving transposition and complex conjugation~\cite{Hamazaki:2020kbp}, proven with non-Hermitian random matrix theories.

A significant technical challenge in complex spectral analysis is the process of unfolding, which is required to normalize the local density of states~~\cite{Akemann:2019aa}. To circumvent this, one can employ the Complex Spacing Ratio (CSR)~\cite{Sa:2020fpf}, a generalization of the Hermitian $r$-parameter to the complex plane~\cite{Atas:2013aa}: see also ~\cite{Fyodorov:1997aa,Li:2021kuv,Li:2024uzg} for the discussion of dissipative spectral form factors. The CSR for a complex eigenvalue $E_k$ is defined as
\begin{align}\label{CSRfor}
   \lambda_{k} = \frac{E^{\text{NN}}_k - E_k}{E^{\text{NNN}}_k - E_k} \,,
\end{align}
where $E^{\text{NN}}_k$ and $E^{\text{NNN}}_k$ are the nearest and next-to-nearest neighbors, respectively. By construction, $\abs{\lambda_k}\leq1$. Chaotic behavior is identified via cubic radial repulsion and angular anisotropy in the distribution of $\lambda_k$, whereas integrable systems exhibit a nearly uniform distribution. Furthermore, we explore the Krylov complexity (KC) as a dynamical probe shortly. In non-Hermitian systems, the bi-Lanczos algorithm allows for the construction of a Krylov basis, yielding three different Lanczos coefficients $a_n$, $b_n$, and $c_n$.

To investigate these statistical signatures, we focus on two primary frameworks: the non-Hermitian Sachdev-Ye-Kitaev (nHSYK) model~\cite{Garcia-Garcia:2021rle,Garcia-Garcia:2022xsh} and general non-Hermitian RMT~\cite{Sa:2020ab,Hamazaki:2020kbp}.

The nHSYK model describes $N$ Majorana fermions in $(0+1)$ dimensions with random $q$-body interactions. It extends the standard SYK model~\cite{Sachdev:1992fk,Kitaev2015Talk} by incorporating a non-Hermitian component. The Hamiltonian is defined as:
\begin{align}\label{SYKMODEL}
H = \sum_{i_1<i_2< \cdots < i_q }^{N}\, (J_{i_1\,i_2\,\cdots\,i_q}+i M_{i_1\,i_2\,\cdots\,i_q})\, \chi_{i_1}\, \chi_{i_2}\, \cdots \chi_{i_q} \,,
\end{align}
where $\chi_i$ are Majorana fermions with $\{\chi_i,\chi_j\} = \delta_{ij}$, and $M$ introduces non-Hermicity. The couplings $J$ and $M$ are Gaussian random variables with zero mean and variance $\langle J^2\rangle = \langle M^2\rangle = (q-1)!/N^{q-1}$. The model possesses involutive symmetries (charge conjugation) that allow the Hamiltonian to be decomposed into independent symmetry blocks, which we utilize to refine our spectral analysis. The parameter $q$ dictates the system's integrability: the $q=4$ case is expected to exhibit many-body chaos, while $q=2$ is a quadratic model that is typically integrable. It is worth noting that symmetry classification of RMT (A, $\text{AI}^\dagger$, and $\text{AII}^\dagger$ classes) can be explored by set of two parameters of nHSYK model: $\{N,\,q\}$~\cite{Garcia-Garcia:2021rle,Garcia-Garcia:2022xsh}, which will be given shortly.

As a benchmark for universality, we also compare the nHSYK results against pure non-Hermitian RMT ensembles. We utilize classes A, AI$^{\dagger}$, and AII$^{\dagger}$  matrices to represent the chaotic limit. To represent the integrable limit, we employ non-Hermitian random matrices with uncorrelated eigenvalues, which should asymptotically match the two-dimensional Poisson statistics.

In the following sections, we will present the results for the spacing distributions $p(s)$, the CSR angular distributions, and the growth of Krylov complexity to evaluate the consistency of these chaos diagnostics in the non-Hermitian quantum systems.

%
\subsection{Bi-Lanczos algorithm and Krylov space}
To investigate Krylov complexity in non-Hermitian many-body systems, we utilize the bi-Lanczos algorithm~\cite{Gaaf_2017,TS2000,Gruning:2011aa}. This approach extends the standard Lanczos procedure~\cite{Lanczos:1950zz} to accommodate the lack of a standard orthonormal basis, which is a hallmark of non-Hermitian dynamics.

The bi-Lanczos algorithm generates two distinct sets of Krylov bases, $\{|\,p_n\rangle\}$ and $\{|\,q_n\rangle\}$, which satisfy the bi-orthonormality condition: $\langle \, p_n \,|\, q_m \rangle = \delta_{nm}$. This procedure yields three sequences of Lanczos coefficients, $\{a_n,b_n,c_n\}$, that govern the dynamics of quantum system in the biorthogonal Krylov subspaces. Starting from a non-Hermitian Hamiltonian $H$, the three-term recurrence relations are defined as
\begin{equation}\label{biL1}
\begin{split}
\!\!\!&|\,r_{n+1}\rangle = (H - a_n) \,|\,q_n\rangle - b_n \,|\,q_{n-1}  \rangle\,, \quad\,\, |\,q_{n}\rangle = c_{n}^{-1}\, |\,r_{n}\rangle \,, \\
\!\!\!&|\,l_{n+1}\rangle = (H^\dagger - a_n^*) \,|\,p_n\rangle - c_n^* \,|\,p_{n-1}  \rangle, \quad |\,p_{n}\rangle = b_{n}^*{^{-1}}\, |\,l_{n}\rangle \,.
\end{split}
\end{equation}
In this basis, $H$ admits a tridiagonal representation, $T$, where the coefficients correspond to the matrix elements:
\begin{align}\label{tridiag}
T=
\begin{pmatrix}
a_0 & b_1 & 0   & \cdots & 0 \\
c_1 & a_1 & b_2 & \ddots & \vdots \\
0   & c_2 & a_2 & \ddots & 0 \\
\vdots & \ddots & \ddots & \ddots & b_{D_{\rm K}-1} \\
0 & \cdots & 0 & c_{D_{\rm K}-1} & a_{D_{\rm K}-1}
\end{pmatrix}\,.
\end{align}
Here, $a_n$ represents the onsite potential (main diagonal), while $b_n$ and $c_n$ act as hopping amplitudes between the superdiagonal and subdiagonal, respectively. In the Hermitian limit, $b_n=c_n$ ensures a symmetric tridiagonal representation of the Hamiltonian. In the non-Hermitian regime, however, the dynamics are distributed across two distinct but interconnected ``ladders" corresponding to the $\{|\,q_n\rangle\}$ and $\{\langle\,p_n\,|\}$ bases, with 
\begin{align}\label{ABC}
\begin{split}
a_n &= \langle p_n| H |\,q_{n}\rangle \,, \quad\,\,\, 
b_{n+1} = \langle p_n| H |\,q_{n+1}\rangle \,, \quad\, \\
c_{n+1} &= \langle p_{n+1}| H |\,q_{n}\rangle \,.
\end{split}
\end{align}
\begin{figure}[t!]
 \centering
{\includegraphics[width=8.0cm]{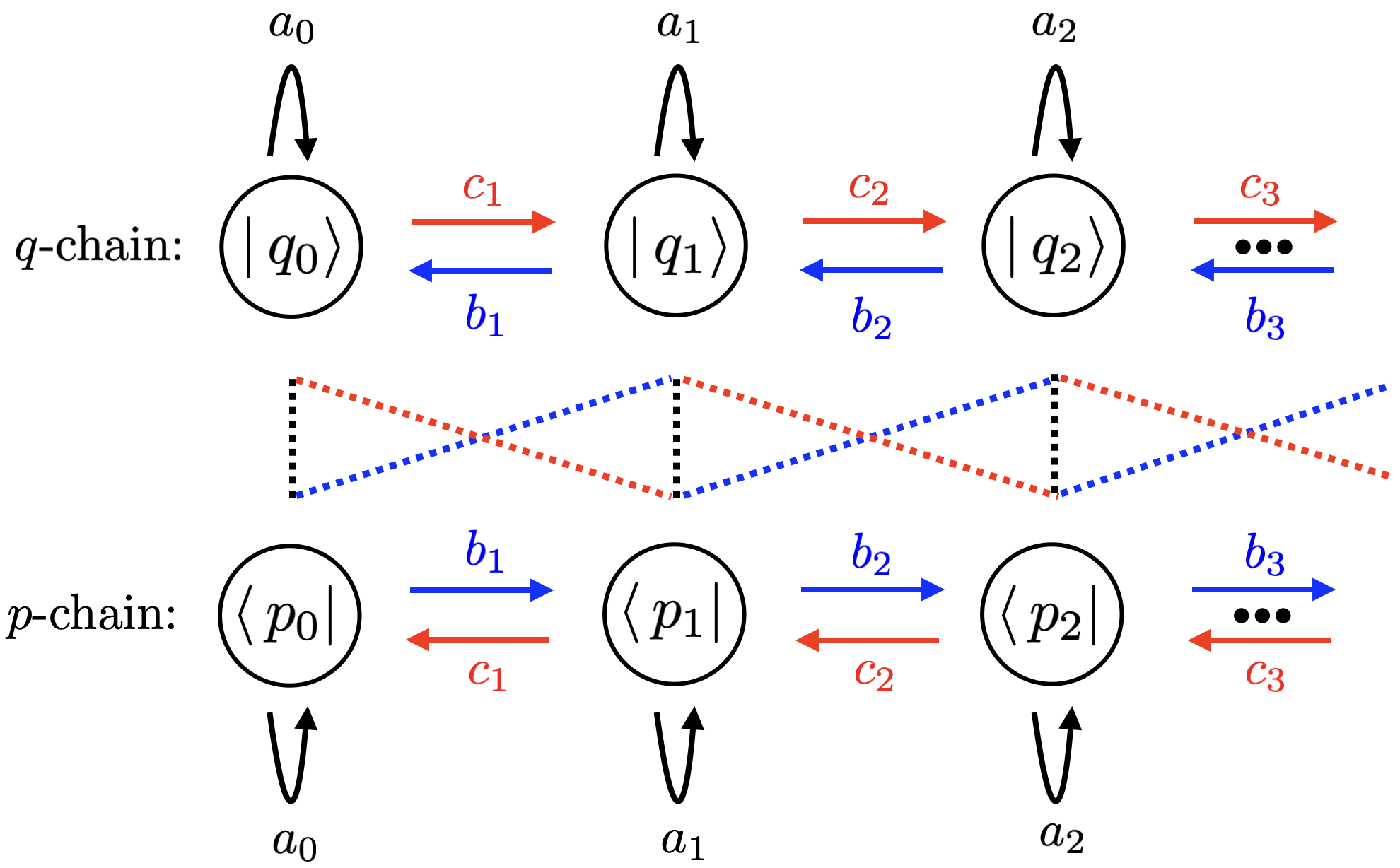}}
\caption{Schematic of the non-Hermitian Hamiltonian action on the bi-orthogonal Krylov chains \eqref{KryChains}. Each site is characterized by an on-site potential $a_n$ (represented by self-loops), with $b_n$ and $c_n$ denoting the leftward and rightward hopping amplitudes, respectively. The dashed lines represent the overlaps $\langle p_n|H|q_m\rangle$, which define the bi-Lanczos coefficients as specified in \eqref{ABC}.} \label{FigChain}
\end{figure}
These bi-Lanczos coefficients characterize the hopping dynamics of the Krylov chains, where $a_n$ acts as an onsite potential and $b_n$, $c_n$ describe nearest-neighbor transitions:
\begin{align}
&H \,|\,q_n\rangle = a_n \,|\,q_n\rangle + b_n \,|\,q_{n-1}  \rangle + c_{n+1} |\,q_{n+1}\rangle \,, \qquad
 \nonumber \\
&H^\dagger \,|\,p_n\rangle = a_n^*  \,|\,p_n\rangle + c_n^* \,|\,p_{n-1}  \rangle + b_{n+1}^* |\,p_{n+1}\rangle    \,,\label{KryChains}
\end{align}
obtained from Eq. \eqref{biL1}. While each relation independently resembles a 1D tight-binding model, the non-Hermitian nature of $H$ necessitates tracking both chains simultaneously to achieve a complete representation. The chains are ``coupled" through the definition of the coefficients as mutual overlaps, $\langle p_n|H|q_m\rangle$. This structure implies that the hopping strength and state evolution within the $q$-chain are intrinsically governed by how the $q$ and $p$ subspaces pair up during the bi-orthogonalization process. See Fig. \ref{FigChain} for a schematic illustration of the coupled Krylov $q-$ and $p-$chains.

To seed the algorithm, we select a normalized TFD state at infinite temperature~\cite{Balasubramanian:2022tpr}. In Hermitian systems, demonstrated in previous sections, the TFD state is a proven diagnostic for quantum chaos~\cite{Balasubramanian:2022tpr,Erdmenger:2023wjg,Camargo:2024deu,Baggioli:2024wbz}, typically manifesting as a pronounced early-time peak in Krylov complexity. To test if this signature persists in the non-Hermitian regime, we set: $|\,p_0\rangle = |\,q_0\rangle = |\,{\rm TFD}\,\rangle$.

For a non-Hermitian Hamiltonian $H$, the bi-orthogonal basis is defined via the right and left eigenvalue equations
\begin{align}
H |n_+\rangle = E_n |n_+\rangle\,, \qquad
H^\dagger |n_-\rangle = E_n^* |n_-\rangle\,,
\end{align}
satisfying the biorthonormality relation $\langle n_-|m_+\rangle=\delta_{nm}$~\cite{Harper:2025lav,Guo:2024wmj}. To initialize the bi-Lanczos recursion, we construct the non-Hermitian TFD state
\begin{align}
|\text{TFD}(\beta)\rangle
= \frac{1}{\sqrt{Z(\beta)}}
\left(e^{-\frac{\beta H}{4}}\otimes e^{-\frac{\beta H^\dagger}{4}}\right)
\sum_n |n_+\rangle_1 |n_+\rangle_2\,,
\end{align}
where $Z(\beta)$ is a normalization constant.

\begin{tcolorbox}[colback=gray!10, colframe=gray!50, title=\textbf{Bi-Lanczos algorithm}, coltitle=white]
Given a normalized initial state, $\ket{\psi_0}$, evolving under a non-Hermitian Hamiltonian, $H$, the bi-orthonormal Krylov basis $\{|\,q_n\rangle\} \,,  \{\langle\,p_n\,|\}$ can be constructed via the bi-Lanczos algorithm:
\begin{enumerate}
\item[\underline{\small{Step 1}}.\,] $b_0 = c_0 := 0\,, \quad |p_{-1}\rangle = |q_{-1}\rangle := 0$ \,,
\item[\underline{\small{Step 2}}.\,] $|p_0\rangle = |q_0\rangle := |\psi_0\rangle\,,\quad a_0=\langle p_0|H|q_0\rangle$,
\item[\underline{\small{Step 3}}.\,] For $n\geq1$:\\
$|\,r_{n}\rangle
= (H - a_{n-1})\,|\,q_{n-1}\rangle - b_{n-1}\,|\,q_{n-2}\rangle, \\
|\,l_{n}\rangle
= (H^{\dagger} - a_{n-1}^*)\,|\,p_{n-1}\rangle - c_{n-1}^*\,|\,p_{n-2}\rangle\,,$
\item[\underline{\small{Step 4}}.\,] Set $b_{n}=c_{n}^{-1}\langle r_{n}\,| \,l_{n}\rangle\,, \quad c_{n} = \sqrt{|\langle r_{n}\,| \,l_{n}\rangle|} \,,$
\item[\underline{\small{Step 5}}.\,] Set $|\,q_{n}\rangle = c_{n}^{-1}\,|\,r_{n}\rangle,\quad |p_{n}\rangle = {b^*_{n}}^{-1}\,|\,l_{n}\rangle\,,$
\item[\underline{\small{Step 6}}.\,] Reassign the element to remove accumulated errors:\\ $|\,q_n\rangle \,\rightarrow\, |\,q_n\rangle 
    - \sum_{l=1}^{n-1} \langle p_l \,|\, q_l \rangle\, |\,q_l\rangle, \\[2pt]
    |\,p_n\rangle \,\rightarrow\, |\,p_n\rangle 
    - \sum_{l=1}^{n-1} \langle q_l \,|\, p_l \rangle\, |\,p_l\rangle \,,$
\item[\underline{\small{Step 7}}.\,] Set $a_{n} = \langle p_{n} \,|\, H\, |\, q_{n} \rangle \,,$
\item[\underline{\small{Step 8}}.\,] If $b_n=0$ stop; otherwise go to {\small{Step 3}}.
\end{enumerate}
\end{tcolorbox}

Starting from the TFD state, the bi-orthogonal Krylov bases and their associated Lanczos coefficients are generated through iterative applications of the Hamiltonian paired with a two-sided Gram-Schmidt-like procedure. Although a non-Hermitian TFD state can be constructed using $H$, $H^{\dagger}$, or a combination of both, the resulting Krylov complexity dynamics are robust and qualitatively independent of the specific eigenvector basis chosen~\cite{Baggioli:2025knt}.
\begin{figure*}[]
\centering
\begin{minipage}{0.45\textwidth}
        \centering
        \includegraphics[width=\textwidth]{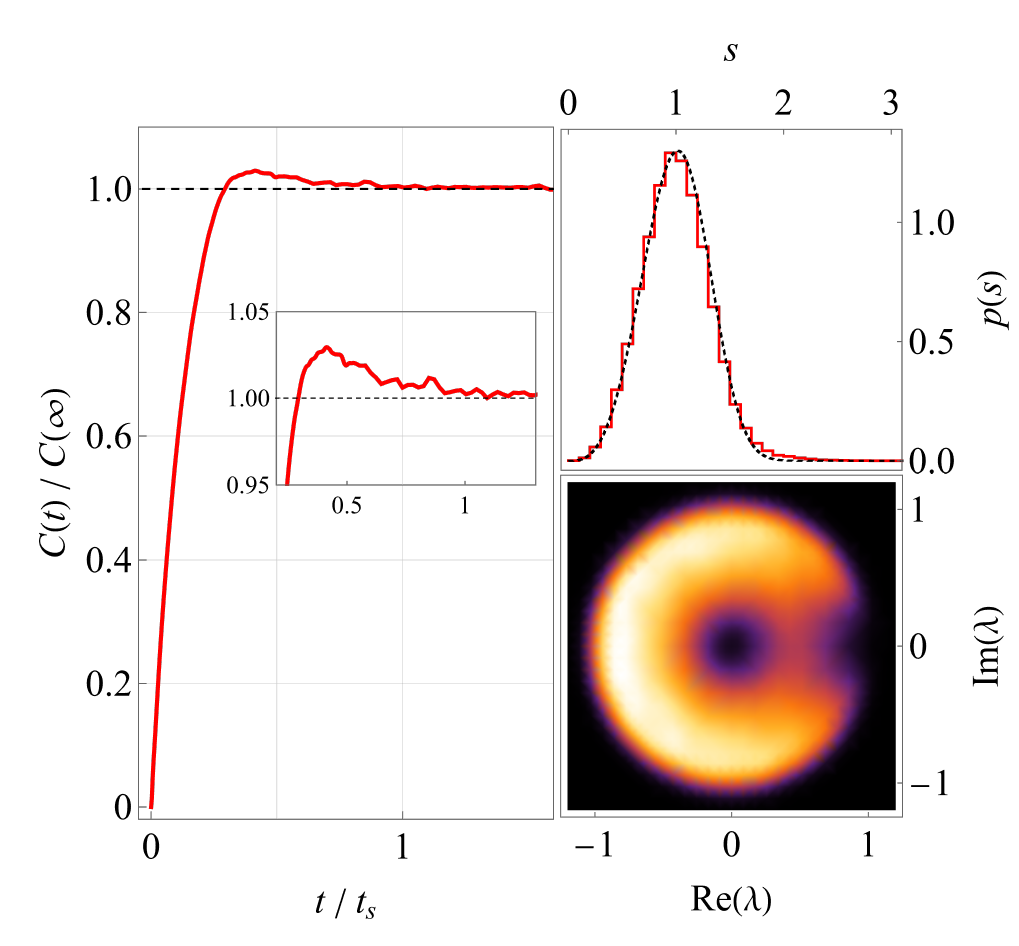}
\end{minipage}
\quad
\begin{minipage}{0.45\textwidth}
        \centering
        \includegraphics[width=\textwidth]{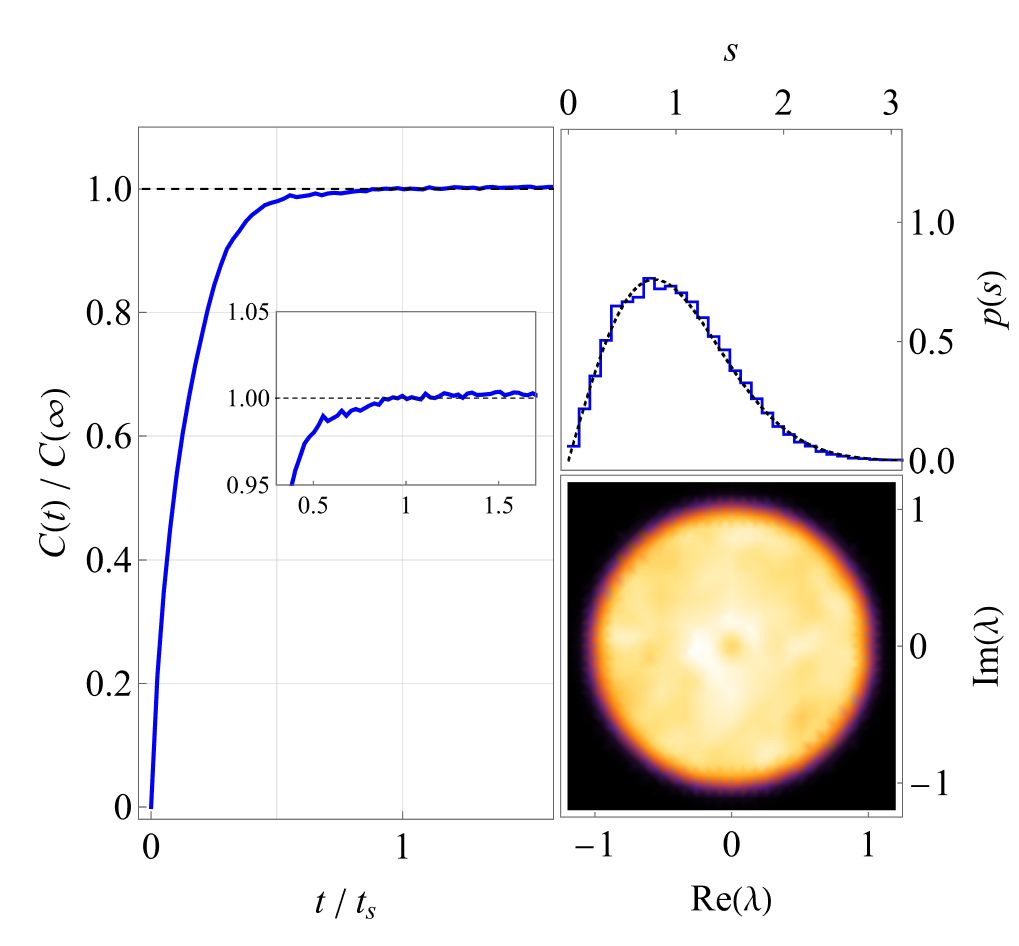}
\end{minipage}
\vspace{-0.3cm}
\caption{Multimodal diagnostics of quantum chaos in the nHSYK model. Comparison of Krylov complexity $C(t)$, complex level spacing distribution $p(s)$, and complex spacing ratio (CSR) for $N=22$. In the chaotic case ($q=4$, left), the system exhibits a pronounced peak in $C(t)$, agreement with GinUE statistics \eqref{GinUEdis} in $p(s)$, and an anisotropic CSR distribution. In the integrable regime ($q=2$, right), the $C(t)$ peak is absent, and spectral statistics \eqref{2DPoidis} $p(s)$ and CSR are consistent with a two-dimensional Poisson distribution and isotropy, respectively.}\label{NHSYK}
\end{figure*}

In practice, maintaining bi-orthogonality is numerically challenging as precision errors can accumulate rapidly. To ensure the stability of the Krylov basis vectors at every iteration, as in Full Orthogonalization algorithm, we perform a full bi-orthogonalization twice (as detailed in Step 6). This double-pass Gram-Schmidt approach effectively mitigates numerical instabilities, preserving the integrity of the tridiagonal representation.

Any time-evolved state $|\psi(t)\rangle$ then can be expanded using either of the biorthogonal Krylov bases, ${|\,p_n\rangle}$ or ${|\,q_n\rangle}$, as follows:
\begin{align}\label{wavefunction}
|\psi(t)\rangle
= \sum_{n=0}^{} \phi_n^p(t) |\,p_n\rangle
=\sum_{n=0}^{} \phi_n^q(t) |\,q_n\rangle \,.
\end{align}
The coefficients $\phi_n^p(t)$ and $\phi_n^q(t)$ allow us to define Krylov complexity in the non-Hermitian systems as 
\begin{align}\label{Krylovcomplexity}
C(t) := \sum_n n\left| \tilde{\phi}_n^{p*}(t) \, \tilde{\phi}_n^q(t) \right|\,.
\end{align}
Here, the tilde indicate dynamically normalized bases, a necessary step to ensure the Krylov probability remains conserved, despite the non-unitary nature of the evolution~\cite{Bhattacharya:2023yec}.

%
\subsection{Robustness of Krylov complexity for chaos}
While well-established in Hermitian physics, Krylov complexity remains relatively uncharted territory in non-Hermitian systems. This gap in the literature is largely due to the technical overhead of the bi-Lanczos algorithm. Unlike the standard Lanczos procedure, bi-Lanczos requires the simultaneous construction and storage of two non-orthogonal bases. This not only doubles memory requirements but also introduces significant numerical instabilities that often demand computationally expensive re-biorthogonalization.

To bypass these hurdles, researchers have often taken one of two paths: (I) Simplification: restricting models to complex-symmetric Hamiltonians to justify using the standard Lanczos algorithm~\cite{Bhattacharya:2023yec,Medina-Guerra:2025rwa,Medina-Guerra:2025wxg}, (II) Scale Reduction: maintaining the full bi-Lanczos approach but limiting the Hilbert space size~\cite{Chakrabarti:2025hsb}. Furthermore, much of the existing work focuses on the quantum Zeno effect or non-Hermitian phase transitions rather than the core signatures of quantum chaos.

A pivotal question is whether the characteristic ``chaos peak" found in Hermitian systems persists in the non-Hermitian quantum systems. A recent study of disordered spin chains~\cite{Zhou:2025ozx} utilized the bi-Lanczos algorithm and suggested that suppressed growth in $C(t)$ could signal a transition from chaos to integrability. However, that study did not observe a peak in either regime, likely because the initial state was not the TFD state.

In contrast, recent work using the TFD state as a seed for the full bi-Lanczos algorithm~\cite{Baggioli:2025knt} has demonstrated that, when applied to the non-Hermitian SYK and random matrix models, Krylov complexity indeed serves as a robust diagnostic of quantum chaos. We review these findings, including the complex level statistics, and the universal peak structure of Krylov complexity.

We focus our primary analysis on the non-Hermitian SYK (nHSYK) model with $N=22$ for both the chaotic ($q=4$) and integrable ($q=2$) regimes. To ensure statistical convergence, observables are averaged over $1,000$ independent realizations. For clarity in comparing dynamical growth, the Krylov complexity is rescaled such that its late-time saturation value $C(\infty)=1$, with the saturation time $t_s$ defined accordingly.

As illustrated in Fig. \ref{NHSYK}, we observe a striking contrast between the two regimes ($q=4$ vs. $q=2$) across all three diagnostics. The system of $q=4$ displays clear signatures of quantum chaos. The Krylov complexity, computed via the bi-Lanczos algorithm, exhibits a pronounced peak before saturation. This dynamical feature is accompanied by a complex eigenvalue spacing distribution $p(s)$ that closely follows the GinUE statistics of non-Hermitian RMT \cite{Grobe:1988zz}, Eq. \eqref{GinUEdis}. Furthermore, the CSR shows strong radial repulsion and angular anisotropy, with $\langle \cos \theta \rangle \approx 0.22$.
\begin{figure}[t!]
 \centering
 \vspace{1.2mm}
     {\includegraphics[width=4.1cm]{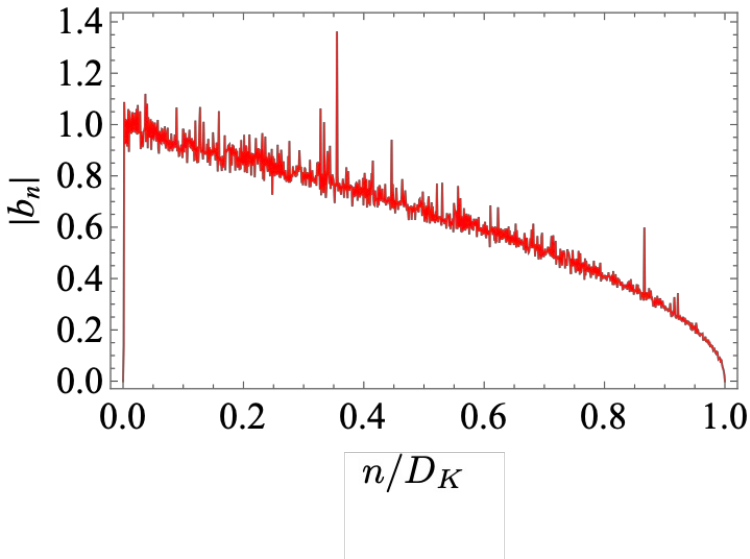}}
     {\includegraphics[width=4.1cm]{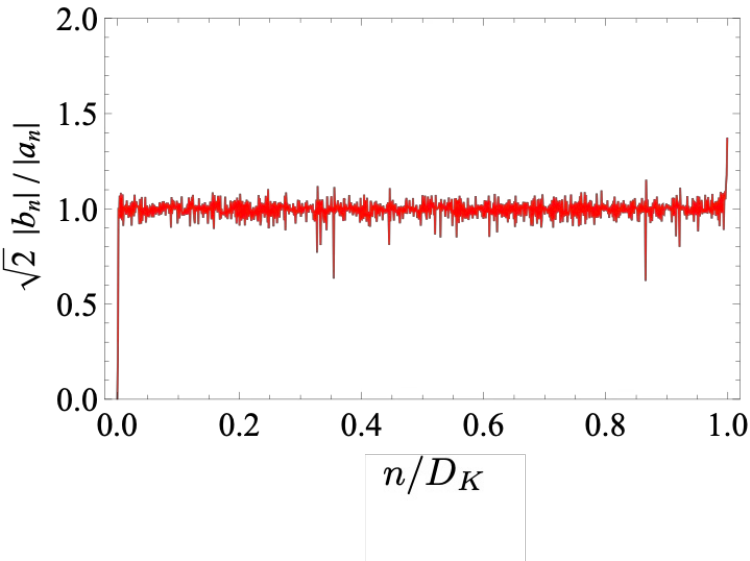}}
 \vspace{1.2mm}
     {\includegraphics[width=4.1cm]{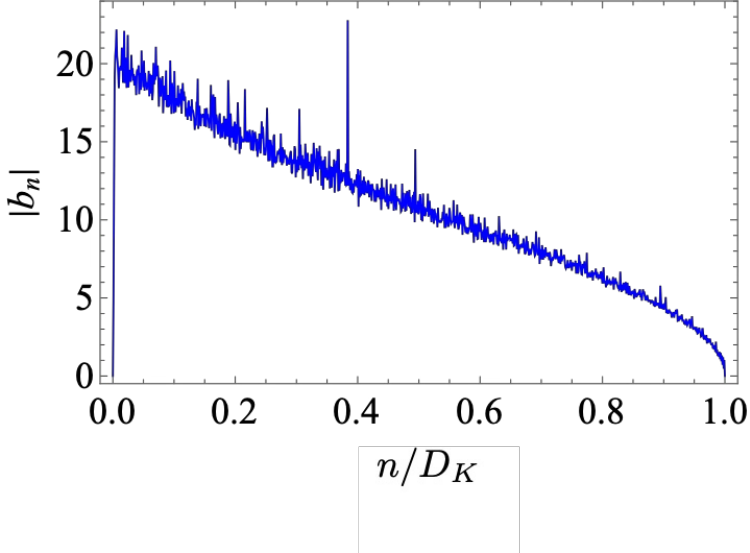}}
     {\includegraphics[width=4.1cm]{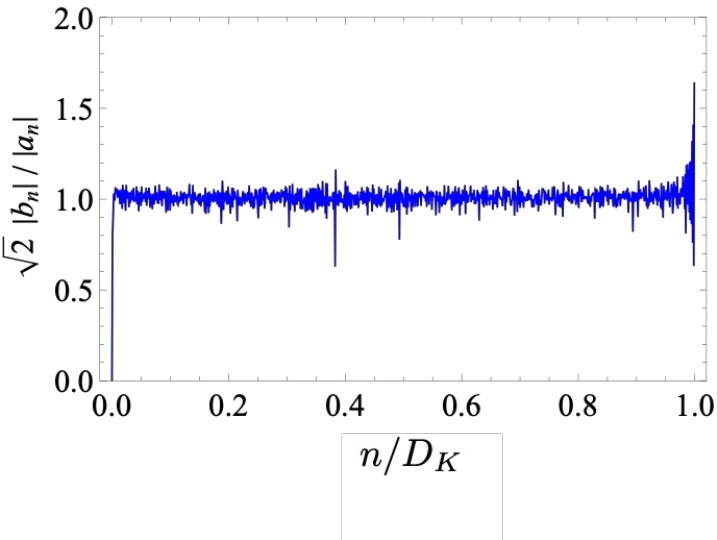}}
\caption{Lanczos coefficients for the nHSYK model. Magnitudes of the bi-Lanczos coefficients are shown for the chaotic ($q=4$, top) and integrable ($q=2$, bottom) cases. The bi-Lanczos coefficients satisfy the empirical relation $1/\sqrt{2})\,|a_n|\approx|b_n|=c_n$, a feature that persists in both chaotic and integrable systems.} \label{Fig.Lanczos}
\end{figure}

In contrast, the $q=2$ case lacks a peak structure. Its spectral statistics align with a two-dimensional Poisson distribution~\cite{Grobe:1988zz}, and the CSR remains nearly isotropic ($\langle \cos \theta \rangle \approx 0.003$), confirming the absence of level repulsion.
\begin{figure*}
    \centering
    \renewcommand{\arraystretch}{1.2} 
    \setlength{\tabcolsep}{3pt}
\begin{tabular}{ccccc}
& \hspace{7mm}$\text{A}$ & \hspace{7mm}$\text{AI}^\dagger$ & \hspace{5mm}$\text{AII}^\dagger$ & \hspace{7mm}Poisson \\[2mm]
        \raisebox{3.0cm}{\rotatebox{90}{}}\hspace{3mm} & 
        \includegraphics[width=3.8cm]{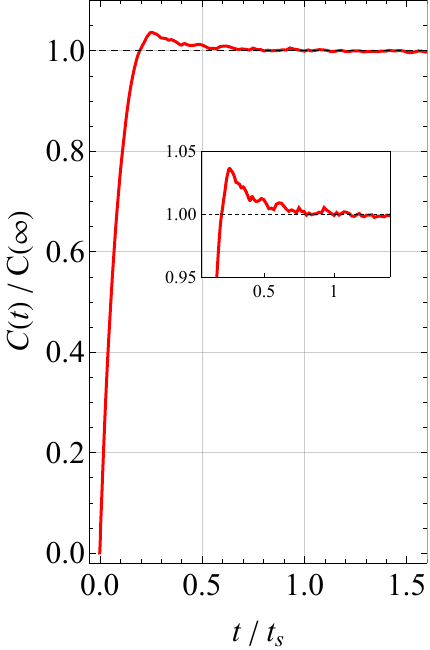} &
        \includegraphics[width=3.8cm]{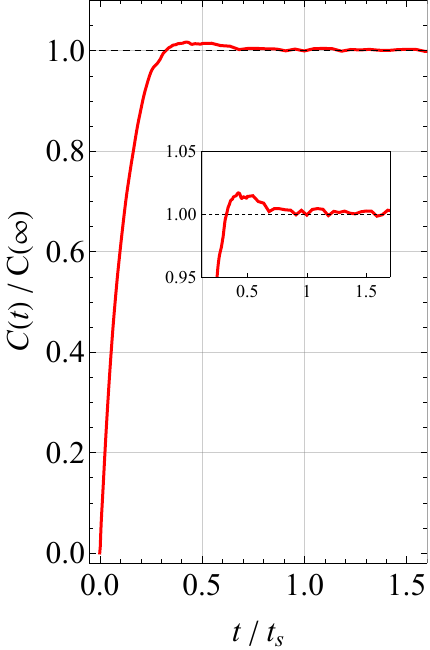} &
        \includegraphics[width=3.8cm]{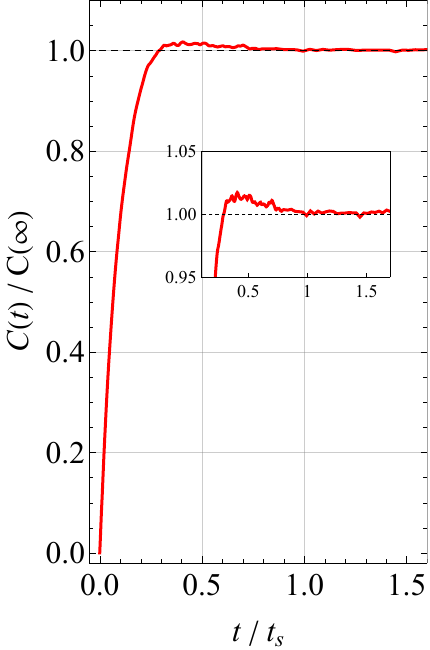} &
        \includegraphics[width=3.8cm]{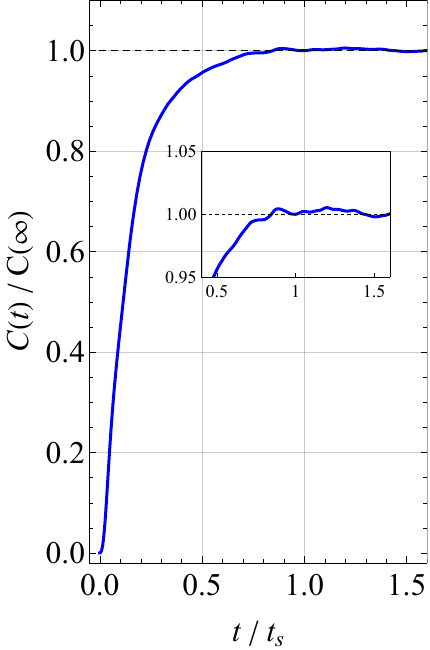}
\end{tabular}
\caption{Krylov complexity in non-Hermitian RMT ensembles. Time evolution of $C(t)$ across the $\text{A}$, $\text{AI}^\dagger$, $\text{AII}^\dagger$ symmetry classes, contrasted with the Poisson limit. The chaotic ensembles exhibit a characteristic early-time peak, which is absent in the featureless growth of the integrable (Poisson) case.}\label{KCRMT}
\end{figure*}

Furthermore, one interesting empirical observation is the emergence of a universal relation among the bi-Lanczos coefficients that is absent in Hermitian systems. One can find that the magnitudes of the coefficients satisfy $(1/\sqrt{2})\,|a_n|\approx|b_n|=c_n$. As shown in Fig. \ref{Fig.Lanczos}, this proportionality persists in the nHSYK model regardless of whether the system is chaotic or integrable.

The origin of this universality likely stems from the structural constraints of the bi-Lanczos recursion. Unlike the Hermitian case, where $a_n$ and $b_n$ are independent, the bi-orthogonality condition $\langle \, p_n \,|\, q_m \rangle = \delta_{nm}$ forces the Hamiltonian's action to be distributed across $a_n$, $b_n$, and $c_n$ within a shared ``normalization budget".

If we interpret $\abs{a_n}^2$ as the intensity of on-site persistence and $\abs{b_n}^2+\abs{c_n}^2$ as the intensity of bidirectional transitions, the relation $\abs{a_n}^2 \approx \abs{b_n}^2+\abs{c_n}^2$ suggests an equipartition of dynamical weights. Since the algorithm inherently satisfies $\abs{b_n}=\abs{c_n}$, this balance naturally converges to the observed ratio. In chaotic models, this relation sharpens in the large-$N$ limit due to self-averaging, whereas in integrable systems, it appears to be a robust geometric consequence of the bi-orthogonal construction.

To test the breadth of these findings, we extend our analysis to non-Hermitian random matrix theories and varying nHSYK configurations~\cite{Garcia-Garcia:2021rle}. By tuning the interaction order $q$ and system size $N$, we explore the full spectrum of non-Hermitian symmetry classes: A, AI$^{\dagger}$ and AII$^{\dagger}$.

As detailed in \cite{Hamazaki:2020kbp}, non-Hermitian random matrices in class A are characterized by a lack of symmetry constraints. Conversely, both classes $\text{AI}^\dagger$ and $\text{AII}^\dagger$ respect transposition symmetry, but are distinguished by the sign of the square of the symmetry operator. Specifically, matrices in class $\text{AI}^\dagger$ satisfy the symmetry condition $H=H^T (\neq H^*)$. In contrast, class $\text{AII}^\dagger$ matrices satisfy $H=\Sigma^y H^T \Sigma^y (\neq \Sigma^y H^* \Sigma^y)$, with the symplectic structure defined by $\Sigma^y = 
\begin{bmatrix}
    0 & -i \, \mathbb{I} \\
    i\, \mathbb{I}  & 0
\end{bmatrix}$, where $\mathbb{I}$ denotes the $N/2 \times N/2$ identity matrix.

As summarized in Table \ref{table}, every chaotic configuration maps onto a specific RMT universality class and consistently exhibits the characteristic peak of Krylov complexity. Conversely, all configurations mapping to the Poisson ensemble lack this feature.
\begin{table}[]
    \centering
    \renewcommand{\arraystretch}{1.5}
    \setlength{\tabcolsep}{6pt}
    \begin{tabular}{cccccc}
        \hline
                & $N = 18$ & $N = 20$ & $N = 22$ & $N = 24$ & Peak \\ 
        \hline
        \hline
        $q = 2$ & Poisson & Poisson & Poisson & Poisson & \xmark \\   
        \hline
        $q = 3$ & $\text{AII}^\dagger$ & $\text{AII}^\dagger$ & $\text{AI}^\dagger$ & $\text{AI}^\dagger$ & \cmark \\        
        $q = 4$ & $\text{A}$ & $\text{AII}^\dagger$ & $\text{A}$ & $\text{AI}^\dagger$ & \cmark \\ 
        $q = 6$ & $\text{A}$ & $\text{A}$ & $\text{A}$ & $\text{A}$ & \cmark \\ 
        \hline
    \end{tabular}
    \caption{Symmetry classification and Krylov complexity's peak signatures in the nHSYK model. Mapping of system parameters ($N,q$) to non-Hermitian RMT universality classes (A, AI$^{\dagger}$, AII$^{\dagger}$) or the Poisson ensemble. The final column indicates the presence (\cmark) or absence (\xmark) of a characteristic peak in the Krylov complexity.}\label{table}
\end{table}

The correspondence between nHSYK and RMT is further reinforced in Fig. \ref{KCRMT}. These models replicate the three-pronged signature of chaos: the KC peak, GinUE-type spacing, and anisotropic CSR. The persistence of the relation $(1/\sqrt{2})\,|a_n|\approx|b_n|=c_n$ across these diverse ensembles suggests that these features are universal properties of non-Hermitian random matrix behavior.

Further discussion on complex spectral statistics and Krylov complexity in non-Hermitian SYK and RMT settings can be found in~\cite{Baggioli:2025knt}. In particular, the Krylov complexity peak is shown also to be a robust diagnostic of chaos in pseudo-Hermitian systems, exemplified by the non-Hermitian random-field XXZ model.

%
\subsection{Singular value decomposition}
An alternative approach to handling the complexities of non-Hermitian spectra involves the use of Singular Value Decomposition (SVD)~\cite{Kawabata:2023aa,Roccati:2023aa,Hamanaka:2024aa,Tekur:2024aa,Nandy:2024aa,Nandy:2025aa}. While this method initially appears to simplify the problem by mapping complex eigenvalues to real singular values, recent research~\cite{Baggioli:2025ohh} using non-Hermitian SYK models has identified significant ``blind spots" where this approach fails to accurately diagnose quantum chaos. This subsection provides a pedagogical overview of the SVD framework and discusses its critical limitations compared to the bi-Lanczos method.

The primary motivation for using SVD in non-Hermitian systems is to avoid the technical difficulties associated with unfolding complex spectra. Any generic non-Hermitian Hamiltonian $H$ can be decomposed as $H = U \Sigma V^{\dagger}$, where $U$ and $V$ are unitary matrices and $\Sigma=\text{diag} \left(\sigma_1,\, \cdots,\, \sigma_d \right)$ contains the singular values. These singular values are the square roots of the eigenvalues of the positive-semidefinite matrix $H^{\dagger} H$.

To study the dynamics and spectral statistics of singular values, one typically employs the Hermitization method. This involves constructing an effective Hermitian Hamiltonian $H_{\text{eff}} = \sqrt{H^\dagger H}$, whose eigenvalues are precisely the singular values $\{\sigma_i\}$ of $H$. By transforming the problem into an effective Hermitian one, one can apply standard diagnostics.

Despite the mathematical convenience of $H_{\text{eff}}$, recent investigations into the non-Hermitian SYK model~\cite{Baggioli:2025ohh} reveal that SVD-based measures can be misleading. Specifically, the SVD approach suffers from two fundamental failures.

The singular value spacing distributions, $p(s_\sigma)$, are expected to mirror the spectral statistics of Hermitian quantum systems. Specifically, integrable regimes follow one-dimensional Poisson statistics, while chaotic regimes are characterized by GUE distributions
\begin{align}\label{sigmaStatistics}
   p(s_\sigma) = e^{-s_\sigma}\,, \qquad 
   p(s_\sigma) = \frac{32}{\pi^2} \, s_\sigma^2 \, e^{-\frac{4}{\pi}s_\sigma^2}\,.
\end{align}
To further characterize the singular spectrum, we employ the singular spacing ratio, $\langle r_\sigma \rangle$, defined as
\begin{align}\label{}
   \langle r_\sigma \rangle = \text{Mean} \left[ \frac{\text{min}\left(s_n,\,s_{n+1}\right)}{\text{max}\left(s_n,\,s_{n+1}\right)} \right] \,, \quad s_n = \sigma_{n+1} - \sigma_{n} \,,
\end{align}
and the singular spectral form factor (SFF)
\begin{align}\label{sigmaSFF}
\begin{split}
\text{SFF}_\sigma = \frac{|Z(t)|^2}{|Z(0)|^2}\,, \qquad Z(t) = \text{Tr} \left[e^{i t H_{\text{eff}}}\right] \,.
\end{split}
\end{align}

\begin{figure}[t!]
 \centering
     {\includegraphics[width=4.2cm]{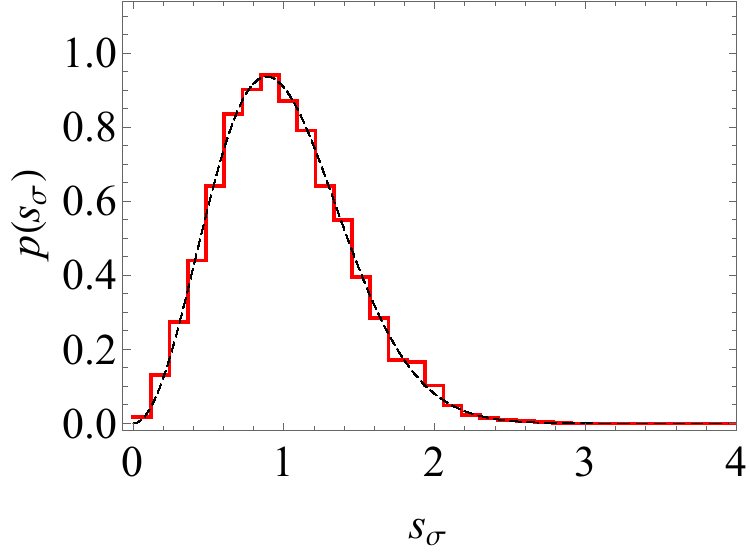}}
     {\includegraphics[width=4.2cm]{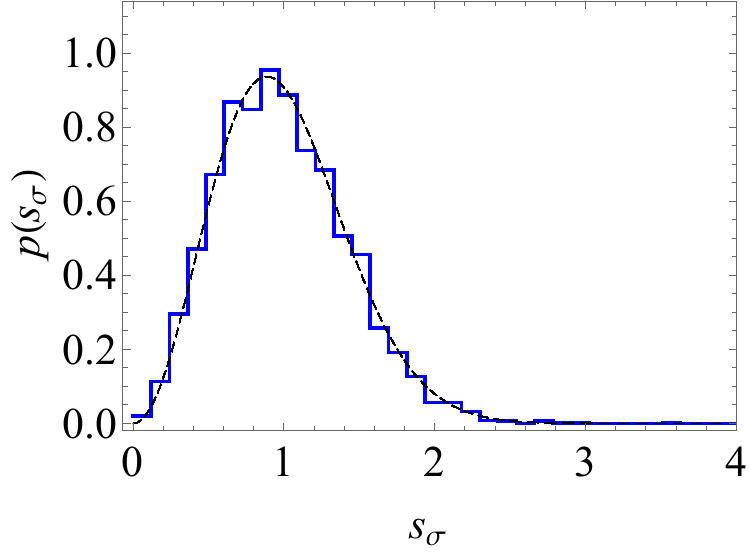}}
     
     {\includegraphics[width=4.2cm]{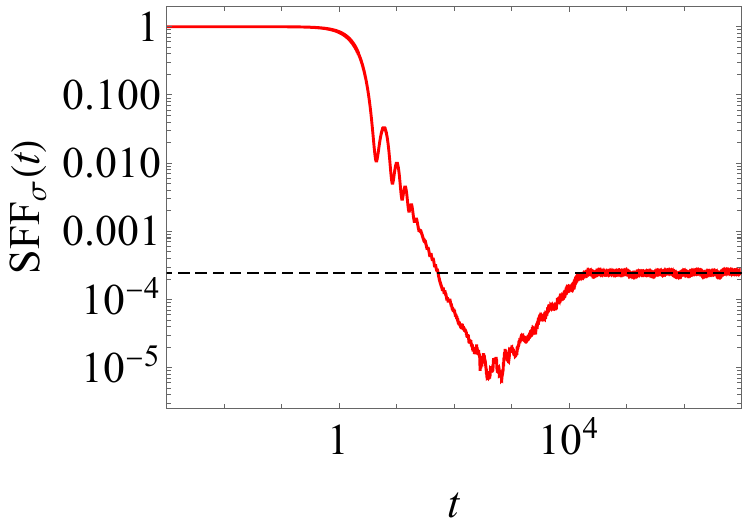}}
     {\includegraphics[width=4.2cm]{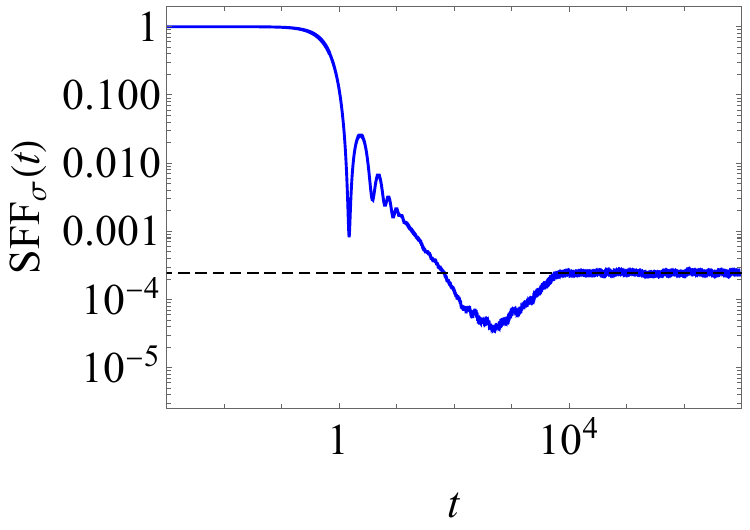}}   
\caption{Top panels: Singular spacing distributions $p(s)$ for $q=4$ (red) and $q=2$ (blue). Bottom panels: Singular spectral form factor SFF$_{\sigma}$, where dashed lines indicate the theoretical plateau at $1/D_{\mathcal H}$ with the system size $D_{\mathcal H}=D_{\rm K}$. A simple moving average has been applied to the SFF$_{\sigma}$ data to suppress short-time fluctuations and highlight the long-term dynamical trend.}
\label{SingularStafig}
\end{figure}

\begin{figure}[t!]
 \centering
     {\includegraphics[width=4.2cm]{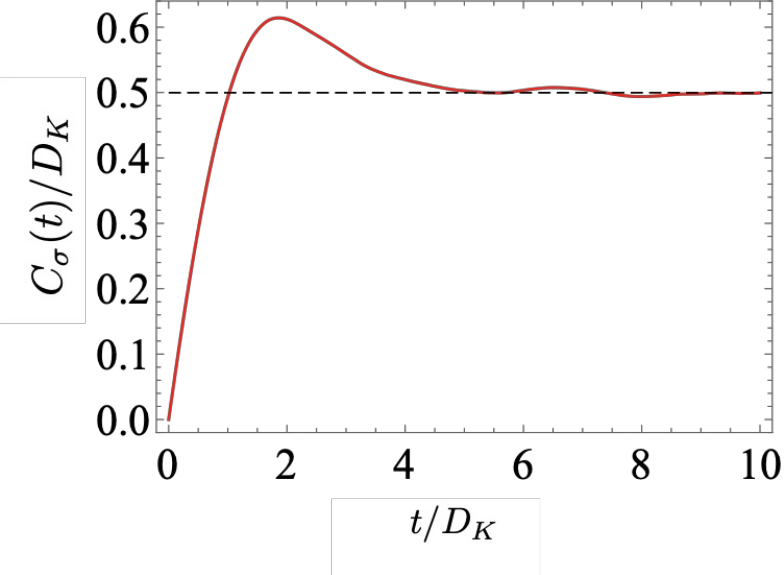}}
     {\includegraphics[width=4.2cm]{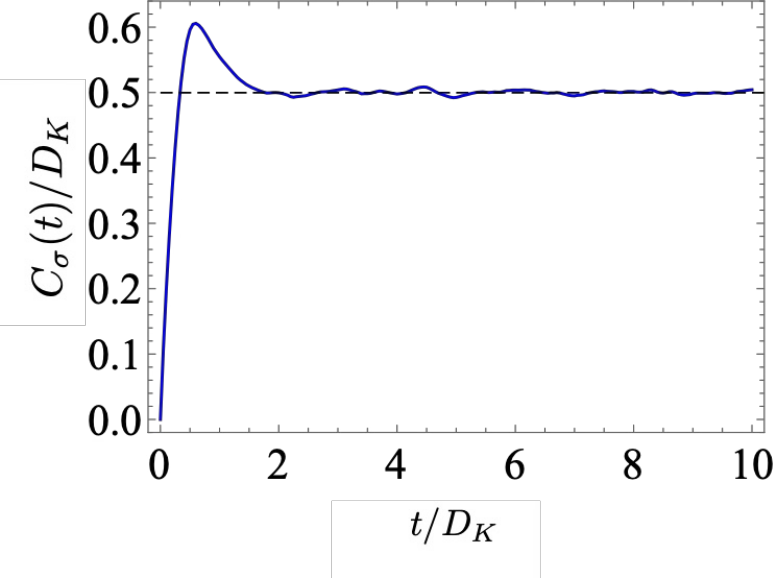}}
\caption{SVD-based Krylov complexity $C_{\sigma}(t)$ for $q=4$ (left) and $q=2$ (right). Both regimes exhibit the characteristic peak typically associated with chaotic systems, demonstrating that SVD Krylov complexity fails to distinguish between integrability and chaos. Dashed lines indicate the saturation value, $C_\sigma(t=\infty) = D_{\rm K}/2$.} \label{SingularKrylovfig}
\end{figure}

In the nHSYK model, complex spectral statistics (CSR) clearly distinguish between the chaotic ($q=4$) and integrable ($q=2$) regimes as shown in Fig. \ref{NHSYK}. However, as shown in Fig. \ref{SingularStafig}, singular value statistics ($\langle r_\sigma \rangle$ and SFF$_\sigma$) identify both models as chaotic, i.e., both $q=4$ and $q=2$ cases exhibit a GUE-like level repulsion (the right in Eq. \eqref{sigmaStatistics}) and a characteristic ramp in the SFF$_\sigma$. This ``false positive" for chaos is also reflected in the Krylov complexity $C_{\sigma}(t)$: see Fig. \ref{SingularKrylovfig}. Both the chaotic and integrable models show a pronounced peak, which is a conjectured hallmark of chaos. This implies that SVD-based complexity tracks the properties of $H^{\dagger}H$ rather than the intrinsic chaotic dynamics of the original non-Hermitian evolution.

Although SVD provides numerical convenience by preserving real and positive quantities, it remains a ``blind" probe that discards the chaos-integrable transition and spectral correlations required to identify true quantum chaos. In contrast, the bi-Lanczos approach circumvents these limitations, providing a robust Krylov complexity signature that is fully consistent with complex spectral statistics.

%
\section{Holographic applications}\label{SECVI}

The connection between quantum information theory and gravity has become one of the most productive directions in modern theoretical physics; see Refs.~\cite{Bousso:2022ntt,Faulkner:2022mlp,Chen:2021lnq} for comprehensive reviews. Within AdS/CFT, black holes provide the canonical setting in which many-body chaos, information scrambling, and emergent geometry meet. Entanglement entropy, through the celebrated Ryu-Takayanagi prescription~\cite{Ryu:2006bv,Nishioka:2009un}, has led to precise notions of spacetime emergence~\cite{Lashkari:2013koa,Faulkner:2013ica,Swingle:2014uza,Haehl:2017sot,Faulkner:2017tkh,Agon:2020mvu,Agon:2021tia} and bulk reconstruction~\cite{Balasubramanian:2013rqa,Balasubramanian:2013lsa,Myers:2014jia,Headrick:2014eia,Czech:2014ppa,Espindola:2017jil,Espindola:2018ozt,Balasubramanian:2018uus,Bao:2019bib,Ahn:2024jkk}.

However, entanglement entropy alone does not capture all aspects of gravitational physics, most notably the late-time growth of black hole interiors~\cite{Susskind:2014moa}. This limitation motivated the formulation of several holographic complexity proposals, including ``complexity=volume''~\cite{Stanford:2014jda}, ``complexity=action''~\cite{Brown:2015bva}, and their ``complexity=anything'' generalizations~\cite{Belin:2021bga,Belin:2022xmt}. These proposals relate boundary notions of complexity to gravitational observables that encode the growth of the black hole interior. The complexity=anything framework, in particular, codifies part of the ambiguities familiar from circuit and Nielsen complexity, where different choices of gates, cost functions, and tolerance thresholds can lead to distinct, but physically motivated notions of complexity. Notably, these proposals have also led to complexity-based perspectives of spacetime emergence~\cite{Czech:2017ryf,Pedraza:2021mkh,Pedraza:2021fgp,Pedraza:2022dqi,Carrasco:2023fcj} and bulk reconstruction~\cite{Hashimoto:2021umd,Xu:2023eof}, in parallel with the entanglement-based formulations.

Krylov complexity provides a new and more microscopic entry in this holographic dictionary. Because it is defined directly from Hamiltonian evolution and the spread of a state in Krylov space, it avoids some of the extrinsic choices that enter circuit or Nielsen complexity, while offering a concrete bridge between spectral data, real-time information spreading, and bulk geometry. Equivalently, when phrased in the language of Nielsen geometry, Krylov complexity amounts to a distinguished choice of gates and metric adapted to the Krylov basis~\cite{Aguilar-Gutierrez:2023nyk,Craps:2025kub}. In this section, we review three related holographic applications. First, we discuss the proposal that, for thermofield double states, Krylov complexity is dual to wormhole length or interior volume, as realized concretely in lower-dimensional holographic settings relating JT gravity and the SYK model. Second, we explain how the rate of Krylov complexity can be related to the proper radial momentum of infalling matter, particularly for states dual to local quenches that create one-particle excitations in the bulk. Third, we review brick-wall models of black holes in AdS/CFT, where stretched-horizon fluctuations are analyzed using spectral diagnostics and Krylov complexity.

%
\subsection{Krylov complexity as wormhole growth}

A central claim made precise in recent years is that, for TFD states in certain holographic settings, Krylov state complexity can be identified with a measure of black hole interior growth. More precisely, it has been shown that the spreading of the boundary state in Krylov space is dual to the growth of the volume inside the Einstein-Rosen bridge connecting the two asymptotic boundaries of the gravitational dual. Krylov complexity therefore provides a microscopic realization of the intuition behind the complexity=volume conjecture~\cite{Rabinovici:2023yex,Heller:2024ldz,Balasubramanian:2024lqk}. The cleanest realization of this relation appears in lower-dimensional models of holography, especially the SYK model and its double-scaled limit \cite{Lin:2022rbf}, together with their gravitational descriptions in JT gravity and sine-dilaton gravity.

In double-scaled SYK (DSSYK), one takes $N,q\to\infty$ while keeping the double-scaling parameter $\lambda\sim q^2/N$ fixed, up to convention-dependent numerical factors. It is often convenient to introduce $p=e^{-\lambda}$. In this limit, the theory admits an exact chord-diagram description, and the Hamiltonian acts on a chord Hilbert space whose basis states $|n\rangle$ are labeled by the chord number $n$~\cite{Berkooz:2024lgq}. The key observation is that, in the TFD sector relevant for the two-sided geometry, this chord basis coincides with the Krylov basis generated by the Lanczos algorithm. Repeated action of the Hamiltonian increases the chord number, so the Krylov index acquires a direct microscopic interpretation.

This identification leads to a direct relation between Krylov complexity and wormhole volume, or length in two-dimensional gravity. For TFD states, the average chord number $\langle n\rangle$ is proportional to the expectation value of the geodesic length $L$ of the Einstein-Rosen bridge. Schematically, one finds $\langle L\rangle \sim 2|\log p|,\langle n\rangle =2|\log p|,C_{\rm DSSYK}(t)$, where $C_{\rm DSSYK}(t)$ is the Krylov complexity in the chord/Krylov basis~\cite{Rabinovici:2023yex,Heller:2024ldz,Balasubramanian:2024lqk}. Thus, the average position of the wave packet on the Krylov chain becomes a bulk measure of wormhole growth.

Historically, this relation was first made sharp for the infinite-temperature TFD state in the triple-scaled SYK/JT setting~\cite{Rabinovici:2023yex}. More recent developments have extended the quantitative correspondence to finite temperature and to the full quantum regime on the gravity side at disk level, using sine-dilaton gravity as a deformation of JT gravity~\cite{Heller:2024ldz}. Nonperturbative extensions further show that the identification between Einstein-Rosen bridge size and spread complexity can be continued to the complete finite-dimensional Hilbert space, where the wormhole size eventually saturates at late times~\cite{Balasubramanian:2024lqk}. In these constructions, the Euclidean preparation of the TFD, or equivalently the two-sided Hartle-Hawking state in the bulk, plays an important role in defining the state whose Krylov complexity is compared with the wormhole geometry. Beyond these two-dimensional settings, however, the correspondence appears to be more subtle: recent analyses of the BTZ black hole find that the spread complexity of the TFD state does not exhibit the expected late-time linear growth of the wormhole volume~\cite{Bhattacharya:2026ybq,Bhattacharyya:2026zpu}. Interestingly, an operator-growth version of Krylov complexity does reproduce this linear behavior~\cite{Bhattacharyya:2026zpu}, extending earlier connections between operator Krylov complexity and bulk geometry in lower-dimensional models~\cite{Kar:2021nbm,Ambrosini:2024sre}.

This picture gives a microscopic account of the dynamics of black hole interiors. At the classical level, the growth of the Einstein-Rosen bridge corresponds to the regime in which the Krylov wave packet moves steadily toward larger Krylov index. At the same time, the discreteness of the chord number suggests how a smooth geometric length can emerge from an underlying quantum-mechanical counting problem. This idea has also been developed from a complementary continuum perspective: in suitable limits, the discrete Krylov-chain dynamics can be mapped to an effective radial wave equation in a near-horizon AdS$_2$ throat, with the Breitenlohner-Freedman bound emerging as a consistency condition for the dual description~\cite{Jeong:2026iac}. Related work has interpreted the Krylov-chain direction, in the context of operator growth, as an emergent radial direction of a black hole scattering problem~\cite{Dodelson:2025rng}.

A further advantage of the Krylov description is that it naturally incorporates late-time saturation. Classical wormholes grow for parametrically long times, but a finite-dimensional boundary Hilbert space cannot support indefinite complexity growth. Nonperturbative extensions of spread complexity make this tension explicit: the auxiliary chord basis agrees with the physical Krylov basis only in an initial regime, while the complete finite-dimensional Krylov basis modifies the very late-time dynamics and leads to saturation of the Einstein-Rosen bridge size~\cite{Balasubramanian:2024lqk}. In this nonperturbative regime, the dynamics can display white-hole-like behavior, with the bridge shrinking from its maximal size toward a late-time plateau. For a thermofield-double state, the nonperturbative late-time saturation of spread complexity sets the finite dimensionality of the black hole Hilbert space, with the saturation plateau determined by the Bekenstein-Hawking entropy~\cite{Balasubramanian:2026azk}.

Beyond the first moment of the Krylov distribution, higher-order Krylov observables have been proposed to encode finer information about bulk geometry. In the DSSYK setting, higher-order Krylov complexities have been argued to capture connected bulk contributions encoded by replica wormholes, while logarithmic Krylov complexity probes the replica saddle structure~\cite{Fu:2025kkh}. The same work also proposes a bulk interpretation of Krylov entropy in a third-quantized description with baby universes: after tracing out the baby-universe sector, the boundary Krylov entropy is identified with the von Neumann entropy of the reduced parent-geometry density matrix, thereby quantifying information flow into baby-universe degrees of freedom. These developments suggest that Krylov-space observables may probe not only semiclassical wormhole growth, but also nonperturbative ensemble-averaged structures of quantum gravity.

%
\subsection{Krylov complexity rate as proper momentum}

The wormhole-growth proposal identifies the value of Krylov complexity for TFD states with a bulk quantity measuring the size of the black hole interior. A closely related proposal concerns its rate of change for states created by localized perturbations, which are represented by infalling matter in the bulk. This setup is commonly referred to as a local quench~\cite{Nozaki:2013wia,Agon:2020fqs}. The complexity=momentum idea states that, for such holographic states, the rate of complexity growth in the boundary theory is dual to the proper radial momentum of the infalling excitation~\cite{Susskind:2020gnl}. This statement was made precise for Krylov state complexity using local operator excitations dual to one-particle states in the bulk~\cite{Caputa:2024sux}. It provides a dynamical version of the Krylov/gravity dictionary: spreading of the state along the Krylov chain is mapped to radial motion in the dual bulk geometry.

The basic intuition originates from the holographic description of operator growth. An initially localized boundary perturbation becomes increasingly complex under time evolution, while its bulk dual moves radially into the interior and, in a thermal state, falls toward the black hole horizon. The complexity=momentum proposal makes this relation quantitative at the level of state complexity: schematically,
\begin{equation}
\dot C_{\rm K}(t)\sim -P_{\rho}(t),
\end{equation}
where $P_{\rho}$ is the momentum conjugate to an appropriate proper radial coordinate $\rho$, up to a normalization. The use of proper distance is essential, since a coordinate-dependent canonical radial momentum does not in general possess an invariant boundary interpretation.

The correspondence was made particularly sharp in AdS$_3$/CFT$_2$ for local quenches, namely states created by local primary operators~\cite{Caputa:2024sux}. In this setting, the survival amplitude entering the Lanczos construction is fixed by CFT two-point functions and admits a bulk description in terms of a massive particle propagating in AdS$_3$. The resulting spread-complexity rate precisely matches the proper radial momentum after the appropriate identification of boundary and bulk scales. At finite temperature, the same construction maps the hyperbolic growth controlled by $\beta$ to radial infall in the BTZ geometry. The simplicity of this correspondence is closely tied to the underlying $SL(2,\mathbb{R})$ structure, which allows the Krylov dynamics to close within a finite-dimensional algebraic framework.

The proposal has subsequently been explored in more general geometries and for a broader range of bulk probes. Within the proper-momentum prescription, general AdS black holes have been argued to reproduce the universal early- and late-time behavior of Krylov complexity, while BTZ provides a special case in which an exact matching with the CFT$_2$ result can be obtained~\cite{Fan:2024iop}. Related work has clarified that the relevant bulk quantity is the radial momentum measured by a stationary observer, extending the correspondence to massless probes~\cite{He:2024pox}. The framework has also been generalized beyond point particles to extended objects, including branes and strings~\cite{Chatzis:2026ekd,Chatzis:2026oou}. Together, these developments indicate that the momentum prescription applies well beyond the original setting.

More recently, this framework has been applied to holographic theories with nontrivial infrared dynamics. In confining geometries, the finite radial extent of the bulk leads to recurrent probe motion and oscillatory Krylov dynamics~\cite{Fatemiabhari:2025usn}. A systematic analysis across several top-down backgrounds found this behavior to be robust, with the characteristic oscillation frequency controlled by the confinement scale and the amplitude depending on both ultraviolet and infrared scales~\cite{Fatemiabhari:2026goj}. This provides a holographic counterpart of oscillatory Krylov dynamics observed in confining theories \cite{Jiang:2025wpj}.

The sensitivity of Krylov dynamics to infrared physics was further explored in the $\mathds{B}_8$ family of gauge theories, which exhibits walking RG flows and interpolates between an infrared conformal fixed point and a confining limit~\cite{Nunez:2026vhw}. Using D0-branes as bulk probes, the corresponding Krylov dynamics was found to exhibit oscillations whose period becomes parametrically large as the flow approaches the intermediate conformal fixed point, providing a characteristic signature of walking. Importantly, these oscillations persist away from the confining limit, where external charges are screened, showing that their presence is more generally associated with the smooth infrared cap and discrete massive spectrum rather than with confinement alone. Their detailed behavior, however, remains particularly sensitive to the approach to genuine confinement.

An important qualification to the momentum-complexity correspondence beyond the semiclassical regime arises in double-scaled SYK and its sine-dilaton gravity dual~\cite{Fu:2025kkh}. In this setting, the identification of the Krylov-complexity growth rate with the proper momentum of an infalling particle was found to hold only at early times, or equivalently in the low-energy Schwarzian regime. Away from this limit, the full DSSYK dynamics no longer closes on the $SL(2,\mathbb{R})$ algebra, and the simple proper-momentum relation receives corrections. These deviations signal sensitivity to ultraviolet physics and to quantum-gravitational effects beyond the leading semiclassical approximation, potentially including finite-$G_N$ and nonperturbative ensemble corrections~\cite{Fu:2025kkh}. The proper-momentum relation should therefore be viewed as a regime-dependent correspondence rather than an exact universal identity valid at arbitrary energies and times.

Taken together, these developments suggest a refined holographic dictionary with a well-defined regime of applicability. In settings with sufficient symmetry and within the appropriate semiclassical regime, the rate of Krylov state complexity can be identified with the proper radial momentum of a bulk excitation. More generally, the momentum prescription should be viewed as an effective geometric description whose range of validity depends on the state, the bulk geometry, and the energy and time scales being probed; deviations from it may themselves encode ultraviolet or quantum-gravitational corrections. At the same time, its extensions to confining and walking theories show that Krylov dynamics can probe detailed infrared structure: while the complexity measures how far the state has spread along the Krylov chain, its growth rate and characteristic oscillation scales encode the motion of bulk probes and the hierarchy of scales encountered along the holographic RG flow.

%
\subsection{Spectral analysis with stretched horizons}

The previous two subsections focused on direct geometric interpretations of Krylov complexity. A complementary route is to probe horizon physics through spectra. The brick-wall model~\cite{tHooft:1984kcu} provides a simple way to obtain a discrete spectrum by placing a stretched horizon, usually implemented as a Dirichlet wall, a small proper distance from the true horizon~\cite{Susskind:1993if}. Originally introduced as a regulator for black-hole entropy, the brick-wall model has recently been repurposed as an effective spectral model for probing quantum-chaotic features of horizons.

In this framework, scalar or fermionic probe fields are quantized in a gravitational background with boundary conditions imposed at the stretched horizon. The resulting normal modes define a discrete probe-field spectrum that can be analyzed using standard tools of quantum chaos: level-spacing distributions, spectral form factors, and Krylov state complexity~\cite{Das:2022evy,Das:2023ulz,Das:2023xjr,Jeong:2024jjn,Jeong:2025jyx,Begines:2026fnx}. The brick wall should not be viewed as a complete microscopic theory of black-hole microstates, but rather as a tractable effective model for studying the spectral structure associated with horizons.

For BTZ black holes, the brick-wall normal modes are closely related to a logarithmic spectrum. Remarkably, even without fluctuations, such a deterministic spectrum can exhibit a linear ramp in the spectral form factor despite the absence of conventional level repulsion~\cite{Das:2023yfj}, illustrating that long-range spectral correlations need not be accompanied by Wigner-Dyson statistics. Introducing Gaussian-distributed fluctuations of the stretched-horizon boundary conditions further leads to clear random-matrix-like signatures across symmetry classes~\cite{Jeong:2024jjn}. Depending on the strength of the fluctuations, the level-spacing distribution can display Wigner-Dyson-like behavior, the spectral form factor develops a dip-ramp-plateau structure, and Krylov complexity shows the characteristic peak before saturation. Importantly, these diagnostics are not redundant: Krylov complexity can remain sensitive to long-range spectral rigidity even when conventional level repulsion is weakened or obscured.

The brick-wall construction has also been extended to higher-dimensional hyperbolic AdS black holes~\cite{Jeong:2025jyx}. In this extended setting, the logarithmic structure characteristic of BTZ is deformed toward a power-law spectrum in higher dimensions, while signatures of chaos persist for moderate dimensions. The persistence of the Krylov peak across these deformations suggests that Krylov complexity is sensitive to robust spectral correlations rather than to a particular form of the spectrum. In the parametrically large-dimensional limit, however, the spectrum becomes highly degenerate, suppressing the usual chaotic diagnostics.

A further test comes from de Sitter and Schwarzschild-de Sitter geometries~\cite{Begines:2026fnx}, where complementary signatures of horizon chaos have also been identified through pole-skipping and shock-wave analyses~\cite{Ahn:2025exp}. In pure de Sitter space, the brick-wall spectrum can exhibit long-range chaotic correlations even when the level-spacing distribution does not take the conventional Wigner-Dyson form. In Schwarzschild-de Sitter space, the presence of both a black-hole and a cosmological horizon leads, in a WKB regime with suppressed tunneling, to two approximately independent near-horizon sectors. The full spectrum is then a superposition of subsequences, which can naturally obscure level repulsion and produce a nonzero value of the level-spacing distribution at $s=0$. Nevertheless, for sufficiently small stretched-horizon fluctuations, the spectral form factor retains an approximately linear ramp and Krylov complexity a pronounced peak, revealing long-range correlations that are not apparent from nearest-neighbor statistics alone.

Taken together, these brick-wall studies show how stretched-horizon spectra provide a concrete arena in which chaos-related features of horizons can be studied using the same tools developed for ordinary quantum systems. They also reinforce a recurring lesson of this review: level statistics, spectral form factors, and Krylov complexity probe complementary aspects of quantum dynamics and should be interpreted together. In particular, the absence of strict Wigner-Dyson level repulsion does not necessarily imply the absence of long-range spectral correlations, which may remain visible in the spectral form factor and Krylov dynamics.

More broadly, the applications reviewed in this section illustrate several complementary roles for Krylov complexity in holography. In suitable regimes, its value can encode wormhole length, its rate can be related to the proper radial momentum of bulk probes, while Krylov dynamics constructed from stretched-horizon spectra can probe spectral correlations associated with horizons. Together, these results suggest that Krylov complexity provides a useful bridge between microscopic quantum dynamics, information spreading, and emergent gravitational physics.

%
\section{Conclusions and outlook}\label{SECVII}
Originally developed as efficient numerical methods for eigenvalue problems, Krylov subspace methods have become a powerful language for describing real-time quantum evolution. This manuscript has provided a pedagogical review of Krylov complexity and its use as a diagnostic of quantum dynamics, with detailed explanations and workable examples aimed at researchers entering the field. More specifically, we have focused on Krylov state complexity, also known as spread complexity, as a diagnostic of quantum chaos and information spreading. By quantifying the average position of a time-evolving state along the Krylov chain, this quantity translates spectral information into a dynamical measure of state spreading. The construction is grounded in the optimal-basis theorem, which identifies the Krylov basis as precisely the basis that minimizes the spread of the state within the relevant class of ordered orthonormal bases.

We examined the behavior of Krylov complexity across a broad range of Hermitian and non-Hermitian systems, from generic random matrix ensembles to specific models, including quantum billiards, quantum spin chains, and variants of the Sachdev-Ye-Kitaev model. Across these examples, Krylov complexity offers a robust way to diagnose chaos dynamically, while remaining closely tied to standard spectral statistics. In Hermitian systems, its dynamical signatures are associated with the defining properties of Wigner-Dyson statistics, such as level repulsion and spectral rigidity; in non-Hermitian systems, they are naturally benchmarked against complex spectral statistics and Ginibre-type universality classes.

A central lesson is that the approach to equilibrium in Krylov space contains important dynamical information. At late times, Krylov complexity generically approaches the midpoint of the Krylov chain, $C(t)\approx D_{\rm K}/2$, where $D_{\rm K}$ denotes the Krylov dimension. However, the route to this value differs sharply between chaotic and integrable models. In the chaotic systems reviewed here, the complexity displays a characteristic overshoot or peak, approaching equilibrium from above, $C(t)\rightarrow (D_{\rm K}/2)^+$. By contrast, integrable systems typically show more constrained spreading and instead approach equilibrium from below, $C(t)\rightarrow (D_{\rm K}/2)^-$. This peak structure therefore provides a useful dynamical marker of chaos, reflecting how chaotic evolution pushes the wave packet deeper into Krylov space before relaxation. Nevertheless, a fully rigorous, model-independent explanation of this behavior remains an important open problem.

Taken together, these results support the view that Krylov complexity provides a coherent dynamical narrative of quantum chaos, from the initial quadratic delocalization to the final ergodic saturation. The characteristic growth-peak-slope-plateau structure appears to be more than a numerical curiosity: it holds across a rich variety of models, ranging from single-particle systems to many-body systems with local and all-to-all interactions. This behavior reflects the interplay between spectral statistics, scrambling, and the effective exploration of Hilbert space. Building on the Lanczos algorithm and its extensions, Krylov methods thus offer a practical and conceptually sharp framework for diagnosing the transition from integrability to chaos, testing chaos indicators beyond traditional Hermitian systems, and probing holographic many-body regimes. Krylov complexity therefore serves as a bridge between real-time dynamics and spectral universality, making it a valuable tool for studying information spreading and quantum dynamics.

The study of Krylov complexity has sharpened our understanding of how quantum information spreads under time evolution, but several frontiers remain open. A pressing direction is to extend the formalism beyond standard time-independent Hamiltonian evolution, including multiseed constructions~\cite{Craps:2024suj}, time-dependent and quenched dynamics~\cite{Takahashi:2024hex}, unitary circuits~\cite{Suchsland:2023cmb}, and Floquet systems~\cite{Nizami:2023aa,Nizami:2024ltk,Yates:2021asz}. In these settings, the Krylov chain is no longer simply a fixed one-dimensional lattice generated by a single static Hamiltonian. For example, the effective hopping amplitudes, basis states, or relevant Krylov sectors can become time-dependent or seed-dependent. Understanding this generalized evolution as propagation or spread on an inhomogeneous, time-dependent Krylov lattice may therefore provide a natural framework for Krylov complexity in driven and nonequilibrium quantum systems.

Krylov methods are also becoming increasingly relevant for quantum algorithms~\cite{Cortes:2021esg,Yoshioka_2025}. Their ability to approximate dynamics within a low-dimensional subspace makes them especially attractive for near-term quantum devices, where full Hilbert-space reconstruction is infeasible. At the same time, experimental probes of quantum chaos, such as spectral form factors~\cite{Vasilyev:2020yxg,Joshi:2021kzb,Dong:2024yaf} and out-of-time-order correlators~\cite{Swingle:2016var,Li:2016xhw,Garttner:2016mqj,Green:2021fcl}, continue to mature. These developments make it increasingly timely to design protocols for extracting Krylov complexity from experimentally accessible data, such as survival amplitudes, correlation functions, or moments of the Hamiltonian, within the capabilities of near-term quantum hardware.

Another major challenge is the transition from quantum-mechanical models to quantum field theory (QFT), where the Hilbert space is infinite-dimensional, locality and UV regularization become essential, and the construction of Krylov bases, Lanczos coefficients, and complexity measures is more subtle. Important progress has already been made in several directions. For example, Krylov complexity has been studied in two-dimensional CFTs, free-field theories, and lattice-regularized QFTs~\cite{Dymarsky:2021bjq,Avdoshkin:2022xuw}; in free and perturbatively interacting massive scalar field theories at finite temperature, where masses, interactions, and UV cutoffs modify the Lanczos coefficients and the resulting Krylov growth~\cite{Camargo:2022rnt}; and in large-$c$ two-dimensional CFTs, where Krylov complexity can depend sensitively on the state and distinguish regimes associated with the Hawking-Page transition in the dual geometry~\cite{Kundu:2023hbk}. Complementary approaches based on symmetry and generalized coherent states have provided a geometric interpretation of Krylov growth~\cite{Caputa:2021ori}, while deformed CFTs and $SL(2,\mathbb{R})$-deformed Hamiltonians provide controlled settings in which to study how integrable deformations, heating, and non-heating regimes affect Lanczos coefficients and Krylov complexity~\cite{Chattopadhyay:2024pdj,Malvimat:2024vhr}. More broadly, a general, regulator-controlled field-theoretic formulation for interacting higher-dimensional QFTs remains open.

Ultimately, Krylov subspace methods have evolved from numerical techniques into a sharp language for understanding and characterizing quantum dynamics. Krylov complexity distills many-body evolution into a simple question: how far does a quantum state spread along its dynamically generated Krylov chain? This question links spectral statistics to real-time information spreading, extends naturally to non-Hermitian dynamics, and reaches into the realm of holography and black hole physics. Krylov complexity is thus not merely another chaos diagnostic; it is a framework for understanding how quantum systems explore Hilbert space, how information scrambles, and how complexity may acquire geometric meaning. Its continued development promises to deepen our understanding of chaos, thermalization, and quantum information across many-body physics, quantum field theory, and quantum gravity.

\vspace{5pt}

%
\textit{Acknowledgments---}
HSJ is supported by an appointment to the JRG Program at the APCTP through the Science and Technology Promotion Fund and Lottery Fund of the Korean Government. HSJ is also supported by the Korean Local Governments -- Gyeongsangbuk-do Province and Pohang City.
JFP is supported by the ‘Atracción de Talento’ program of the Comunidad de Madrid under grant 2020-T1/TIC-20495, the Spanish Agencia Estatal de Investigación through grants CEX2025-001574-S, PID2021-123017NB-I00, and PID2024-156043NB-I00, funded by MCIN/AEI/10.13039/501100011033, and ERDF, EU. The research presented in this publication falls within the research line Strings and Quantum Gravity.

%
\bibliographystyle{apsrmp4-2}
\bibliography{Refs}

\end{document}